\documentclass[12pt]{article}
\usepackage{amsthm,amsfonts
,
}
\title{Uniform Lieb-Thirring inequality for the three dimensional
Pauli operator with a strong non-homogeneous
magnetic field}
\author{L\'aszl\'o Erd\H os
 \thanks{Partially supported by NSF grants DMS-9970323, DMS-0200235,
and by the Erwin Schr\"odinger Institute, Vienna}
\\ School of Mathematics, GeorgiaTech and MaPhySto \\
and \\
Jan Philip Solovej \thanks{Work partially supported
     by the Danish Natural Science Research Council, by MaPhySto -- A network in
Mathematical
     Physics and Stochastics, funded by a
     grant from The Danish National Research Foundation, and by the EU
research network HPRN-CT-2002-00277} 
\\ Department of Mathematics, University of Copenhagen}

\date{Feb. 20, 2004}

\newtheorem{theorem}{Theorem}[section]
\newtheorem{proposition}[theorem]{Proposition}
\newtheorem{corollary}[theorem]{Corollary}
\newtheorem{lemma}[theorem]{Lemma}
\newtheorem{definition}[theorem]{Definition}

\newcommand{\bbp}{{\bf p}}
\newcommand{\rd}{{\rm d}}
\newcommand{\be}{\begin{equation}}
\newcommand{\ee}{\end{equation}}
\newcommand{\bey}{\begin{eqnarray}}
\newcommand{\eey}{\end{eqnarray}}
\newcommand{\beys}{\begin{eqnarray*}}
\newcommand{\eeys}{\end{eqnarray*}}

\newcommand{\bB}{{\bf B}}
\newcommand{\bA}{{\bf A}}
\newcommand{\bZ}{{\bf Z}}
\newcommand{\bp}{-i\nabla}

\newcommand{\bw}{{\bf w}}
\newcommand{\bsigma}{\mbox{\boldmath $\sigma$}}
\newcommand{\bomega}{\mbox{\boldmath $\omega$}}

\newcommand{\cU}{{\cal U}}

\newcommand{\bR}{{\bf R}}
\newcommand{\bC}{{\bf C}}

\newcommand{\bn}{{\bf n}}
\newcommand{\bm}{{\bf m}}
\newcommand{\bv}{{\bf v}}
\newcommand{\bN}{{\bf N}}

\newcommand{\ov}{\overline}

\newcommand{\e}{\varepsilon}
\newcommand{\tD}{\widetilde D}
\newcommand{\hD}{\widehat D}

\renewcommand{\a}{\alpha}
\renewcommand{\t}{\theta}
\newcommand{\Tr}{{\rm Tr}}
\newcommand{\tr}{{\rm tr}}
\newcommand{\sfrac}[2]{{\textstyle \frac{#1}{#2}}}
\newcommand{\wh}{\widehat}
\newcommand{\wt}{\widetilde}

\newcommand{\bPi}{\mbox{\boldmath $\Pi$}}

\newcommand{\cO}{{\cal O}}
\newcommand{\cY}{{\cal Y}}
\newcommand{\cS}{{\cal S}}

\newcommand{\cF}{{\cal F}}
\newcommand{\cA}{{\cal A}}
\newcommand{\cB}{{\cal B}}
\newcommand{\cE}{{\cal E}}
\newcommand{\cP}{{\cal P}}
\newcommand{\cD}{{\cal D}}
\newcommand{\cV}{{\cal V}}
\newcommand{\cW}{{\cal W}}
\newcommand{\cK}{{\cal K}}

\newcommand{\cM}{{\cal M}}
\newcommand{\cN}{{\cal N}}
\newcommand{\cR}{{\cal R}}
\newcommand{\cH}{{\cal H}}

\newcommand{\D}{{\cal D}}
\newcommand{\bD}{{\bf D}}

\newcommand{\Om}{\Omega}

\renewcommand{\t}{\theta}

\newcommand{\tdel}{\wt\Delta}
\newcommand{\tcD}{\wt \D}
\newcommand{\tPi}{\wt\Pi}
\newcommand{\tbPi}{\wt \bPi}
\newcommand{\tcK}{\wt\cK}
\newcommand{\tcM}{\wt\cM}
\newcommand{\tcW}{\wt\cW}
\newcommand{\tcU}{\wt\cU}

\newcommand{\res}{\frac{1}{\cD^2+P}}
\newcommand{\tres}{\frac{1}{\tcD^2 + P}}
\newcommand{\tPijres}{\frac{\tPi_j}{\tcD^2 + P}}
\newcommand{\tdres}{\frac{\tcD}{\tcD^2 + P}}
\newcommand{\twres}{\frac{\wt\cW}{\tcD^2 + P}}
\newcommand{\tures}{\frac{\wt\cU}{\tcD^2 + P}}

\newcommand{\ttres}{\frac{\tPi_3}{\tcD^2 + P}}
\newcommand{\tpres}{\frac{\tPi_j}{\tcD^2 + P}}
\newcommand{\dres}{\frac{\D}{\D^2+P}}

\newcommand{\tri}{| \! | \! |}

\newcommand{\ad}{\Big(\cdots \Big)^*}

\begin{document}
\maketitle

\begin{abstract}
The Pauli operator describes the energy of a nonrelativistic
quantum particle  with spin $\sfrac{1}{2}$
in a magnetic field  and an external potential.
A new Lieb-Thirring type inequality on the sum of the negative
eigenvalues is presented. The main feature compared to
earlier results is that in the large field regime the present
estimate grows with the optimal (first) power
of the strength of the magnetic field. As a byproduct of the method,
we also obtain an optimal upper bound on the pointwise density of zero energy 
eigenfunctions of the Dirac operator.
The main technical tools are: 

(i) a new localization scheme
for the square of the resolvent
of a general class of second order elliptic operators;

(ii) a geometric construction of a Dirac operator
with a constant magnetic field that approximates the original Dirac
operator in a tubular neighborhood of a fixed field line.
The errors may depend on the regularity of the magnetic field 
but they are uniform in the field strength.
\end{abstract}



\medskip\noindent
{\it Running title:} Lieb-Thirring inequality for Pauli operator

\tableofcontents

\section{Introduction}\label{sec:intro}

\subsection{Notations}

Let $\bB\in C^4(\bR^3;\bR^3)$ be a magnetic field, $\mbox{div} \,\, \bB =0$,
 and $V\in L^1_{loc}(\bR^3)$ a real valued potential function.
Let $\bA:\bR^3\to\bR^3$ be a vector potential generating the 
magnetic field, i.e. $\bB=\nabla\times\bA$.
 The 3-dimensional
Pauli operator is the following operator acting
on the space of $L^2(\bR^3 ;  \bC^2)$ of spinor-valued functions:
\be
        H=H(h,  \bA, V): = [\bsigma\cdot (-ih\nabla +\bA)]^2 + V
        = (-ih\nabla +\bA)^2 + V(x) + h\bsigma\cdot \bB(x)\; ,
\label{eq:Pauli}
\ee
where $\bsigma=(\sigma^1,\sigma^2,\sigma^3)$ is the vector of the
Pauli spin matrices, i.e.,
$$
        \sigma^1=\left(\matrix{0&1\cr1&0}\right),\
        \sigma^2=\left(\matrix{0&-i\cr i&0}\right),\
        \sigma^3=\left(\matrix{1&0\cr0&-1}\right).
$$
The spectral properties of $H$ depend only on $\bB$ and $V$ and
do not depend on the specific choice of $\bA$. We shall be concerned
only with gauge invariant quantities therefore we can
always make the Poincar\'e gauge choice.
In particular, we can always assume that $\bA$ is
at least as regular as $\bB$.
The operator $H=H(h,  \bA, V)$ is defined as the Friedrichs'
extension of the corresponding quadratic form from $C_0^\infty(\bR^3; \bC^2)$.

The Pauli operator describes the motion of a
non-relativistic electron, where the electron spin is
important because of its interaction with the magnetic field.
For simplicity we have not included any physical parameters
(i.e., the electron mass, the electron charge, the speed of
light, or Planck's constant $\hbar$)
in the expressions for the operators. In place of
Planck's constant we have the semiclassical parameter
$h$ and in most of the paper we also set $h=1$.

The last identity in (\ref{eq:Pauli})
can easily be checked. If we define the three dimensional
Dirac operator
\be
        \D:= \bsigma\cdot(-ih\nabla + \bA (x))\; ,
\label{eq:dirac}
\ee
then we recognize
the last identity in (\ref{eq:Pauli}) as the Lichnerowicz' formula.

The eigenvalues of $H$ below the essential spectrum are of special interest.
They determine the possible bound states
of a non-relativistic electron subject to the magnetic field
$\bB$ and the external potential $V$. Under very general conditions
on $V$ and $\bB$ one
can show that the bottom of the essential spectrum
for the Pauli operator is at zero (see \cite{HNW}).
This is in sharp contrast to the case of the
spinless magnetic Schr\"odinger operator, $(-ih\nabla +\bA)^2 + V(x)$,
whose essential spectrum is not known in general even for
decaying potentials.

Therefore we shall restrict our attention to the negative eigenvalues, 
$e_1(H)\leq e_2(H)\leq \ldots \leq 0$ of $H$.
It is known that under very general conditions there are infinitely many
negative eigenvalues even for constant magnetic field 
\cite{Sol}, \cite{Sob-86}, however their sum is typically
finite. We recall that the sum of the eigenvalues below the
essential spectrum is equal to the ground state energy
of the noninteracting fermionic gas subject to $H$.

The sum of the negative eigenvalues, $\sum_j e_j(H)$,
has been extensively studied 
 recently. In order to find the asymptotic
behavior of the ground state energy
of a large atom with interacting electrons,
 one needs, among other things,  a semiclassical asymptotics
for $\sum_j e_j(H)$ as $h\to0$.

The semiclassical formula for the sum of the negative
eigenvalues is given as
\begin{equation}
        E_{scl}(h,  \bB, V):= - h^{-3} \int_{\bR^3} P(h |\bB(x)|,
         [V(x)]_-)\rd x
\label{ESC}\ee
with
\be
        P(B, W):= \frac{B}{3\pi^2}\left( W^{3/2} + 2\sum_{\nu =1}^{\infty}
        [2\nu B - W]_-^{3/2}\right) =
        \frac{2}{3\pi}\sum_{\nu =0}^{\infty} d_{\nu}B[2\nu B - W]_-^{3/2}
\label{press}\ee
being the pressure of the three dimensional
Landau gas ($B, W \geq 0$). Here $[x]_-=\max\{ 0, -x\}$ refers to the
negative part of $x$, $d_0 := (2\pi)^{-1}$ and $d_{\nu}
:=\pi^{-1}$ if $\nu \geq 1$. Observe that if $\|\bB\|_\infty=o(h^{-1})$
then $E_{scl}$ reduces to leading order to the standard Weyl term,
$-2(15\pi^2)^{-1}h^{-3}\int_{\bR^3}[V]_-^{5/2}$,
as $h\to0$. The main feature of the semiclassical formula is that
it behaves linearly with the field strength in the strong field
regime.  

For the proof that $\sum_j e_j(H)$ is asymptotically equal to $E_{scl}$
as $h\to0$, first one must  establish a non-asymptotic 
bound on the sum of the negative eigenvalues to control various
error terms from the non-semiclassical regions.  Such estimates for general
Schr\"odinger type operators 
are often referred to  as Lieb-Thirring (LT) type estimates
\cite{LT1}. 
The bound must behave like the semiclassical formula in all 
relevant physical parameters; in this case, in particular, it
should grow linearly in
the field strength.
A weaker apriori estimate typically leads to a semiclassical
asymptotics that is not uniform in the field strength \cite{Sob-98},
\cite{ES-II}.

\subsection{Summary of previous results}

A non-asymptotic LT bound for the Pauli operator has first been
established in \cite{LSY-II} for the case of the constant magnetic
field, $\bB = const.$,
\be
        \sum_j |e_j(H)| \leq (const.)
        \Big( \int [V]_-^{5/2}  + \int  |\bB|[V]_-^{3/2}
        \Big)
\label{LSYnaive}\ee
with $h=1$ and this bound was used to prove that $E_{scl}$  gives
the correct asymptotics for the sum of the negative eigenvalues.

The first generalizations of such estimates for non-homogeneous
magnetic fields were given in \cite{E-1995}. The first
general bound was of the form $(const.)
\Big( \int [V]_-^{5/2}  + \|\bB\|_\infty^{3/2} \int [V]_-\Big)$,
then the main focus was to study unbounded fields.
It was observed, that (\ref{LSYnaive})
 cannot hold in general. There are two  problems
in connection with (\ref{LSYnaive}) for nonhomogeneous field.

Firstly, even when $\bB$ has  constant direction in $\bR^3$
 (\ref{LSYnaive}) is correct only if $|\bB(x)|$
is replaced by an effective field strength, $B_{\rm eff}(x)$,
 obtained by averaging  $|\bB|$ locally on the magnetic
lengthscale, $|\bB|^{-1/2}$.

Secondly, the existence of the celebrated Loss-Yau zero modes
\cite{LY} contradicts  (\ref{LSYnaive}). Indeed, for certain magnetic
fields with  nonconstant direction
the Dirac operator $\D$ has a nontrivial $L^2$-kernel.
In this case a small potential perturbation of $\D^2$ 
shows that  $\sum_j |e_j(H)|$  behaves
as $\int n(x) [V(x)]_- \rd x$, i.e. it is linear in $V$.
Here $n(x)$ is the
density of zero modes, $n(x)= \sum_j |u_j(x)|^2$,
where $\{ u_j\}$ is an orthonormal basis in $\mbox{Ker} \, \D$.
Thus an extra term linear in $V$ must be added
to (\ref{LSYnaive}). It turns out that in order to
estimate $n(x)$ by the magnetic field it is again
important to replace $|\bB(x)|$ by an effective
field.

The problem of the effective field was first succesfully
addressed by Sobolev, \cite{Sob-96},  \cite{Sob-97}
and later by Bugliaro et. al. \cite{BFFGS} and Shen
\cite{Sh}. In particular, the $L^2$-norm of
the effective field, $\| B_{\rm eff}\|_2$, is comparable to $\|\bB\|_2$
in \cite{BFFGS}, and the same holds for any $L^p$-norm in
Shen's work. In a very general bound 
 proved in \cite{LLS} the second term in (\ref{LSYnaive}) is replaced
with $\| \bB \|_2^{3/2} \| V \|_4$.

In the  works \cite{E-1995}, \cite{Sob-97}, \cite{Sh}, \cite{LLS},
\cite{BFFGS} on three dimensional
magnetic Lieb-Thirring inequalities,
the density $n(x)$ is estimated by 
a function that behaves quantitatively as $| \bB(x)|^{3/2}$.
In particular, in the strong field regime these estimates
are not sufficient to prove  semiclassical asymptotics
uniformly in the field strength, they typically
give results up to $\|\bB\|_\infty \leq (const.)h^{-1}$ \cite{Sob-98}.

We remark that the bounds in \cite{LLS}
and \cite{BFFGS} have nevertheless been very useful
in the proof of magnetic stability of matter.
In this case the magnetic energy, $\int |\bB|^2$, is
also part of the total energy to be minimized, therefore
even the second moment  of the magnetic field is controlled.
We also remark that if the field has a {\it constant direction},
then no Loss-Yau zero modes exist, $n(x)\equiv 0$. 
In this case Lieb-Thirring type bounds that grow linearly with
$|\bB|$ have been proved in \cite{E-1995} and \cite{Sob-96},
\cite{Sob-97}. This
problem is technically very similar to the two dimensional 
case.

Since $n(x)$ scales like $(length)^{-3}$ and 
$|\bB(x)|$ scales like $(length)^{-2}$, a simple dimension
counting shows that $n(x)$ cannot be estimated in general
by the first power of $|\bB(x)|$ or by any smoothed version
$B_{\rm eff}(x)$.
However, if  an extra lengthscale
is introduced, for example certain derivatives of the field are
allowed in the estimate, then it is possible to give
a bound on the eigenvalue sum that grows  slower than $|\bB|^{3/2}$
in the large field regime.
There are only two results so far in this direction.

The work \cite{BFG} uses a lengthscale 
on which $\bB$ changes. The estimate eventually
scales like $b^{17/12}$, if the magnetic field
is rescaled as $\bB(x) \mapsto b \bB(x)$,  $b\gg1$.
As far as local regularity is concerned,
 only $\bB \in H^1_{loc}$ is required.
However, $n(x)$ is estimated by a quantity that 
depends {\it globally} on $\bB(x)$ not just in a neighborhood
of $x$. On physical grounds one expects the following {\it locality
property}: the zero modes of $\cD$ are supported near the support
of the magnetic field.

We prove a stronger locality property, namely that the size of $|\bB|$
away from a compactly supported negative potential will be irrelevant
for the estimate on the sum of the negative eigenvalues.
The result of \cite{BFG} does not give such bound for an important
technical reason. In order to produce an effective field 
strength $B_{\rm eff}$,
the $|\bB|$ is averaged out by a 
convolution function $\varphi$ that must satisfy 
$|\nabla \varphi|\le (const.) \varphi$, i.e. $\varphi$ must have a long tail.
For $\bB\in L^2$ the effective magnetic field has 
a comparable $L^2$-norm, but it is not true for the
 localized $L^2$-norms.

Our earlier work \cite{ES-I} had a different approach to reduce
the power $3/2$ of $|\bB|$
in the estimate of $n(x)$. We introduced two {\it global} lengthscales,
$L$ and $\ell$ respectively,
 to measure the variation scale of the field strength $|\bB|$
and the unit vector 
$\bn: =\bB/|\bB|$ that determines the geometry of the field lines.
 This required
somewhat more regularity on $\bB$ than \cite{BFG} and it
also involved the unnatural  $W^{1,1}$-norm  of $V$.
The estimate behaved like $b^{5/4}$
in the large field regime, if we rescaled $\bB\mapsto b \bB$, $b\gg 1$.
For fields with a  nearly constant direction, $\ell\gg 1$,
the bound was actually better, it behaved
like $b+ b^{5/4}\ell^{-1/2}$. This indicates
that it is only the variation of $\bn$ and not that of  $\bB$
that is responsible for the higher $b$-power.

Due to the improvement in the $b$-power 
from $3/2$ to $5/4$ in the Lieb-Thirring estimate we could also
prove the semiclassical eigenvalue asymptotics 
in the regime $b \ll h^{-3}$ for potentials in $W^{1,1}$
\cite{ES-II}.
This bound turned out to be sufficient to show
that the Magnetic Thomas-Fermi theory exactly
reproduces the  ground state energy of a large atom with
nuclear charge $Z$ in the semiclassical regime,
i.e. where $b \ll Z^3$, $Z\to\infty$ \cite{ES-II}.
The condition $b\ll Z^3$ is optimal as far as the
semiclassical theory is applicable 
as the results of \cite{LSY-I} show for super-strong ($b \ge Z^3$)
constant magnetic fields.

Despite the successful application of the bound in \cite{BFG}
to the stability of matter with quantized electromagnetic field
with an ultraviolet cutoff \cite{BFrG}, and despite that the Lieb-Thirring
inequality given in \cite{ES-I} fully covered the semiclassical
regime of the large atoms, it is still important to
establish a uniform Lieb-Thirring type bound with the correct
power in the magnetic field. Such bound will likely be the key
to generalize the analysis of the super-strong field regime
of \cite{LSY-I} to non-homogeneous magnetic fields.
In this paper we present a Lieb-Thirring bound that 

\begin{itemize}
\item  grows
linearly in the field strength;

\item depends on the potential $V$ in a natural way;

\item has the locality property in the sense discussed above.

\end{itemize}

\noindent
 We also state the corresponding semiclassical result
in Theorem \ref{thm:sc} but its details,
that are similar to \cite{ES-II}, will be published separately.

A simpler proof of a Lieb-Thirring estimate with both the linear
dependence in the field strength and the correct behavior in $V$
is given in \cite{ES-IV}. This approach, however, does not
give the locality property.

\subsection{Density of zero modes}

As a byproduct, we also obtain a bound on the density of the
zero modes, $n(x)$, that behaves
optimally  in the field strength in case of regular fields.
Actually, we control the density of all low lying 
states by giving an estimate for the diagonal element of
the spectral projection kernel $\Pi (\D^2\le c)(x,x)$
that grows linearly with the strength of the magnetic field
 for any fixed constant $c$.

We remark that the zero modes of the Dirac operator for
particular classes of magnetic fields are well understood.
The surprising first examples
were due to Loss-Yau \cite{LY} and later the present
authors gave a more systematic geometric construction \cite{ES-III}.
This construction, in particular, gives examples that show
that the density can grow at least linearly in the magnetic field
strength. Other generalizations of the original construction of
Loss-Yau are also available \cite{AMN}, \cite{El-1}.
However, there is no complete understanding of
all magnetic fields with zero modes yet. 

It is also known that magnetic fields with zero modes form a
slim set in the space of all magnetic fields (\cite{BE}, \cite{El-2})
but no quantitative result  is available in the general case. 
Our result (Corollary \ref{cor}) is the first general estimate on the
density of zero modes that scales optimally (i.e., linearly) in the field
strength. This result is formulated as a corollary, since it easily follows
from the main theorem, but we shall prove
it first on the way to the proof of our main theorem (Theorem \ref{thm:main}).

It is amusing to note that it takes a considerable
effort to show that zero modes exist at all, but
it is even more difficult to give an optimal  upper bound
on their densities for strong regular fields.
This is actually the main technical achievement of the present paper.

\subsection{Organization of the proof}

In Section \ref{sec:var}
we introduce what we call
 the {\it combined lengthscale} $L_c(x)$ of a given magnetic field
$\bB(x)$.
 This is a local variation lengthscale on which the magnetic field
does not change substantially. 
More precisely, this is the case only in the regime where
the magnetic field is strong; where the field is weak,
the combined lengthscale is simply chosen to be 
of the order of the magnetic lengthscale, $|\bB|^{-1/2}$.

In Section \ref{sec:mainthm} we formulate 
our main result on the new Lieb-Thirring inequality 
(Theorem \ref{thm:main})
and its corollary on the density of zero modes.
We also state a semiclassical result  (Theorem \ref{thm:sc})
whose proof will be published separately.

The proof starts in Section \ref{sec:structure} 
with a separation of the contributions from the low and the 
high energy regimes. The cutoff threshold is space dependent,
it is at a level $P(x) \sim L_c(x)^{-2}$.
Technically it is done by inserting $P(x)$ into the resolvent
in the Birman-Schwinger kernel and using a resolvent expansion.
We will call the two regimes the zero mode regime and the positive energy
regime, respectively, because the separation is dictated
by the need for a special treatment of the zero modes.
The basic estimates on the contribution
from these regimes are given in Theorem
\ref{thm:spl}.
We remark that to ensure ultraviolet convergence in the zero mode regime,
squares of resolvents need to be estimated as well 
(\cite{BFFGS}).

 In both regimes we perform a two-scale
localization, like in \cite{ES-I}. For both
localizations, however, the approaches used here
are substantially improved, as we explain below.

\bigskip

The first localization is isotropic and its lengthscale is determined
by $L_c(x)$. This is constructed in Section \ref{sec:cover}.
The main difference between the current isotropic
localization and the corresponding one
in \cite{ES-I} is that in our earlier paper we assumed a
universal positive bound on the combined lengthscale,
therefore we could use a regular grid of congruent cubes. 
In order to ensure the locality property, in this paper
 we need to use a covering argument to select
localization domains 
of different sizes and with  a finite overlap.
In  domains where
the magnetic field is relatively weak ($|\bB|\leq (const.)L_c^{-2}$),
we shall neglect all magnetic effects.

In Sections \ref{sec:pos1} and \ref{sec:zero1} we show how 
to localize the eigenvalue estimates onto the isotropic
domains. In the positive energy regime we
apply a version of the IMS localization formula
for the resolvent (Proposition \ref{prop:pullup}) that was already
used in \cite{BFFGS}. However, the same formula does
not hold for the square of the resolvent which is needed in the zero
mode regime. A new localization
scheme is developed in Proposition \ref{prop:pullin}
to localize the square of the resolvent of a second
order elliptic operator. 
 The localized versions of the necessary estimates 
in the positive energy regime and in the zero mode regime
are stated in Propositions \ref{prop:pos} and \ref{prop:zero}, 
respectively.

Typically it is not hard to localize resolvents of second order
elliptic operators
onto cubes of size $\ell$ at the expense of an error $\ell^{-2}$.
However localizing the square (or higher powers) of the
resolvent requires off-diagonal estimates on the resolvent
kernel (see Proposition 7.1).
 While these are typically easily available for
scalar elliptic operators without spin,
 we {\it do not know any apriori
off-diagonal control on the resolvent of $\D^2$}.
If the original Pauli operator is estimated by a constant field
Pauli operator, then {\it aposteriori} we can extract
off-diagonal estimates, but without comparison
with the constant field problem, we do not have
off-diagonal control. This is the main reason why we are unable
to extend the elegant and short method of \cite{ES-IV}
to give any locality properties.

Starting from Section \ref{sec:cyl} a  second localization is performed onto
 curvilinear cylindrical
domains with a transversal lengthscale 
$|\bB(x)|^{-1/2}$  along the field lines.
The geometry of the cylindrical domains and the coordinate system
are explained in Section \ref{sec:cylcord},
and  a new partition
of unity subordinated to the cylindrical domains is  constructed
in Section \ref{sec:cylpart}.

Within each cylindrical domain
the magnetic field is approximated by a field $\beta_c$, given
as a 2-form, that is constant in the appropriate cylindrical coordinates
and after a conformal change of the metric (Definition \ref{def:constfield}). 
The Dirac operator $\D_c$ with a magnetic field $\beta_c$  
(Definition \ref{def:constdirac}) will be used to approximate the
original Dirac operator $\D$ in the corresponding cylindrical domain.
Section \ref{sec:gendirac} is devoted to the construction of
$\D_c$ and it  uses the geometric structure behind
the Dirac operator on a non-flat manifold outlined, for example,
in  \cite{ES-III}.

\bigskip

The second main difference between \cite{ES-I} and the current
work lies in the cylindrical localization. In \cite{ES-I}
we considered straight cylinders to approximate
tubular neighborhoods of magnetic field lines
and we approximated the field by a constant one
within each cylinder. The curving of the magnetic
field was not respected by the approximation
hence the error was not uniform in the field strength.
This is the main reason why the Lieb-Thirring inequality in
\cite{ES-I} does not have the optimal $|\bB|$-power.

In the new construction the cylindrical localization domains 
are curved in such a way
as to follow a field line and we also construct appropriate spinor
coordinates. This geometric approach enables us to control
$\D^2$ with errors that are uniform in the field strength although
they depend on the combined lengthscale of $\bB$. 
This eliminates the $|\bB|$-dependent error in the large field
regime. The near-zero
energy states, in particular the zero modes of $\D$, need to be controlled
with such a precision in order not to overestimate
their contribution to the negative eigenvalues of $\D^2-V$.
The proof of the $|\bB|$-independent control is quite involved
and it relies heavily on the intrinsic geometric properties
of the Dirac operator.

Section \ref{sec:pos} completes the proof of the positive energy regime.
In this regime errors that are independent
of $|\bB|$ can be absorbed into the local energy shift $P(x)$.
Proposition \ref{prop:cylpos} contains the necessary spectral estimate
localized onto the cylindrical domains in the original coordinates.
We first translate the estimate
into cylindrical coordinates. In these coordinates the
approximating field is constant and we can use 
the magnetic localization formula (\ref{eq:loc})
from \cite{ES-II}. This method  yields a $|\bB|$-independent
cylindrical localization error.
Finally, having constructed the approximating local Dirac operators
with constant fields,  we can use the Lieb-Thirring inequality for the Pauli
operator with a constant field obtained in \cite{LSY-II}.

In Section \ref{sec:zero} we complete the estimate of
the zero mode regime. We need to estimate the density of
the near-zero energy states of $\D^2$. This is given by the diagonal
kernel of the spectral projection operator, $\Pi(\D^2 \leq P_0)(x,x)$
where $P_0$ is the typical value of the regularly varying
function $P$ around $x$.
This operator can be bounded by the resolvent, but the
diagonal element of the resolvent is infinite
because of the ultraviolet divergence.
 Therefore we need to control $\Pi(\D^2 \leq P_0)$ by the square
of the resolvent, $(\D^2 + P_0)^{-2}$. For regions  with weak 
magnetic fields the magnetic field can be neglected and we can simply
use the diamagnetic inequality. 
The problem thus can be reduced to estimating  the
resolvent square of the free Laplacian (Section \ref{sec:weakmag}).

For regions with a strong field (Section \ref{sec:strongmag})
we again use the approximating
constant field operators.
 However, the magnetic localization
formula (\ref{eq:loc}) is not valid for the square of the Pauli operator, so 
localizing onto cylindrical domains is more complicated. 
Fortunately, at this stage we do not need operator inequalities,
we need to estimate only the diagonal element 
of the square of the resolvent at each fixed
point $u$. First we transform
the problem into the new coordinates associated
with the field line through $u$ (estimates 
(\ref{eq:sqre}) and (\ref{eq:sqdre})).
Then we use resolvent expansions extensively
to approximate $\D$ by $\D_c$. Since we estimate
the square of the resolvent,
we need to control
the offdiagonal elements of the resolvent itself. 
For a constant field, the offdiagonal decay is Gaussian
on a magnetic lengthscale $|\bB|^{-1/2}$. A similar feature is proved
for the resolvent with a nonconstant field via the constant
field approximation.

In the Appendix we collected the proofs of several 
Propositions and Lemmas which can be skipped at a
first reading.

It is amusing to note that the most complicated part
of the proof (Sections \ref{sec:zero1} and \ref{sec:zero})
controls the possible ultraviolet regime of near zero energy
states. On physical grounds this regime should be irrelevant
if we knew that low energy eigenstates of $\D^2$ have transversal
momentum of order $|\bB|^{1/2}$ and parallel momentum independent of
the field strength. The main difficulty is to
obtain such information on the low lying states.

\medskip
{\it Convention:} Throughout the proof
universal constants are denoted by a general $c$
whose value can be different even within the same equation.
Constants depending on numbers $a, b,\ldots$ are denoted by $c(a,b,\ldots)$.
Integration over $\bR^3$ with respect to the Lebesgue measure,
 $\int_{\bR^3}\rd x$, is simply denoted by $\int$.
We shall say that two positive numbers $a, b$ are {\bf comparable}
if $\sfrac{1}{2}\leq a/b \leq 2$.

\section{Lengthscales of the magnetic field}\label{sec:var}
\setcounter{equation}{0}

Let $\bB\in C^4(\bR^3,  \bR^3)$ be a  magnetic field and let 
$\bn:=\bB/|\bB|$ be the unit vectorfield in the direction
of the magnetic field at all points where $\bB$ does not vanish.

For any $L\ge0$ and $x\in \bR^3$ we define
\be
    B_L(x): = \sup\{ |\bB(y)|\; : \; |x-y|\leq L \}
\label{eq:BL}
\ee
and
\be
    b_L(x): =  \inf\{ |\bB(y)|\; : \; |x-y|\leq L \}
\label{eq:bL}
\ee
to be the supremum and the infimum of the magnetic field strength
on the ball of radius $L$ about $x$. These functions are continuous
in both the $L$ and $x$ variables.

The Pauli operator will be localized on different
lengthscales determined by the magnetic field. We now define
these scales.

\begin{definition}[Lengthscales of a magnetic field]\label{def:mag} 
Given a $C^4$-magnetic field $\bB$. We define   the 
{\bf magnetic  lengthscale}  of $\bB$ as
\be
       L_m(x):= \sup\{ L>0 \; : \; B_L(x) \leq L^{-2}\} \; .
\label{eq:magnscale}
\ee
The {\bf variation lengthscale} of $\bB$ is given by
\begin{equation}\label{eq:Lv}
  L_v(x): =\sup\Big\{ L\geq0 \; : \; L^\gamma \sup\Big\{ \Big|\nabla^\gamma
  \bB(y) \;\Big| \; : \; |x-y|\leq L\Big\} \leq b_L(x),
  \; \gamma=1,2,3,4 \Big\} \; 
\end{equation}
Finally we set
\be
       L_c(x): = \max \{ L_m(x), L_v(x) \}
\label{def:Lc}
\ee
to be the {\bf combined lengthscale} of $\bB$ at $x$.
\end{definition}

A magnetic field $\bB:\bR^3\to\bR^3$ determines two
local lengthscales. The magnetic lengthscale, $L_m$, is
comparable with 
$|\bB|^{-1/2}$. The lengthscale $L_v$ determines the 
scale on which the field $\bB$ {\it varies}. One may think of $L_v$ as the smaller of the 
two lenghtscales describing the variation of 
the field strength, i.e., the variation scale of $\log |\bB|$, and the variation scale of the field lines $\bn$.

For weak magnetic fields the magnetic effects can be neglected
in our final eigenvalue estimate, so the variational
lengthscale becomes irrelevant. This idea is reflected
in the definition of $L_c$; we will not need to localize
on scales shorter than the magnetic scale $L_m$.

Note that for any $\bB\in C^4(\bR^3, \bR^3)$ we have that
$0< L_c(x)\leq \infty$ for all $x\in \bR^3$. If $L_c(x)=\infty$
for some $x\in\bR^3$, then $\bB$ is constant on $\bR^3$. Moreover
the value $L_c(x)$ at any $x$ does not depend on $\bB$ outside the
ball centered at $x$ with radius $L_c(x)$. If follows in particular,
that if $\bB$ vanishes in a ball of radius $\delta$ around $x$,
then $\delta\leq L_c(x)$.

\section{Main Theorem}\label{sec:mainthm}
\setcounter{equation}{0}

We are ready to state our main results.

\begin{theorem}[Uniform Lieb-Thirring inequality]\label{thm:main}
 We assume that the magnetic field is $\bB \in C^4(\bR^3, \bR^3)$.
Let $\bA\in C^4(\bR^3, \bR^3)$ be a vector potential, 
 $\nabla\times
\bA=\bB$, and let $\cD : = \bsigma\cdot (\bp+\bA)$ be the free Dirac
operator with magnetic field $\bB$
on the trivial spinorbundle over $\bR^3$, that can be identified with 
$L^2(\bR^3,\bC^2)$.
Let $V$ be a scalar potential.
Then the sum of the negative eigenvalues, $e_j$, of the Pauli
operator $H:= \cD^2+V= [\bsigma\cdot (\bp+\bA)]^2 + V$ satisfies
\be
       |\Tr \; H_-|= \sum_j |e_j| \leq c \int [V]_-^{5/2}
       + c\int |\bB|[V]_-^{3/2} + c\int (|\bB|+L_c^{-2})L_c^{-1} [V]_-
\label{eq:main}
\ee
with universal constants. 
\end{theorem}

{\it Notation:} For any self-adjoint operator $H$ we let
$H_- : = \sfrac{1}{2}[ |H| - H ]$ denote its negative part.

\begin{corollary}[Density of zero modes]\label{cor}
Given a magnetic field $\bB\in C^4(\bR^3, \bR^3)$ with a combined
lengthscale $L_c$, the density of zero modes of the free Dirac operator
$\D$ with magnetic field $\bB$ satisfies
\be
      n(x):= \sum_j |u_j(x)|^2 \leq c(|\bB(x)|+L_c^{-2}(x))L_c^{-1}(x)
\label{eq:nest}
\ee
with a universal constant, where $\{u_j\}$ is an orthonormal basis
in the kernel of $\D$.
\end{corollary}

{\it Remarks.} (i) The density function $n(x)$ was also estimated
in \cite{BFG}. In the strong field limit $\bB\mapsto
b\bB$, $b\gg 1$, the estimate behaved as $b^{17/12}$.
Moreover, unlike in \cite{BFG}, our estimate on $n(x)$ uses only local
information on $\bB(x)$ as explained in Section \ref{sec:var}.
For example, if $\bB$ vanishes inside a ball centered at $x$ with
radius $\delta$, then $n(x)\leq c\delta^{-3}$.

(ii) The bound (\ref{eq:nest}) is optimal as far as the strength
of the field $|\bB|$ is concerned. This fact follows from
the construction of Dirac operators with kernels of high multiplicity
following the method of \cite{ES-III}. For example, the density
of Aharonov-Casher zero modes for a constant magnetic field of strength 
$B\gg 1$ on $S^2$ is of order $B$. The geometric procedure of \cite{ES-III}
allows one to construct  a Dirac operator on $\bR^3$ whose
zero energy eigenfunctions are obtained from the
eigenfunctions on $S^2$ by an explicit transformation.
The density of these states remain comparable to the strength
of the magnetic field at least away from infinity.

(iii) Notice that the Lieb-Thirring inequality of \cite{LSY-II} for a
 {\it constant} field is recovered  in Theorem \ref{thm:main}.

(iv) The uniform Lieb-Thirring bound for a {\it constant direction}
field, \cite{Sob-97}, \cite{ES-I}, does not directly follow from
our main theorem as it is stated. On one hand, (\ref{eq:main})
contains a term linear in $V$ that is unnecessary for a constant
direction field. On the other hand, we assume high regularity on
$\bB$.  This regularity is needed only to construct
the appropriate curvilinear cylindrical localization, which
is unnecessary for a field with  constant direction.

However, our present technique to estimate squares of the resolvents 
can improve these results in another aspect.
For example, if the support of $\bB$ and the support of $V$
are separated, our Lieb-Thirring estimate depends only on
the separation distance whereas all previous bounds scale with the
magnitude of $\bB$. As a byproduct
of such a result one can also improve the estimates
on the ground state density of the two dimensional Pauli
operator given in \cite{E-93}.

\bigskip

Armed with a uniform Lieb-Thirring inequality, the following
semiclassical asymptotics may be proved by combining the techniques
of the current paper and \cite{ES-II}. The details
of the proof will be published separately.

\begin{theorem}\label{thm:sc} We assume that $\bB \in C^4(\bR^3, \bR^3)$
and $V\in L^{5/2}(\bR^3)\cap L^1(\bR^3)$. Then the sum of the
negative eigenvalues, $e_j(b,h)$, of
the Pauli operator $[\bsigma\cdot (-ih\nabla + b \bA)]^2 + V$
is asymptotically given by
\be
        \lim_{h\to0} \frac{\sum_j e(b, h)}{E_{scl}(h, b \bB, V)}\;
        =1 
\label{eq:sc}
\ee
where the limit is uniform in the field strength $b$.
\end{theorem}
 
{\it Remark.} This result was obtained for a homogeneous magnetic field 
in \cite{LSY-II}. Analogous results  
for $d=2$ were obtained in \cite{ES-II} and \cite{Sob-98}.
The latter work also extends the two dimensional analysis
to obtain (\ref{eq:sc}) for three dimensional
 magnetic fields with constant direction.
For a general three dimensional magnetic field the limit (\ref{eq:sc})
is proven up to $b\ll h^{-3}$ for $V\in W^{1,1}$
 in \cite{ES-II}. With a different method Sobolev also
obtains (\ref{eq:sc}) up to $b\leq (const.)h^{-1}$
without assumptions on the derivatives
of $\bB$ and $V$ \cite{Sob-98}.

\section{Proof of the Main Theorem \ref{thm:main}}\label{sec:structure}
\setcounter{equation}{0}

\subsection{Tempered lengthscale}

Since localization errors decrease with the localization
length, we would optimally like to choose the biggest
possible scale, i.e. $L_c(x)$, for
our localization scale. However,  neighboring localization
domains must be comparable in size so that the localization
errors could be reallocated. This forces us to require
a tempered behavior on the localization scales,
which may result in choosing a localization scale smaller than $L_c$.
Proposition \ref{prop:L} below shows that this 
technical requirement can be met at the expense
of a factor $\sfrac{1}{2}$ and this justifies the
introduction of the tempered lengthscale $L:=\sfrac{1}{2}L_c$. 
 Before the precise statement
we need the following definition:

\begin{definition}\label{def:temp} Let $\e>0$ be a positive number.
A positive function $f(x)$ on $\bR^3$ is called {\bf $\e$-tempered} if
\be
        |x-y|\leq \e^{-1}f (x)
         \Longrightarrow {1\over 2}\leq {f (y)
        \over f (x)}\leq 2 \qquad \forall x, y\in\bR^3\; .
\label{eq:elltempered}
\ee
If $\e=1$, then a $1$-tempered function will be simply called
{\bf tempered}.
\end{definition}

\begin{proposition}[Existence of  tempered lengthscale] \label{prop:L}
For any not identically constant
 magnetic field, $\bB\in C^4(\bR^3, \bR^3)$, 
$L(x)= \sfrac{1}{2} L_c(x)$ is finite and defines a tempered
function.
Moreover,  if $B_{L(x)}(x)>L(x)^{-2}$, 
then $b_{L(x)}(x)>0$, and for $\gamma=1,2,3,4$
\be
       L(x)^\gamma \sup\Big\{ \Big|\nabla^\gamma
       |\bB(y)| \;\Big| \; : \; |x-y|\leq L(x)\Big\} \leq b_{L(x)}(x),
\label{eq:tsupnablaB}
\ee
and
\be
        L(x)^\gamma \sup\{ |\nabla^\gamma
       \bn(y)|\; : \; |x-y|\leq L(x) \} 
       \leq 1.
\label{eq:tsupnablan}
\ee
 For a constant magnetic field  $\bB=const$ we have $L_c=\infty$ and
we set $L(x):=\infty$.
\end{proposition}
The  proof is given in Section \ref{sec:varsc}.

\bigskip

For  a constant magnetic field $\bB$ the tempered scale 
has been defined to be
infinity for the transparent formulation of our theorem.
The estimate (\ref{eq:main}) for this case has been proven
in \cite{LSY-II}. It is possible to apply our proof
to this case as well, but setting $L=\infty$ directly
may require minor remarks along the proof. 
In order to avoid this inconvenience, we can choose
$L$ to be any fixed real number for which the proof goes
through without changes and finally let $L\to\infty$ 
in the final result (\ref{eq:main}).

We introduce a universal constant $0<\e < \sfrac{1}{1000}$
that has to be chosen small enough for the proof to work
but we shall not keep track of the exact numerical value needed.
We consider it fixed throughout the proof.

Let $L(x)$ be the tempered lengthscale of $\bB$. Introduce
 $\ell(x): = \e L(x)$, then the properties of $L(x)$ set in
  Definition \ref{def:mag} and Proposition \ref{prop:L}
 are translated into $\ell$ as follows:
 
\begin{itemize}
\item $\ell(x)$ is $\e$-tempered.

\item For all $x\in \bR^3$ such that $\sup\{ |\bB(y)| \; : \;
  |x-y|\leq \e^{-1}\ell\} \ge \e^2\ell^{-2}$ and for
 $\gamma=1,\ldots , 4$
we have
\be
        \ell(x)^\gamma\; \sup \Big\{ \Big| \; \nabla^\gamma|\bB(y)| \; \Big| \; : \;
        |x-y|\leq \e^{-1}\ell(x) \Big\}
        \leq \e^\gamma \inf  \Big\{ |\bB(y)| \; : \; 
        |x-y|\leq \e^{-1}\ell(x) \Big\}\; .
\label{eq:supnablaB}
\ee
and
\be
        \ell(x)^\gamma\; \sup \Big\{\; \Big| \nabla^\gamma \bn(y)\Big|
         \; : \;
        |x-y|\leq  \e^{-1}\ell(x) \Big\}\leq \e^\gamma \; .
\label{eq:supnablan}
\ee
\end{itemize}

We define 
$$
        P(x): = \e^{-5}\ell(x)^{-2}
$$
and
for any positive function $f> 0$ we introduce the notation
$$
        R_f = R(f) = (\cD^2  + f)^{-1}\; .
$$

\subsection{Separation into low and high energy regimes}

We shall prove Theorem \ref{thm:main}
with $L_c$ replaced by $L$.
Since $\D^2 + V \ge \D^2 - [V]_-$, we may consider
only non-positive potentials. For convenience we change
the sign and we will work with the operator $H:= \D^2 - V$
with $V\ge 0$. 
 
By the Birman-Schwinger principle
\be
        |\Tr \; H_-| = \int_0^\infty  \,\,  n\Bigg( 
        V^{1/2} R_E
        V^{1/2}, 1\Bigg) \rd E
\label{eq:sneg}
\ee
where $n(A, \mu)$ is the number of eigenvalues of
the operator $A$ greater than or equal to $\mu$. 
For any $E>0$ we have,  by the resolvent identity, that
$$
        R_E = R_{P+E} + R_{P+E}PR_E =
         R_{P+E} + R_{P+E}P R_{P+E} + R_{P+E}PR_EPR_{P+E} \; .
$$
Using that $P\leq \cD^2 + P + E$ and $R_E\leq E^{-1}$, we obtain
$$
        R_E \leq 2  R_{P+E} + E^{-1}  R_{P+E}P^2R_{P+E} \; .
$$
For any positive operators $X_1, X_2$, 
\be
        n(X_1+X_2, e_1+e_2)\leq n(X_1,e_1) + n(X_2, e_2) \; ,
\label{eq:nsum}
\ee
hence (\ref{eq:sneg}) is estimated as
\be
          |\Tr \; H_-|  \leq \int_0^\infty  \,\,  n\Bigg( 
        V^{1/2} R_{P+E}
        V^{1/2}, \sfrac{1}{4}\Bigg) \rd E
        +  \int_0^\infty  \,\,  n\Bigg( 
        2   V^{1/2}  R_{P+E}P^2R_{P+E}
        V^{1/2}, E\Bigg) \rd E  \; .
\label{spl}
\ee
The second term carries the contribution of the near zero energy 
eigenfunctions of the free Pauli operator $\D^2$. This will be called 
the {\it zero mode  regime}.

For the first term
we notice that
\be
         \int_0^\infty  \,\,  n\Bigg( 
        V^{1/2} R_{P+E}
        V^{1/2}, \sfrac{1}{4}\Bigg) \rd E =  \Big|\Tr(\D^2 + P - 4V)_-\Big|
\label{eq:BS}
\ee
by the Birman-Schwinger principle. This term contains
the contribution from free eigenfunctions with energy at least $O(P)$
and it will be called the {\it positive energy regime}.

The following Theorem estimates the two terms in (\ref{spl}) and
it completes the proof of the Main
Theorem by choosing $\e$ sufficiently small.
$\;\;\;\Box$

\begin{theorem}\label{thm:spl} For a sufficiently
small universal $\e$ and with the notations above we have
\bey
          \Big|\Tr(\D^2 + P - 4V)_-\Big|
        &\leq&
        c(\e)\int \Big( V^{5/2} + |\bB| V^{3/2}\Big) \; , 
\label{eq:Iest} \\
              \int_0^\infty  \,\,  n\Bigg( 
        2   V^{1/2}  R_{P+E}P^2R_{P+E}
        V^{1/2}, E\Bigg) \rd E
        &\leq&
         c(\e) \int VP^{1/2}(|\bB|+P) \; .
\label{eq:IIest}
\eey
\end{theorem}
The proof of Theorem \ref{thm:spl} is given in the rest of the
paper.

\medskip

{\it Convention about operator kernels:} If $A$ is a Hilbert-Schmidt
operator on  a Hilbert space
of the form $L^2(\rd \mu)\otimes \bC^N$, $N\in \bN$,
we denote by $A(x,y)$ its 
$N\times N$-matrix valued integral
kernel which is  $L^2$ on the product space.
If, in addition, $A$ is of trace class, we can even define
its diagonal kernel which,  by a slight abuse of notation,
will be denoted  by $A(x,x)$. One possible way to define it
is to write $A$ as a product of two Hilbert-Schmidt operators, $A=HK$,
and $A(x,x): = \int H(x,y)K(y,x) \rd \mu (y)$. This is
an $L^1$ matrix valued 
function of $x$ and as such it is independent of the choice
of $H$ and $K$.

{\it Convention about traces:} We shall denote by $\Tr$
the trace on $L^2(\rd\mu)\otimes \bC^N$ and by $\tr$ the trace
on $\bC^N$. If $A$ is of trace class on $L^2(\rd \mu)\otimes \bC^N$,
then $\tr A(x,x)$ is in $L^1(\rd\mu)$.

\section{Isotropic geometry of the first localization}\label{sec:cover}
\setcounter{equation}{0}

In this section we construct the domains for the first localization.
The construction is determined by the function $\ell(x)$.
We shall construct a discrete 
set of points $\{ x_i \}$.
 The localization domains will be balls about $x_i$ with radii
 $\ell(x_i)$ 
and they will have finite overlap. Moreover, the
magnetic field will not change much in each localization ball
since $\ell(x)$ determines the local scale of variation of $\bB$.
Outside of this domain the field will be replaced by a constant field.
This procedure will apply to balls with relatively strong fields.
On balls where $\bB$ is small we neglect magnetic effects and
replace the Pauli operator by the free Laplacian.

\subsection{Regular fields}

\begin{definition}\label{def:Dreg}
 Given $\ell, K>0$ and a ball $D$ of radius $\ell$ centered at $z_0\in\bR^3$.
A magnetic field $\bB$ is called {\bf $D$-strong} if
$| \bB(z_0) |\ge \e^{-2} \ell^{-2}$, otherwise it is called
{\bf $D$-weak}. A $D$-strong
magnetic field is called
{\bf $(D, K)$-regular}  if for $\gamma=1,\ldots ,4$

(i) $\Big| \nabla^\gamma |\bB| \Big| \leq K\e^\gamma\ell^{-\gamma}
 | \bB(z_0)|$ on $D$;

(ii) $|\nabla^\gamma \bn |\leq K\e^\gamma\ell^{-\gamma}$ on $D$ 
 (with $\bn:= \bB/|\bB|$);
\medskip

A $(D,K)$-regular field $\bB$ is called {\bf extended 
$(D,K)$-regular}
if it is continuous on the whole space and
 $\bB$ is constant outside of $D$. The value of $\bB$ outside of $D$
is denoted by $\bB_\infty$ and for
$\bB_\infty\neq0$ we set $\bn_\infty: = \bB_\infty/|\bB_\infty|$.
\end{definition}

A $(D,K)$-regular field $\bB$ clearly 
has a small total variation on $D$:
\be
        \Big| |\bB(x)|-|\bB(z_0)|\Big| \leq 2K\e |\bB(z_0)|
\label{eq:Btem}\ee
for any $x\in D$. For an extended $(D,K)$-regular
field (\ref{eq:Btem}) is valid for any $x\in \bR^3$,
and
\be
        \|\bn(x)-\bn_\infty\|\leq K\e\; .
\label{eq:ntem}
\ee

The following statement follows from the definitions above: 

\begin{lemma}\label{lemma:Bisreg}
 Let $\bB(x)$ and $\ell(x)$ satisfy the conditions
(\ref{eq:elltempered}), 
(\ref{eq:supnablaB}) and
(\ref{eq:supnablan}).  Let $\tD$ be the ball of radius $10\ell(z_0)$
about 
some $z_0\in \bR^3$.

(i) If  $\bB$ is $\tD$-strong, then
 $\bB$ is $(\tD, 1)$-regular and $|\bB(x)|\ge
 \e^{-1}\ell(z_0)^{-2}$
for any $x\in \tD$.

(ii) If $\bB$ is $\tD$-weak,  then  $|\bB(x)|\le
 \e^{-2}\ell(z_0)^{-2}$
for any $x\in \tD$.
\end{lemma}

\noindent
{\it Proof.} (i) For $\e\leq \e_0$ we obtain that for any $x\in \tD$
$$
        \Big|\; |\bB(z_0)|- |\bB(x)| \; \Big|
        \leq 10\e \inf  \Big\{ |\bB(y)| \; : \; 
        |x-y|\leq \e^{-1}\ell(z_0) \Big\}
        \leq 10\e |\bB(z_0)|
$$
using (\ref{eq:supnablaB}). In particular, $|\bB(x)|\ge
(1-10\e) |\bB(z_0)| \ge 
\e^{-1}\ell(x)^{-2}$ for any $x\in \tD$ because $\bB$ is $\tD$-strong,
$|\bB(z_0)|\ge \e^{-2}(10\ell(z_0))^{-2}$,
and $\ell(z_0)\leq 2 \ell(x)$.
Properties (i) and (ii) in Definition \ref{def:Dreg} follow
from (\ref{eq:supnablaB}) and
(\ref{eq:supnablan}).

(ii) Suppose that for some $x\in \tD$ we have $|\bB(x)| >
\e^{-2}\ell(z_0)^{-2}$. Using that $|z_0-x|\leq 10\ell(z_0) \leq 20
\ell(x)< \e^{-1}\ell(x)$ if $\e\leq \sfrac{1}{20}$, we obtain
$$
        \Big|\; |\bB(z_0)|- |\bB(x)| \; \Big|
        \leq 10\e \inf  \Big\{ |\bB(y)| \; : \; 
        |y-x|\leq \e^{-1}\ell(x) \Big\}
        \leq 10\e |\bB(z_0)|
$$
which contradicts to $|\bB(z_0)|< \e^{-2}(10\ell(z_0))^{-2}$.
 $\;\;\;\Box$

\subsection{Covering lemma and cutoff functions}\label{sec:cutoff}

Let $B(x, r)$ denote the closed ball centered at $x$ with
radius $r$. We introduce the following notations for any $x\in \bR^3$
$$
        \hD_x : = B\Big( x, {\ell(x)\over 10}\Big),
        \quad
        D_x : = B\Big( x, \ell(x)\Big),
        \quad
        \tD_x : = B\Big( x, 10\ell(x)\Big) \; .
$$

\begin{definition}\label{def:unif}
 Let $\ell(x)$ be an $\e$-tempered function
and let $I$ be a countable index set.
The discrete set of points $\{ x_i\}_{i\in I}$
 is called an $\ell$-{\bf uniform} set of points 
with intersection constant $N$ if

(i) $\bR^3\subset\bigcup_{i\in I} \hD_{x_i} $;

(ii) Any ball $\tD_{x_j}$ intersects no more than $N$ other balls
from the collection $\{ \tD_{x_i} \}$.
\end{definition}

The proof of the following covering Lemma is  
given in Section \ref{section:coverproof}.

\begin{lemma}\label{lemma:cover} Let $\ell(x)$ be $\e$-tempered,
then there exists an $\ell$-uniform set of points $\{ x_i\}_{i\in I}$
with some universal intersection constant $N$.
\end{lemma}

In the rest of the proof we fix such a collection of points 
$\{ x_i \}$, determined by the magnetic field via $\ell(x)$. For brevity
we shall use $\ell_i: = \ell(x_i)$, $\wh D_i:=\wh D_{x_i}$,
$D_i:=D_{x_i}$ and
 $\wt D_i := \wt D_{x_i}$.

\begin{definition}
An index $i\in I$, the corresponding point $x_i$ and
 ball $D_i$ are called 
{\bf strong} ({\bf weak}) if $\bB$ is $D_i$-strong (weak).
\end{definition}

The following Lemma is an application of
 Lemma \ref{lemma:Bisreg}:

\begin{lemma}\label{lemma:strongcubes}
 Let $x_i$ be a strong point, then $\bB$ is $(\tD_i, 1)$-regular
and $\inf_{\tD_i} |\bB| \ge \e^{-1}\ell_i^{-2}$.
If $x_i$ is a weak point,  then $\sup_{\tD_i}|\bB|\leq \e^{-2}\ell_i^{-2}$.
\end{lemma}

\bigskip

Given an $\e$-tempered function $\ell(x)$ and
an $\ell$-uniform set of points $\{x_i\}_{i\in I}$,  for each $i\in I$
we  choose smooth
functions $\theta_i$, $\wh\chi_i$, $\chi_i$ and $\wt\chi_i$
with values between 0 and 1,
such that the following hold:

\begin{itemize}
\item $\sum_i \theta_i^2(x) \equiv 1$, $\mbox{supp}(\theta_i)\subset
D_i$ and $\|\nabla\theta_i\|_\infty\leq c\ell_i^{-1}$;
\item  $\wh\chi_i\equiv 1$ on $B(x_i, 3\ell_i)$, 
$\mbox{supp}(\wh\chi_i)\subset
 B(x_i,4\ell_i)$, $\|\nabla\wh\chi_i\|_\infty \leq 2\ell_i^{-1}$;
\item  $\chi_i\equiv 1$ on $B(x_i, 4\ell_i)$,
 $\mbox{supp}(\chi_i)\subset B(x_i, 5\ell_i)$,
 $\|\nabla\chi_i\|_\infty \leq 2\ell_i^{-1}$;
\item  $\wt\chi_i \equiv 1$ on $B(x_i, 6\ell_i)$,
$\mbox{supp}(\wt\chi_i)\subset B(x_i, 7\ell_i)$,
$\|\nabla^\gamma\wt\chi_i\|_\infty \leq (2\ell_i)^{-\gamma}$,
$\gamma=1,\ldots, 4$.
\end{itemize}
Such choice is possible since the balls $\hD_i$ cover.
Notice that $\nabla\wh\chi_i$ is  supported 
on the annulus
$$
        A_i : =  B(x_i, 4\ell_i)\setminus B(x_i, 3\ell_i)\; .
$$
Finally we choose functions $\{ \varphi_i \}_{i\in I}$ such that
$\varphi_i \equiv 1$ on $ A_i$,
$\mbox{supp}(\varphi_i) \subset  
B(x_i, 5\ell_i)\setminus B(x_i, 2\ell_i)$ and
$|\nabla\varphi_i|\leq 2\ell_i^{-1}$.

\bigskip

\subsection{Approximate magnetic fields and Pauli operators}
\label{sec:appro}

\bigskip

We define approximate vector potentials $\bA_i$
and magnetic fields $\bB_i:=\nabla\times
\bA_i$, $i=1,2,\ldots$,
subordinated to the balls $D_i$. The definition
is different for weak and strong indices $i\in I$.

If $i\in I$ is a {\bf weak} index, then let $\wh \bA_i$ be the
Poincar\'e gauge of $\bB$ with base point $x_i$, in particular
$\nabla\times\wh\bA_i=\bB$ on $\bR^3$ and 
$|\wh \bA_i| \leq c\ell_i \sup_{\tD_i} |\bB| \leq c\e^{-2} \ell_i^{-1}$
 holds true on $\tD_i$ by  Lemma \ref{lemma:strongcubes}. We define
$\bA_i : = \bA - (1-\wt\chi_i)\wh\bA_i$, then $\bB_i = \wt\chi_i\bB
+\nabla\wt\chi_i\times \wh\bA_i$. Clearly  
$\bA(x) = \bA_i(x)$ for all $x\in B(x_i, 6\ell_i)$ and
\be
        \|\bB_i\|_\infty\leq c \sup_{\tD_i} |\bB|\leq c\e^{-2}\ell_i^{-2} \; 
\label{eq:AiA}\ee
with $\mbox{supp} \; \bB_i \subset \tD_i$.

If $i\in I$ is a {\bf strong} index,  then $\bA_i$ is
 given by the following lemma.

\begin{lemma}[Choice of the local field on strong balls]
  \label{prop:appr}
Assume that $\bB$ is $(\tD_i,1)$-regular, then 
there exists a vector potential $\bA_i$  such that 
$\bA_i\equiv \bA$
 on $B(x_i, 6\ell_i)$ and the magnetic field $\bB_i =\nabla\times
\bA_i$ satisfies
\be
        \bB_i(x) \equiv \bB (x), \qquad x\in B(x_i, 6\ell_i)
        \qquad \mbox{and} \quad \bB_i(x)\equiv \bB(x_i) \quad 
        x\in\bR^3\setminus B(x_i, 7\ell_i) \; .
\label{eq:bi1}\ee
Moreover, $\bB_i$ is extended $(\tD_i,  100)$-regular, in 
particular for $\gamma=1,\ldots, 4$
\bey
        \ell_i^\gamma \big\| \nabla^\gamma |\bB_i| \, \, \big\|_\infty
        &\leq&
         100 \e^\gamma |\bB_i(x_i)|\; ,
\label{eq:bi3} \\
     \ell_i^\gamma  \big\| \nabla^\gamma \bn_i \, \, \big\|_\infty 
        &\leq &
        100\e^\gamma \; ,
\label{eq:bi4}\\
              |\bB_i(x) - \bB(x_i)|
        &\leq &
         100\e |\bB(x_i)|\; 
\label{eq:bi2}
\eey
for any $x\in \bR^3$.
\end{lemma}

Armed with these definitions of $\bA_i$, we define

\be
        \cD_i: = \bsigma\cdot (\bp+ \bA_i)
\label{def:cdi}
\ee
to be the approximating Dirac operator  associated with $\tD_i$.
The operator $\cD_i$ coincides with $\cD$ on
$B(x_i, 6\ell_i)$ because $\bA=\bA_i$ in this domain, in particular
\be
        \cD\chi_i = \cD_i\chi_i \; .
\label{eq:deqd}
\ee

\noindent
{\it Proof of Lemma \ref{prop:appr}.}
Since $\bB$ is $(\tD_i,  1)$-regular,
from (\ref{eq:Btem}) we obtain that
\be
        |\bB(x)-\bB(x_i)|\leq \e |\bB(x_i)|, \qquad x\in \wt D_i\; .
\ee
 Let $\bA^\#_i$ be the Poincar\'e gauge on $\tD_i$
of the magnetic field $\bB-\bB(x_i)$, then $\nabla\times\bA^\#_i=
\bB-\bB(x_i)$ and
\be
        |\bA^\#_i(x)|\leq \e\ell_i |\bB(x_i)|
\label{eq:Ai}
\ee
for any $x\in \wt D_i$. We then define
\be
        \bB_i: = \nabla\times (\wt\chi_i \bA^\#_i) + \bB(x_i)
\label{def:bstari}
\ee
Easy calculations show that this field is $(\tD_i,  100)$-regular
and (\ref{eq:bi3})--(\ref{eq:bi2}) hold.

The gauge  $\bA^\#_i 
+ \sfrac{1}{2} \bB(x_i)\wedge (\cdot -x_i)$ generates $\bB$, hence
$$
        \bA=\bA^\#_i + \sfrac{1}{2} \bB(x_i)\wedge (\cdot -x_i)
        +\nabla\phi_i
$$
with some $\phi_i: \bR^3\to\bR$. Since
$(\wt\chi_i \bA^\#_i) 
+ \sfrac{1}{2} \bB(x_i)\wedge (\cdot -x_i)$ generates $\bB_i$,
we define
\be
        \bA_i: = (\wt\chi_i \bA^\#_i) 
+ \sfrac{1}{2} \bB(x_i)\wedge (\cdot -x_i) +\nabla\phi_i\; .
\label{def:Astar}\ee
Then $\nabla\times \bA_i = \bB_i$.
 $\;\;\;\Box$

\section{Positive energy regime: proof of (\ref{eq:Iest}) in Theorem
\ref{thm:spl}}\label{sec:pos1}
\setcounter{equation}{0}

We recall the set $\{ x_i \}$ constructed 
in Section \ref{sec:cover} and let
$$
        \ell_i : = \ell(x_i), \quad  P_i: = P(x_i) =
        \e^{-5}\ell_i^{-2},
        \quad
        b_i : = |\bB(x_i)|\; .
$$
We also recall that 
for any positive function $f$ we  denote the resolvents by
\be
        R_f := R[f]:= (\cD^2 + f)^{-1}\; .
\label{eq:resolv}
\ee
Note that in general $R_f$ and $R_g$ do not commute.
For simplicity we also introduce
\be
         R_i[f]:  = (\cD^2_i + f)^{-1}.
\label{eq:iresolv}
\ee

\begin{proposition}[Pull-up proposition]\label{prop:pullup}
Let $I$ be a countable index set
and let $g_i$, $i\in I$, be a family of nonnegative smooth functions such
that $0<\sum_{i\in I} g_i^2(x) <\infty$ for every $x\in \bR^3$.
Let $A_i$, $i\in I$ be a family of positive invertible self-adjoint operators
on $L^2(\bR^3, \bC^2)$. Then
\be 
    \Big(\sum_{i\in I}g^2_i \Big)
    {1\over \sum_{i\in I} g_i A_ig_i}
     \Big(\sum_{i\in I}g^2_i \Big) \leq \sum_{i\in I}
     g_i \; {1\over A_i} \; g_i\;.
\label{eq:pullup}
\ee
\end{proposition}

\noindent
{\it Proof of Proposition \ref{prop:pullup}.} This proof is basically
given in \cite{BFFGS}, we repeat it here for completeness.
All positive self-adjoint operators below are interpreted as quadratic
forms. We start with the operator inequality
\be 
      J^*J {1\over J^* A^{-1} J} J^*J \leq J^*AJ
\label{eq:proj}
\ee
for any positive self-adjoint operator $A$ and any operator $J$.

We define a map $J: L^2(\bR^3, \bC^2) \mapsto \bigoplus_i L^2(\bR^3,\bC^2)
=: \cH$ as $J: \psi \mapsto \{ g_i\psi\}$.
We define an operator
$\wt A$ on $\cH$ as $\wt A : \{ \psi_i\} \mapsto \{ A_i \psi_i\}$.
It is easy to check that 
$$
        J^* \wt A J = \sum_ig_i A_i g_i
        \qquad \mbox{on} \quad L^2(\bR^3,\bC^2)\; ,
$$
$J^*J= \sum_i g_i^2$
and that $(\wt A)^{-1} = \wt{A^{-1}}$.
Thus
$$
        \Big( \sum_i g^2_i\Big) {1\over \sum_i g_i A_i g_i}
        \Big(\sum_i g_i^2\Big) = J^*J {1\over J^*\wt A J} J^*J
        \leq J^* (\wt A)^{-1} J = \sum_i g_i {1\over A_i} g_i\;. \qquad \Box
$$

The following proposition is the localized version of
(\ref{eq:Iest}) for strong balls
 and its proof is given in  Section \ref{sec:pos}.

\begin{proposition} [Positive energy regime] \label{prop:pos}
Let $D$ be a ball of radius $\ell$ and let $K>0$ be a positive number.
Let $\bB$ be extended $(D,  K)$-regular, let 
the function $0\leq\chi\leq 1$ be supported
on $D$.
Then for any positive numbers $M,\mu>0$ there
exists  a constant $\e(M, K,\mu)$ such that for any $\e\leq
\e(M, K,\mu)$ we have
\be
        \Big| \Tr( \D^2 + \mu\e^{-5}\ell^{-2} -M\chi^2 V)_-\Big|
        \leq c(M,K,\e)\int_{ D} \Big( V^{5/2} + |\bB| V^{3/2}\Big)\; .
\label{eq:posprop}
\ee
\end{proposition}

\medskip

Armed with these two Propositions, we can finish the estimate
 (\ref{eq:Iest}) in Theorem \ref{thm:spl}.
Using the finite overlap property of $\tD_i$'s (Lemma \ref{lemma:cover}),
and that $\theta_i\leq \chi_i \leq 1$,
we see that
$$
        1\leq \Xi(x):=\sum_{i\in I} \chi_i^2(x) \leq N \; .
$$
Moreover, by the localization estimate,
$$
        \int |\cD\psi|^2 \ge \sfrac{1}{N} \sum_{i\in I}\int |\chi_i \cD\psi|^2
        \ge\sfrac{1}{2N}\sum_{i\in I} \int |\cD\chi_i\psi|^2 - \sfrac{2}{N}
        \sum_{i\in I} \langle \psi, (\nabla\chi_i)^2\psi\rangle \; ,
$$
hence
$$
        \cD^2 \ge \sfrac{1}{2N} \sum_{i\in I}\chi_i\cD^2\chi_i
        - \sfrac{8}{N}\sum_{i\in I}\ell_i^{-2}{\bf 1}( \tD_i)\;,
$$
where ${\bf 1}(\cdot )$ is the characteristic function.
Using (\ref{eq:deqd}) 
we may simply replace $\D^2$ by $\D^2_i$ on the support of $\chi_i$.
If $\e$ is sufficiently small, we obtain
\be
        \cD^2 + P+E \ge \sfrac{1}{4N} \sum_{i\in I} 
        \chi_i(\cD^2_i+P_i +E)\chi_i
\label{eq:IMS}
\ee
using the finite overlap property and
 that  $P$ is comparable to $P_i=\e^{-5}\ell_i^{-2}$ on $\tD_i$.
The resolvent can be estimated by
$$
        R_{P+E} \leq {4N\over  \sum_i \chi_i(\cD_i^2+P_i+E )\chi_i }
        \leq 4N\Xi^{-1}  \Big(\sum_{i\in I} \chi_i R_i[P_i+E]
         \chi_i \Big)\Xi^{-1}
$$
using Proposition \ref{prop:pullup}.
Hence, by the Birman-Schwinger priciple
\bey
        \Big|\Tr(\D^2 + P -4V)_-\Big|
        &=& \int_0^\infty n\Big( V^{1/2} R_{P+E} V^{1/2},
        \sfrac{1}{4}\Big)\rd E
        \nonumber\\
        &\leq&  \int_0^\infty  \,\,  n\Bigg( 
        \Xi^{-1}V^{1/2} \Big(\sum_{i\in I} \chi_i  R_i[P_i+E]
         \chi_i \Big)
        V^{1/2}\Xi^{-1}, \sfrac{1}{16N}\Bigg) \rd E
        \nonumber\\
        &\leq& \int_0^\infty  \,\,  n\Bigg( 
        \sum_{i\in I} V^{1/2}  \chi_i R_i[P_i+E]
         \chi_i 
        V^{1/2}, \sfrac{1}{16N}\Bigg) \rd E\;.
\label{eq:BSp}
\eey
Here we used 
\be
        n(ABA,e) = n(B^{1/2} A^2 B^{1/2},e)
\label{eq:switch}\ee
for any nonnegative operator $B$ and arbitrary operator $A$
 with the choice $A=\Xi^{-1}$
and we estimated $\Xi^{-2}\leq 1$.

We also use a strengthening of (\ref{eq:nsum}). If the positive
self-adjoint operators $A, B$ are disjointly supported, i.e.,
there exists an orthogonal projection $\Pi$ such that $\Pi A \Pi = A$
and $(I-\Pi)B (I-\Pi)= B$, then
\be
        n(A+ B, e)= n(A, e)+ n(B, e)\;.
\label{eq:ndissum}\ee
The proof is trivial.

In order to use  Proposition \ref{prop:pos}, we have to pull 
the summation out in (\ref{eq:BSp}). We split this sum
into a few infinite sums so that each contain disjointly supported
terms.
 Since the balls $\{\tD_i\}$ have uniformly finite overlap with constant $N$
(see (ii) of Lemma \ref{lemma:cover}),
there exists a partition of the index set $I = I_1 \cup I_2\cup \ldots
\cup I_{N+1}$ such that if $j,j'\in I_k$, for any $1\leq k\leq N+1$,
then $\tD_j\cap \tD_{j'} =\emptyset$. Such a partition
can be obtained by a greedy algorithm. We order the index set $I$
in some way
and we put each index one by one  into one of the sets.
We always put the new index into one of the sets where it has no
conflict 
with the indices already put into this set.
A new index $j$ is said to be in conflict with a previously
placed index $i$ if $\tD_i\cap \tD_j\neq\emptyset$.
Since every index can have a conflict with at most $N$ other indices,
each index can be placed somewhere at each step of the placement.

\bigskip

Hence, using (\ref{eq:nsum}) first, then (\ref{eq:ndissum}), we have
\bey
        n\Bigg( \sum_{i\in I} V^{1/2}\chi_i \;
         R_i[P_i+E] \chi_i V^{1/2}, \sfrac{1}{16N} \Bigg) 
        &=& n\Bigg( \sum_{k=1}^{N+1} \sum_{i\in I_k} V^{1/2}\chi_i \;
        R_i[P_i+E] \chi_i V^{1/2}, \sfrac{1}{16N} \Bigg)
        \nonumber\cr
        &\leq&  \sum_{k=1}^{N+1} n\Bigg( \sum_{i\in I_k} V^{1/2}\chi_i \;
        R_i[P_i+E] \chi_i V^{1/2}, \sfrac{1}{16N( N+1)} \Bigg) 
        \nonumber\cr
        &=& \sum_{k=1}^{N+1} \sum_{i\in I_k} n\Bigg(
        V^{1/2}\chi_i \;
        R_i[P_i+E] \chi_i V^{1/2},  \sfrac{1}{16N( N+1)} \Bigg)
        \nonumber\cr
        &=& \sum_{i\in I}   n\Bigg(
        V^{1/2}\chi_i \;
        R_i[P_i+E] \chi_i V^{1/2},  \sfrac{1}{16N( N+1)} \Bigg)\;,
\nonumber
\eey
so combining this estimate with (\ref{eq:BSp}) and applying
the Birman-Schwinger principle in the opposite direction, we obtain
\be
         \Big|\Tr(\D^2 + P -4V)_-\Big|  
         \leq \sum_{i\in I} \Big|\Tr( \D_i^2 + P_i -
        M\chi_i^2 V)_-\Big|
\label{eq:BSfin}
\ee
with $M:= 16N(N+1)$.

We then apply Proposition \ref{prop:pos} for each strong index $i$ 
for the ball $D=\tD_i$, radius $\ell=10\ell_i$, 
 and the magnetic field $\bB_i$ that is
extended $(\tD_i, K=100)$-regular. For small enough $\e$ we obtain
\bey
        \Big| \Tr( \D_i^2 + P_i -M\chi_i^2 V)_-\Big|
        &\leq &c(\e, M)\int_{\tD_i} \Big( V^{5/2} + |\bB_i|
        V^{3/2}\Big)
        \nonumber\\
       &\leq & c(\e, M)\int_{\tD_i} \Big( V^{5/2} + |\bB|
        V^{3/2}\Big) 
        \qquad \mbox{for}\;\;\ i \;\; \mbox{strong}, \nonumber
\eey
where the last inequality follows from (\ref{eq:Btem}).
For  the weak indices $i$  we use $\D_i^2 = (\bp +\bA_i)^2
+\bsigma\cdot\bB_i$
and $\|\bsigma\cdot\bB_i\| \leq c\e^3P_i$ (see (\ref{eq:AiA})) and
we obtain
\bey
        \Big| \Tr( \D_i^2 + P_i -M\chi_i^2 V)_-\Big|
        &\leq& \Big| \Tr( (\bp+\bA_i)^2 -M\chi_i^2 V)_-\Big| 
        \nonumber\cr
        &\leq& c(M)\int  V^{5/2}\chi_i^5
\nonumber
\eey
by the usual Lieb-Thirring inequality for magnetic Schr\"odinger
operators without spin.
Summing up these estimates we obtain from (\ref{eq:BSfin}) that
\bey
         |\Tr (\D^2 + P -4V)_-|  
        &\leq& c(\e)\sum_{i\in I} \int_{\wt D_i}
         \Big( V^{5/2} + |\bB| V^{3/2}\Big) 
        \nonumber\cr
        &\leq& c(\e) \int
         \Big( V^{5/2} + |\bB| V^{3/2}\Big) \; ,
\nonumber
\eey
again by the finite overlap
property of $\bB$.
This completes the estimate (\ref{eq:Iest}). $\;\;\;\Box$.

\section{Zero mode regime: proof of (\ref{eq:IIest}) in Theorem 
\ref{thm:spl}}\label{sec:zero1}
\setcounter{equation}{0}

The estimate   (\ref{eq:IIest}) essentially involves
estimating the square of the resolvent of $\cD^2$. However, the analog
of Proposition \ref{prop:pullup} does not hold for the 
square of the resolvent, i.e.
$$
        \Big( \sum g_i^2\Big) \Phi\Big( \sum g_i A_i g_i\Big)
                \Big( \sum g_i^2\Big)
        \leq \sum g_i \Phi(A_i)g_i
$$
with $\Phi(t)=t^{-2}$
is {\it not} true in general. Here is a 2 by 2 matrix counterexample
with $g_1=g_2=2^{-1/2}$:
$$
        A_1 =\pmatrix{1 & 1 \cr 1 & 2}\; ,
        \qquad A_2 = \pmatrix{2 & 1 \cr 1 & 2}\;.
$$

Without such an inequality, we have to use a resolvent expansion.
In addition to the square of the localized resolvent, we need to
control  offdiagonal terms. Such an estimate is given in the following
Proposition, which is a general statement about squares of resolvents of
second order differential operators.
The proof is given in Section \ref{sec:pullinpr}.

\begin{proposition}
 [Pull-in proposition]\label{prop:pullin} Given an $\e$-tempered
 function $\ell(x)$ and a  function $F(x)>0$ satisfying
\be
        \frac{1}{2}\leq {F(x)\over F(y)}\leq 2
\label{eq:Freg}
\ee
for all $|x-y|\leq \e^{-1}\ell(x)$. Set $P(x)=\e^{-5}\ell(x)^{-2}$.
Let $A= \cA \cdot \nabla + \cB$ be a first order differential
 operator acting on $\bigoplus_k L^2(\bR^3)$,
$1\leq k <\infty$, with smooth coefficients, i.e.
 $\cA(x)$ is a vector of $k\times k$ matrices,
$\cB(x)$ is a $k\times k$ matrix, all smoothly depending on $x$.
We assume that 
\be
        \sup_x\| \cA(x) \| \leq c_0 \; .
\label{eq:Abound}
\ee
Let  $T= A^*A$.

Given an $\ell$-uniform set of points
 $\{ x_i\}_{i\in I}$ as  in Lemma \ref{lemma:cover}.
Let  $\chi_i,\theta_i, \varphi_i$ be chosen as in Section
\ref{sec:cutoff}. 
We assume that for every $i$
 there exists a first order differential operator $A_i
= \cA_i \cdot \nabla + \cB_i$ 
 on $\bigoplus_k L^2(\bR^3)$ such that
\be
       \cA = \cA_i, \quad \cB = \cB_i \quad \mbox{on} \; 
       \mbox{supp}(\chi_i), 
\label{eq:coinc}
\ee
and let $T_i = A_i^*A_i$. 
Then there exists an $\e_0$ depending only on $c_0$ in (\ref{eq:Abound})
such that for any $\e\leq \e_0$ and $\mu\ge 0$ we have
\be
        {1\over T+P+\mu} F^2 {1\over T+P+\mu} \leq c
        \sum_{i\in I} F_i^2  \theta_i^2 \Big( P_i^{-1} {1\over T_i + P_i}
        A_i^*\varphi_i^2 A_i {1\over T_i + P_i}
         +{1\over (T_i + P_i)^2}\Big)\theta_i^2
\label{eq:pullin1}\ee
where $F_i: = \sup \{ F(x)\; : \; x\in \tD_i\}$
and $P_i = P(x_i)$.
\end{proposition}

{\it Remark.}
If $\cA$, $\cB$ are well-behaved,  then $T$ looks like an elliptic
constant coefficient differential operator on short scales. In this
case the
estimate localizes the square of the resolvent  in such a way
that the diagonal element of  the operator kernels on the right hand
side of (\ref{eq:pullin1}) remain finite.
This is clear for the second term on the RHS since 
the estimate is integrable in the ultraviolet regime (behaves like $p^{-4}$
in the momentum $p$). The first term behaves only as $p^{-2}$, but the
supports of  $\theta_i$ and $\varphi_i$ are well separated, which
makes the diagonal element finite.

The diagonal kernels of the localized operators are
estimated in the following Proposition whose proof 
is given in Section \ref{sec:zero}.

\begin{proposition}[Zero mode regime] \label{prop:zero}
Let $D$ be a ball of radius $\ell>0$ with center $z_0\in \bR^3$ and $K>0$. 
We assume that either

(i)  $\|\bB\|_\infty\leq c\e^{-2}\ell^{-2}$ and $\bB$ is supported
on the ten times bigger ball $\wt D=B(z, 10\ell)$; or

(ii)  $\bB$ is extended $(D, K)$-regular.

\noindent
Let $\D$ be any Dirac operator with magnetic field $\bB$.
Set $P:=\e^{-5}\ell^{-2}$, $R[P]:= (\D^2+P)^{-1}$,
  let $0\leq \varphi\leq 1$
be a function with $\mbox{dist}(z_0,\mbox{supp}\; \varphi) \ge 2\ell$.
 If $\e\leq\e(K)$, then
the following estimates hold for any $u\in D$
\bey
        \tr\; R[P]^2(u,u) &\leq& c(|\bB(u)|P^{-3/2}+P^{-1/2})
\label{eq:squarediag} \\
        \tr \Big(  R[P]\cD \varphi^2 \cD
        R[P]\Big)(u,u)
        &\leq& c(|\bB(u)|P^{-1/2}+P^{1/2})\; ,
\label{eq:sqderdiag}
\eey
where recall that $\tr := \tr_{\bC^2}$ stands for the trace in the spin space.
\end{proposition}

{\it Remark.} The diagonal elements in (\ref{eq:squarediag}),
(\ref{eq:sqderdiag}) are gauge invariant, i.e, they do not 
depend on the choice of the vector potential in the 
Dirac operator.

\medskip

Using these Propositions, we can complete the proof of
 the estimate  (\ref{eq:IIest})
in Theorem \ref{thm:spl}.
We use that
the function $F(x):=P(x)$ and the operators  $A:=\cD$, $A_i:= \cD_i$
 satisfy the conditions of Proposition
\ref{prop:pullin} by using (\ref{eq:deqd}) .
 Setting $\mu=E$ in (\ref{eq:pullin1})
we obtain
\bey
\lefteqn{
        \int_0^\infty  \,\,  n\Bigg( 
        2   V^{1/2}  R_{P+E}P^2R_{P+E}
        V^{1/2}, E\Bigg) \rd E }
\label{eq:IIsplit}\\
       & \leq& 2\int_0^\infty n\Bigg(  c V^{1/2}\sum_{i\in I} P_i^2 
        \theta_i^2 \Big( P_i^{-1}
        R_i[P_i]\cD_i\varphi_i^2 \cD_i R_i[P_i]
         + R_i^2[P_i]\Big)\theta_i^2 V^{1/2}, E \Bigg)\rd E \nonumber\\
       & =& c \sum_{i\in I} \Tr \;  V   \theta_i^4 
        \Big( P_i
        R_i[P_i]\cD_i\varphi_i^2 \cD_i  R_i[P_i]
         + P_i^2  R_i^2[P_i]\Big) \nonumber
\eey
using that $\int_0^\infty n(T, E)\rd E = \Tr \; T$ for any positive
operator $T$.
This sum of traces can be estimated by
\be
          c \sum_{i\in I}\int  V(x) \theta_i^4(x)  
        \Big( |\bB_i(x)| P_i^{1/2} + P_i^{3/2} \Big)\rd x \leq c
        \int VP^{1/2}(|\bB|+P)
\label{eq:II2}\ee
using Proposition \ref{prop:zero}
with $D=D_i$, $\ell=\ell_i$,  $\bB=\bB_i$, $K=100$ and for sufficiently
small $\e$.
The construction of $\bB_i$ for both weak and strong indices in Section
\ref{sec:appro} guarantees that either (i) or (ii)
holds true in Proposition \ref{prop:zero}.
 We also used that $|\bB_i|\leq c(|\bB|+P_i)$ and
 $P_i \leq c P(x)$ 
 on the support of $\theta_i$ (see
 (\ref{eq:Btem}), (\ref{eq:AiA}) and (\ref{eq:bi2})), moreover that
 $\sum_i\theta_i^4 \leq \sum_i\theta_i^2 =1$.
This completes the proof of (\ref{eq:IIest}).

\section{Cylindrical geometry of the second localization}\label{sec:cyl}
\setcounter{equation}{0}

Throughout this section  $\bB=(B_1, B_2, B_3)$ is
 an extended $(D, K)$-regular magnetic
field. Let $\bB_\infty$ be
the value of $\bB$ outside $D$, we set $b:=|\bB_\infty|$
and $\bn_\infty: = \bB_\infty/b$. The corresponding magnetic 2-form $\beta$
is given by
\be
     \beta: = B_3 \rd x_1 \wedge \rd x_2 +  B_1 \rd x_2 \wedge \rd x_3
     +  B_2 \rd x_3 \wedge \rd x_1\; .
\label{eq:2f}
\ee
Let $z_0$ be the center of $D$ and define the {\bf supporting plane}
of $\bB$,
$$
        \cP: = \{ z\in\bR^3\; : \; (z-z_0)\cdot \bn_\infty = -\ell\}\; ,
$$
to be the plane that is orthogonal to the parallel field lines outside $D$.
We fix an orthonormal basis $p_1, p_2$ in $\cP$ such that $p_1, p_2,
\bn_\infty$ is positively oriented. Any point $z$ in $\cP$ can
be identified with a point 
$\hat z =(\hat z_1, \hat z_2)\in \bR^2$ via $z-z_0= 
\hat z_1 p_1 +\hat z_2 p_2$, i.e. $\hat z_i= p_i\cdot (z-z_0)$, $i=1,2$.
We will use these coordinates to parametrize $\cP$.

For any $z\in\cP$ we denote
by $\varphi_z(\tau)$ the field line through $z$ with 
arc length parametrization $\tau$, i.e. 
$$
        \dot{\varphi_z}(\tau)
        =\frac{\rd}{\rd\tau}\varphi_z(\tau) 
         = \bn(\varphi_z(\tau))\; , \qquad \varphi_z(0)=z\; .
$$
Since $\bB$ is extended $(D,  K)$-regular,
$\dot\varphi_z$ is constant
outside $D$. If $\e$ is small enough  (depending on $K$),
we can assume that the length of the field line within $D$ is at most
$4\ell$ using (\ref{eq:ntem}). Therefore
 $\dot\varphi_z(\tau)$ is constant for $|\tau|\ge 4\ell$.

Every field line intersects $\cP$
since $\bn$ nowhere vanishes
and $\bn\cdot\bn_\infty \ge 1- \|\bn-\bn_\infty\|^2\ge \sfrac{1}{2}$ if 
$\e$ is sufficiently small. Therefore
the field lines $\{ \varphi_z(\tau)\; : \; z\in\cP\}$
form a foliation of $\bR^3$ and for each $x\in\bR^3$
we denote by $\pi(x)\in\cP$ the unique point such that 
$x=\varphi_{\pi(x)}(\tau)$ for some $\tau\in\bR$.

\subsection{Coordinates and conformal factor}\label{sec:cylcord}

The following lemma will be used to
 introduce coordinates, $\xi=(\xi_1, \xi_2, \xi_3)$, on $\bR^3$
associated with the field line $\varphi_z(\tau)$, $z\in\cP$,
which will be called the {\bf central field line}.
The field line will be characterized by $\xi_1=\xi_2=0$.
The point $z\in\cP$  will be called the {\bf base} of the coordinate
system.
The coordinates are functions of $x\in \bR^3$
and the inverse function will be denoted by $x(\xi):\bR^3\to\bR^3$.
We may also use the notation $\xi^z(x)$ and $x^z(\xi)$
to indicate the dependence on the base.
For notational convenience we sometimes use 
 $\xi_\perp: = (\xi_1, \xi_2)$.

In order to treat different error terms we introduce a notation
similar to the standard ``big-oh'' notation.

\begin{definition}\label{def:oxi} Let $k$ and
$\alpha$ be nonnegative integers and let $\ell>0$ be
a real number.
We say that a complex  function $f(\xi)$
is of class $\cO_k^\ell(|\xi_\perp|^\alpha)$ 
if there exists a constant $C$ such that
\be
        |\partial^\bm_\xi f(\xi)|\leq C \ell^{-|\bm|}
        \Big[ \min\Big( \frac{|\xi_\perp|}{\ell}, 1\Big)\Big]^{(\a - m_1-m_2)_+}
\label{eq:oxi}
\ee
for any multiindex $\bm=(m_1, m_2, m_3)$ with $|\bm|: =
m_1+m_2+m_3\leq k$. The definition can be extended to matrix
valued functions and to forms by replacing the absolute value
with any matrix or form norm on the left hand side of (\ref{eq:oxi}).
For $\ell=1$ we set $\cO_k(|\xi_\perp|^\a):=\cO^1_k(|\xi_\perp|^\a)$,
for $k=0$ we  set $\cO^\ell (|\xi_\perp|^\a) : =\cO_0^\ell
 (|\xi_\perp|^\a)$ and for $\a=0$ we set $\cO_k^\ell(1):=\cO_k^\ell(|\xi_\perp|^0)$.
\end{definition}

{\it Remark.} With a slight abuse
of  notation $\cO_k(|\xi_\perp|^\alpha)$ will be used
to denote not only the class of these functions
but any element of this class, similarly to the way $\cO(1)$
is used.

\bigskip

We also would like the magnetic field to be of constant strength
along the central field line which is achieved by a conformal
change of metric with a factor $\Om$. Let $\rd s^2$ be
the standard Euclidean metric and $\rd s_\Om^2 : = \Om^2 \rd s^2$
be a conformally equivalent one.
The following lemma describes the necessary information about
the new metric and coordinates. The proof is given in Section
\ref{sec:metricproof}.

\begin{lemma}[New metric and coordinates] \label{lemma:metric}
Given  positive numbers $K,\ell$, a ball $D$ of radius $\ell$,
and center $z_0$, 
 an extended $(D, K)$-regular magnetic field $\bB$, 
an orthonormal basis
$p_1,p_2$ in the supporting plane $\cP$ such that $p_1, p_2, \bn_\infty$
is positively oriented and the coordinate identification
$z\in \cP\leftrightarrow \hat z = (p_1\cdot (z-z_0), p_2\cdot (z-z_0))\in
\bR^2$.
If $\e$ small enough depending on $K$, then for any $z\in \cP$ there exist
coordinate functions $\xi=\xi^z(x)=
(\xi_1,\xi_2, \xi_3)
= (\xi_\perp, \xi_3)$, and positive functions $\Om(\xi), h(\xi)
 \in C^2(\bR^3)$ 
with the following properties:
\be
        \xi_3^z(x)=0 \quad \mbox{and}\quad \xi_\perp^z(x)
        = \hat x - \hat z \qquad \mbox{for}
        \quad x\in\cP\; ,
\label{eq:x=xi}
\ee
\be
        \xi_\perp^z(\varphi_z(\tau))= 0  \qquad \forall\tau \; .
\label{eq:xizero}
\ee
The function $(\hat z,x)\in \bR^5 \mapsto \xi^z(x) \in \bR^3$
and the inverse function  $(\hat z, \xi)\to x^z(\xi)$
belong to $C^3(\bR^5)$.
Moreover, if $D\xi$ and $Dx$
 denote the Jacobians of these functions,
then for
$\gamma=1,2,3$  we have
\be
        \big\| D^\gamma x \big\|, \; 
        \big\| D^\gamma \xi \big\| \leq c(K)\e^\gamma\ell^{-\gamma+1} \; .
\label{eq:jacder}
\ee
The metric
$\rd s_\Om^2:= \Om^2 \rd s^2$ can be expressed as
\be
        \rd s_\Om^2 = \sum_{i,j=1}^2 a_{ij}\rd\xi_i\rd\xi_j + 
        h^2\rd\xi_3^2 \; ,
\label{eq:dsomtran}
\ee
where $a_{ij}\in C^2(\bR^3)$ satisfies
\be
        \sup \Big\{ | a_{ij}(\xi) - \delta_{ij}| \; : \;
        \ell\leq |\xi_\perp|\leq 10\ell \Big\}  \leq c(K)\e \; .
\label{eq:abound}\ee
Moreover, $a_{ij} =\delta_{ij}$ away from the set 
$\ell\leq |\xi_\perp|\leq 10\ell$, i.e.
\be
        \rd s_\Om^2 = \rd\xi_1^2 + \rd\xi_2^2 +
        h^2\rd\xi_3^2 \qquad \mbox{on the domain} \qquad |\xi_\perp|\leq \ell
        \;\;\;\mbox{or} \;\; |\xi_\perp|\ge 10\ell \; .
\label{eq:dsom}
\ee
The functions $\Om$, $a_{ij}$ and $h$ also satisfy 
\be
        \Om \equiv a_{ij}\equiv h\equiv 1 \quad \mbox{on the domain}\quad
        |\xi_3|\ge 3\ell \; ,
\label{eq:Omhup}
\ee
\be
        \Om = f(\xi_3)( 1 + \e \cO_2^\ell(|\xi_\perp|)), \quad
        \mbox{and}\quad 
        h =   1 + \e \cO_2^\ell(|\xi_\perp|) 
\label{gttbound}
\ee
with
\be
        f(\xi_3): = \Bigg( \frac{ |\bB (x(0,\xi_3))|}{ b}
         \Bigg)^{1/2} \; ,
\label{def:f}
\ee
and
\be
        \Om \equiv 1, \quad \partial_{\xi_\perp}h\equiv 0,
          \quad \mbox{on the domain}\quad |\xi_\perp| \ge 10\ell \; .
\label{eq:omout}
\ee
Globally, the following bounds hold
\be
        \| h -1\|_\infty, \; \; 
        \|\Om-1\|_\infty \leq c(K)\e \; ,
\label{eq:homtemp}
\ee
\be
         \|\nabla^\gamma a_{ij}\|_\infty, \;  \; \|\nabla^\gamma h\|_\infty, 
        \;\; \|\nabla^\gamma \Om\|_\infty \leq c(K)\e^\gamma\ell^{-\gamma} \; ,
        \quad \gamma=1,2 \; .
\label{eq:homdertemp}
\ee
Moreover, there exists an orthonormal basis $\{e_1, e_2, e_3\}$
in the $\rd s_\Om^2$ metric such that $e_3 = h^{-1}\partial_{\xi_3}$
everywhere, and $e_j = \partial_{\xi_j}$, $j=1,2$,
apart from the region $\{ \xi\; : \;
\ell\leq |\xi_\perp|\leq 10\ell, |\xi_3|\leq 4\ell\}$.
\end{lemma}

{\it Remark.} 
The estimates in Lemma \ref{lemma:metric} actually depend only
on the variational lengthscale of the direction of
the magnetic field, $\bn$, and they are independent of the variational 
lengthscale of its strength. The proof given in  Section
\ref{sec:metricproof} uses a construction
that relies only on the field line structure and
on the logarithmic gradient of $|\bB|$ along the field
line. However, 
$$
    \frac{\nabla_\bn |\bB|}{|\bB|} = -\mbox{div} \, \bn
$$
since  $\bB =|\bB|\bn$
is divergence-free, therefore derivatives of $\bn$ alone
control the errors.
 
\bigskip

We define the spin-up projection associated with a field line
through a given point.  

\begin{definition}\label{def:pup}
Given $z\in\cP$, the field line $\varphi_z(\tau)$,
 the associated coordinates $\xi^z(x)$ and the inverse function $x^z(\xi)$
as defined in Lemma \ref{lemma:metric}.
Then the {\bf spin-up projection} associated with $\varphi_z(\tau)$
 is given by
a 2 by 2 matrix
\be
        P_z^\uparrow(x): = \frac{1}{2}\Big[ 1 + \bsigma\cdot
        \bn\Big( x^z(0, \xi_3^z(x))\Big)\Big]\; 
\label{eq:pup}
\ee
at any point $x\in\bR^3$.
\end{definition}
 Note that $P_z^\uparrow$ is constant on
the level sets of $\xi^z_3$.

\subsection{Cylindrical partition of unity and grid of field lines}
\label{sec:cylpart}

We start with a technical lemma.

\begin{lemma}
Given $y\in\cP$ and the associated coordinates
$\{ \xi_k^{y}\}$, $k=1,2,3$,
as constructed in Lemma \ref{lemma:metric}, then for
any sufficiently small $\e\leq \e(K)$ and
 any $z\in\bR^3$
\be
        \frac{1}{2} \leq \frac{|\xi_\perp^y(z)|}{|y-\pi(z)|}
        \leq 2\; ,
\label{eq:yz}
\ee
\be
        \| P^\uparrow_y(z)-
         P^\uparrow_{\pi(z)}(z)\|\leq cK\e\ell^{-1}|y-\pi(z)|\; ,
\label{eq:yzspin}
\ee
where $\|\cdot\|$ denotes the standard norm of  2 by 2 matrices.
\end{lemma}

\noindent
{\it Proof.} Denote $u=\pi(z)$ and set
$q(\tau):  = \xi^y(\varphi_u(\tau))
-\xi^y(\varphi_y(\tau)) \in\bR^3$
and let $r(\tau): = q_\perp(\tau)= (q_1(\tau), q_2(\tau))$.
We have $|q(0)|=|r(0)|=|u-y|$ and by $\dot\varphi(\tau)=\bn(\varphi(\tau))$
and Lemma \ref{lemma:metric} we
can estimate
\bey
        |\dot q(\tau)|&\leq& \|D_x\xi\|_\infty |\dot\varphi_u(\tau)
        -\dot\varphi_y(\tau)| + \| D^2_x\xi\|_\infty
        |\varphi_u(\tau)-\varphi_y(\tau)|\nonumber\\
        &\leq&  ( \|D_x\xi\|_\infty \|\nabla\bn\|_\infty +
        \|D^2_x\xi\|_\infty)
        |\varphi_u(\tau)
        -\varphi_y(\tau)| \nonumber\\
        &\leq&  (\|D_x\xi\|_\infty\|\nabla\bn\|_\infty +
        \|D^2_x\xi\|_\infty) \| (D_x\xi)^{-1}\|_\infty
         |q(\tau)| \nonumber \\
        &\leq& c(K)\e\ell^{-1}|q(\tau)|
\label{eq:grom}
\eey
for $|\tau|\leq 4\ell$ and $\dot q(\tau) \equiv 0$ for $|\tau|\ge 4\ell$.
Therefore $\sup_\tau |q(\tau)|\leq |u-y| e^{c(K)\e}$ by Gromwall's inequality
and
$$
        \sup_\tau |\dot r(\tau)| \leq \sup_\tau |\dot q(\tau)|
\leq cK\e\ell^{-1} e^{c(K)\e}|u-y| \; .
$$
Combining this with $|r(0)|=|u-y|$ we obtain
$\sfrac{1}{2} |u-y|\leq \sup_\tau |r(\tau)|\leq 2|u-y|$
if $\e$ is sufficiently small.
Note that for some $\tau$
$$
        |\xi_\perp^y(z)|
        = \Big| \xi_\perp^y(\varphi_{\pi(z)}(\tau)) -
        \xi_\perp^y(\varphi_{y}(\tau))\Big| = |r(\tau)|\; ,
$$
which concludes the proof of (\ref{eq:yz}).

For the proof of (\ref{eq:yzspin}) we again set $u=\pi(z)$ and 
by Definition \ref{def:pup} and
Lemma \ref{lemma:metric} we estimate
\bey
        \|P^\uparrow_y(z) -P^\uparrow_u(z)\| 
        &\leq& \|\nabla\bn\|_\infty \Big| x^y(0, \xi_3^y(z)) - 
        x^u(0,\xi_3^u(z))\Big| \nonumber\\
        &\leq& \|\nabla\bn\|_\infty \Big( 
        \Big| x^y(0, \xi_3^y(z)) - x^y(0,\xi_3^u(z))\Big|
        + \Big| x^y(0, \xi_3^u(z)) - x^u(0,\xi_3^u(z))\Big|\Big) \nonumber\\
        &\leq &
        \|\nabla\bn\|_\infty (  2\| D_u\xi^u\|_\infty +\|D_ux^u\|_\infty)
        |y-u|\nonumber\\
        &\leq& cK\e\ell^{-1}|y-u| \nonumber\; ,
\eey
using $|x^y(\xi)-x^y(\xi')|\leq 2|\xi-\xi'|$ that follows
from (\ref{eq:jacder}) if $\e$ is sufficiently small.
 This completes the proof of 
(\ref{eq:yzspin}). $\;\;\Box$

We construct a grid of field lines.
Choose a  square lattice
$\cY: =\{ y_j \; : \; j\in \bZ^2\}$
 on $\cP$ with spacing $b^{-1/2}$, i.e., $|y_j-y_k|= b^{-1/2}|j-k|$,
$j,k\in \bZ^2$.
Applying Lemma \ref{lemma:metric} to each field line $\varphi_{y_j}$,
we construct 
conformal factors $\Om_j$, orthonormal bases 
$\{e_1^{(j)}, e_2^{(j)}, e_3^{(j)} \}$ and
coordinate functions $\xi^{(j)}=(\xi^{(j)}_{1}, \xi^{(j)}_{2},
 \xi^{(j)}_{3})$. 
We now construct a set of Gaussian localization functions
with a transversal lengthscale of order $b^{-1/2}$
that are essentially supported around the field lines  $\varphi_{y_j}$.
Let $\eta\leq \sfrac{1}{4}$ be
  a small positive number to be specified later and we
define
\be
        v_j(x) =  \exp{\Big( - \frac{\eta b}{4}
          [\xi_\perp^{(j)}(x)]^2\Big)} \; .
\label{eq:gauss}
\ee
We set $P_j^\uparrow(x):=P_{y_j}^\uparrow(x)$ to be
 the 2 by 2 spin-up projection matrix associated with the field
line through $y_j$ (see Definition \ref{def:pup}).

\begin{lemma}\label{lemma:moments}  If $\e$ is sufficiently small
 depending only on $K$, then 
 for any $\gamma>0$, $\kappa \ge 0$ we have
\bey
         \sum_{j\in\bZ^2} 
        (\eta b)^\kappa [\xi^{(j)}_\perp(x)]^{2\kappa} v_j(x)^\gamma
  &\leq &
         c(\gamma,\kappa)\eta^{-1} \; ,
\label{eq:moments} \\
          \sum_{j\in\bZ^2} 
         v_j(x)^\gamma
 &\ge& c(\gamma)\eta^{-1} \; ,
\label{eq:alsomoments} 
\eey
uniformly in $x\in \bR^3$.
Moreover,  there is a universal
constant $C_0$ and for any $0<\lambda<1$ there exists $0 < \eta(\lambda)\leq 
\sfrac{1}{4}$
such that for any $\eta\leq \eta(\lambda)$
\be
        \sum_{j\in\bZ^2} v_j^4(x) \Bigg[
        b\Big( \lambda - \eta^2 b [\xi_\perp^{(j)}(x)]^2\Big)
        P_j^\uparrow(x)  + C_0\ell^{-2}\Bigg]\ge 0\; . 
\label{eq:spinalloc}
\ee
\end{lemma}

\noindent
{\it Proof.} Since $(\eta b \xi_\perp^2)^\kappa \exp (- \sfrac{\gamma
\eta b}{8}\xi_\perp^2) \leq c(\gamma,\kappa)$ uniformly in $\xi_\perp$,
it is sufficient to estimate $\sum_j v_j^{\gamma/2}$ for the
proof of (\ref{eq:moments}).
Using (\ref{eq:yz}) we obtain
$$
        \sum_{j\in\bZ^2} v_j^{\gamma/2}(x) 
        \leq \sum_j \exp\Big( - \frac{\gamma\eta b}{16}
        |y_j-\pi(x)|^2\Big) \leq c(\eta\gamma)^{-1}
$$
since $y_j$ runs through a square grid with spacing $b^{-1/2}$.
The proof of (\ref{eq:alsomoments}) is similar.

For the proof of (\ref{eq:spinalloc}) we define $k\in \bZ^2$
to be an index such that $|y_k-\pi(x)|\leq b^{-1/2}$.
Then by (\ref{eq:yz}) and Schwarz' inequality
$$
        |y_k-y_j|^2\leq 2b^{-1} + 2|\pi(x)-y_j|^2
        \leq 2b^{-1} + 4|\xi_\perp^j(x)|^2 \; .
$$
Combining this estimate with (\ref{eq:yzspin})
and using that $(P^\uparrow)^2=P^\uparrow$ we have 
$$
        P_j^\uparrow \ge \sfrac{1}{2}P_k^\uparrow
        - 2\|P_j^\uparrow-P_k^\uparrow\|^2
        \ge \sfrac{1}{2}P_k^\uparrow - \ell^{-2}|y_k-y_j|^2
        \ge  \sfrac{1}{2}P_k^\uparrow - 4\ell^{-2}|\xi_\perp^j|^2
        - cb^{-1}\ell^{-2}
$$
and
$$
        P_j^\uparrow\leq 2P_k^\uparrow+ 2(P_j^\uparrow-P_k^\uparrow)^2
        \leq 2P_k^\uparrow + 4\ell^{-2}|\xi_\perp^j|^2 +cb^{-1}\ell^{-2} 
$$
if $\e$ is sufficiently small.
We omitted the $x$ argument for brevity.
Therefore  we can use (\ref{eq:moments}) and (\ref{eq:alsomoments})
 to estimate
\bey
\lefteqn{
        \sum_j v_j^4 b(\lambda-\eta^2b |\xi^j_\perp|^2)P_j^\uparrow} 
        \nonumber\\
        &\ge& \frac{b}{2} 
        \sum_j v_j^4 (\lambda -4\eta^2 b |\xi_\perp^j|^2) P_k^\uparrow
        - c\sum_j v_j^4\Big( \lambda b |\xi^j_\perp|^2 + c\lambda +
        \eta^2b^2 |\xi^j_\perp|^4 + \eta^2 b|\xi^j_\perp|^2\Big)\ell^{-2}
        \nonumber\\
        &\ge& \frac{b}{2} (c\eta^{-1}\lambda -c)P_k^\uparrow
        - c\eta^{-1}\ell^{-2}\nonumber\\
        & \ge& - \sum_j v_j^4(x) c\ell^{-2}\nonumber
\eey
if $\eta$ is sufficiently small. We can choose $C_0$ 
to be the universal constant $c$ in the last formula and
the proof of (\ref{eq:spinalloc}) is completed.
 $\;\;\Box$

\section{Dirac operator on $\bR^3$ with a general metric}\label{sec:gendirac}
\setcounter{equation}{0}

The following sections  summarize basic information
about the Dirac operator over a non-flat manifold.
More details are found in \cite{ES-III}
(the sign of $\bA$ is chosen to be the opposite in this paper).
The presentation here is simplified because
the spinor bundle is trivial and we can work
in a global orthonormal basis.

Throughout this section
we shall consider $\bR^3$ with a general Riemannian metric $g =
(\cdot, \cdot)$
and we shall consider the Dirac operator for this particular
Riemannian manifold. The Dirac operator will be an unbounded
self-adjoint operator in $L_g^2(\bR^3)\otimes \bC^2$ (the subscript
$g$ refers to the fact that the measure is the volume form of $g$).

Let $\{e_1, e_2, e_3\}$ be a global orthonormal basis of vectorfields
and let $\{ e^1, e^2, e^3\}$ be the dual basis. If $X$ is a vectorfield
on $\bR^3$, we denote by
\be
      P_X: = \frac{1}{2}\Big[ 1 + \sum_{j=1}^3 (X, e_j)\sigma^j\Big]
\label{def:PX}
\ee
the {\it spin projection} in direction $X$ with respect to the basis
$\{e_1, e_2, e_3\}$.

We also introduce a {\it covariant derivative} on $L^2_g(\bR^3)\otimes\bC^2$ by
\be
        \nabla_X: = \partial_X + \sfrac{i}{2} \bsigma\cdot\bomega(X)
        \; .
\label{eq:covdef}
\ee
Here we define
\be
        \bomega(X): = \Big( (\nabla_X e_3, e_2),\;  (\nabla_X e_1, e_3),
        \; 
        (\nabla_X e_2, e_1)\Big)\; ,
\label{eq:bomega}
\ee
where $\nabla_X$ refers to the Levi-Civita connection
on vectorfields for the metric $g$ on $\bR^3$.

If $\a$ is a (real) 1-form we define the corresponding covariant 
derivative on $L^2_g(\bR^3)\otimes\bC^2$
(see Proposition 2.9 in \cite{ES-III})
\be
        \nabla_X^\a:=\nabla_X + i\a(X)\; .
\label{eq:nablaalp}
\ee
The magnetic 2-form is $\beta:=\rd \a$.
We define the {\it Dirac operator} by
\be
       \D^\a :=
      \sum_{j=1}^3 \sigma^j(-i\nabla^\a_{e_j})\; .
\label{def:dirac}
\ee
It is a symmetric operator in $L^2_g(\bR^3)\otimes\bC^2$
(Theorem 3.2 in \cite{ES-III}).
Note that $\D^\a$ also depends on the metric $g$ and the choice of
$\{ e_1, e_2, e_3\}$ but this fact will usually be suppressed in
the notation.

 For notational convenience 
 we introduce the
following {\it vector of covariant derivatives}
\be
      \bPi^\a : = (-i\nabla_{e_1}^\a, 
        -i\nabla_{e_2}^\a, -i\nabla_{e_3}^\a)\; .
\label{eq:momvector}
\ee
With this notation we may write $\D^\alpha = \bsigma\cdot \bPi^\a$.
Note that the components of $\bPi^\a$ are not self-adjoint,
however the components of the vector
\be
        \bD^\a = (D_1^\a, D_2^\a, D_3^\a):= \bPi^\a -\frac{i}{2}
        \Big( \mbox{div}\; e_1, \mbox{div}\; e_2,  \mbox{div}\; e_3\Big)
\label{def:D}
\ee
are self-adjoint operators.

For any one form $\lambda=\lambda_1e^1+ \lambda_2e^2+\lambda_3e^3$ 
we define $\sigma(\lambda):=\lambda_1\sigma^1+ \lambda_2\sigma^2
+\lambda_3\sigma^3$.
The Lichnerowicz' formula (see, e.g., Theorem 3.4 in \cite{ES-III})
states that
\be
        [\D^\a]^2 = [\bPi^\a]^*\cdot\bPi^\a + \frac{1}{4} R +
         \sigma(\star\beta)\; ,
\label{eq:lich}\ee
where $R$ is the scalar curvature of $g$ and $\star$ denotes the Hodge dual.

In terms of $D_j^\a$ operators, the Lichnerowicz' formula reads as
\be
        [\D^\a]^2 = [\bD^\a]^2  + \frac{1}{4}  R + \frac{1}{4}  
        \sum_{j=1}^3 [\mbox{div}\; e_j]^2
        +\frac{1}{2} \sum_{j=1}^3 \partial_{e_j} (\mbox{div} \; e_j)
        + \sigma(\star\beta)\; .
\label{eq:lichD}\ee

\medskip

For the flat Euclidean metric with the standard orthonormal
basis we have $\bomega\equiv0$. In this case if $a$ denotes
 the 1-form  dual to the vector field $\bA=(A_1, A_2, A_3)$ then
$\bPi^a = \bbp_\bA$ where $\bbp_A$ is the vector of operators
 $(-i\partial_1 + A_1,
-i\partial_2 + A_2, -i\partial_3 + A_3)$. Therefore we 
 obtain
the usual Dirac operator $\D=\D^a$  defined in Section \ref{sec:intro}
with $\beta=\rd a$ given by (\ref{eq:2f}).
Moreover, $P_X = P_z^\uparrow$ if $X(x)= \bn(x^z(0,\xi^z(x)))$
from (\ref{eq:pup}) and (\ref{def:PX}).

\subsection{Gauge transformation}\label{sec:gauge}

In the previous construction $\D^\a$ and $\bPi^\a$ depend on $\a$
and also on
 $\{e_1, e_2, e_3\}$. Up to a unitary equivalent gauge transformation,
 however, $\D^\a$ and $\bPi^\a$ depend only on the metric $g$ and the 
magnetic 2-form $\beta$. Similarly, the
 spin projection $P_X$ defined in (\ref{def:PX}) is gauge-invariant.

 More precisely, given another 1-form $\a'$ 
with $\rd\a' =\beta$ and another orthonormal basis  $\{e_1', e_2',
e_3'\}$
with the same orientation, 
we denote the corresponding
operators by $\D'$ and $\bPi'$ and let $P_X'$ be
the spin projection. 
There exist a real valued function $\phi(x)$
and a continuous function $R(x)\in SO(3)$ on $\bR^3$ such that
$\a' = \a + \rd\phi$ and $\sum_k w_k e_k' = \sum_k (R\bw)_k e_k$
for any $\bw\in \bR^3$. Let $U_R(x)\in SU(2)$ denote the
image of $R(x)$ under the isomorphism $SO(3)\to SU(2)/\{\pm 1\}$.
The requirement that $U_R(x)$ be a continuous function of $x$
determines $U_R$ uniquely up to a global sign.
In particular 
\be
       U_R (\bsigma\cdot\bv) U_R^* = \bsigma\cdot (R\bv)
\label{eq:spinrot}
\ee
for any $\bv\in\bR^3$,
i.e. $ R(\psi,\bsigma\psi) = (U_R\psi, \bsigma U_R\psi)$
for any $\psi\in \bC^2$, where $(\psi,\bsigma\psi)$
denotes the vector $\Big( (\psi, \sigma^1\psi), (\psi, \sigma^2\psi),
(\psi,\sigma^3\psi)\Big)\in\bR^3$.

We define the unitary operator of the form
\be
          [\cU_{R,\phi}\psi](x) = e^{i\phi(x)} 
          U_R(x)\psi(x)\; ,
\label{eq:gauge}
\ee
then
\be
        P_X' = \cU_{R,\phi}^* P_X \cU_{R,\phi}
\label{eq:pupgauge}
\ee
and
\be
        \D'  = \cU_{R,\phi}^* \D^\a \cU_{R,\phi}\; ,
        \qquad \mbox{and} \quad \bw\cdot\bPi'  = 
        (R \bw)\cdot \cU^*_{R,\phi}\bPi^\a \cU_{R,\phi}
\label{eq:dgauge}
\ee
for any $\bw\in \bR^3$.
In particular, the spectrum of $\D^\a$ and the functions
$$
     \tr \Big({1\over ([\D^\a]^2 +c)^2}(x,x)\Big) \qquad \mbox{and}\qquad
     \tr \Big({1\over [\D^\a]^2 + c}\D^\a \varphi^2\D^\a {1\over [\D^\a]^2 +
       c}(x,x)\Big)
$$
depend only on $g$ and $\beta$, where $\varphi$ is any function on
$\bR^3$
and $c>0$ is   a constant.

\subsection{Change of the Dirac operator under a conformal
change of the metric}\label{sec:confch}

 Let $\Om$ be a positive real function on $\bR^3$ and let $g_\Om: =
\Om^2 g$ be a metric which is conformal to $g$. 
Consider the $(f_1, f_2, f_3):=(\Om^{-1} e_1,  \Om^{-1} e_2, \Om^{-1} e_3)$
 orthonormal basis in $g_\Om$.  Given  a 1-form $\a$ 
 we let $\nabla_X^{\a, \Om}$ and $\D_\Om^\a$ 
denote the corresponding covariant derivative and
Dirac operator. With the notation
\be
        \bPi_\Om^\a:= \Big(-i\nabla_{f_1}^{\a,\Om},
        -i\nabla_{f_2}^{\a,\Om}, -i\nabla_{f_3}^{\a,\Om}
        \Big)\; 
\label{eq:momvec}
\ee
we have $\D_\Om^\a = \bsigma\cdot \bPi_\Om^\a$.
Then from Section 4 of \cite{ES-III}  
\be
      \D_\Om^\a = \Om^{-2}\D^\a\Om \; 
\label{eq:confdirac}
\ee
and
\be
        \nabla_X^{\a,\Om} =\nabla_X^\a +\sfrac{1}{4} \Om^{-1}
        [\sigma(X^*), \sigma(\rd\Om)]
\label{eq:nablaom}
\ee
for any vector $X$, where $X^*$ refers to the 1-form which is
dual to the vector $X$ relative to the metric $g$,
and $\sigma(X^*), \sigma(\rd\Om)$ are computed in the $\{ e_1, e_2, e_3\}$
basis. In particular,
\be
        \bPi^\a_\Om= \Om^{-1}\bPi^\a - \sfrac{i}{4}\Om^{-2} 
        \Big( [\sigma^1, \sigma(\rd\Om)],
        [\sigma^2, \sigma(\rd\Om)], [\sigma^3, \sigma(\rd\Om)]\Big)\; .
\label{eq:piaom}
\ee

\subsection{Constant approximation of the magnetic field
along a field line}\label{sec:apprfield}

The goal of this section is to express the Dirac operator
with a non-homogeneous regular magnetic field  as a sum of a constant field
Dirac operator  and some error terms in a neighborhood of a given
field line. This can be done if the original
Dirac operator is already written in an appropriate orthonormal basis
and with a carefully selected vector potential. The basis and the 
vector potential are determined by the local magnetic field.

Given an extended $(D,K)$-regular field $\bB$. Consider the
corresponding 2-form $\beta$ and a fixed field line.
Let the coordinates $(\xi_1, \xi_2, \xi_3)$, 
the  new metric $g_\Om=\Om^2\rd s^2$ 
with a conformal factor $\Om$ and the orthonormal basis
 $\{ e_1, e_2, e_3\} $ 
be as constructed in Lemma \ref{lemma:metric}, 
associated with the given field line.
Let $\a$ denote a vector potential, $\rd\a=\beta$, to be chosen later.
Let $\bPi^\a$ be  given by (\ref{eq:momvector})
and let $\D^\a:=\bsigma\cdot\bPi^\a$.

On the central line 
and in the regime $|\xi_\perp|\ge 10\ell$
the magnetic field  $\beta$ is constant in
the $\rd s_\Om^2$ metric:
$$
        \beta (e_1, e_2) = \Om^{-2} \beta( \Om e_1,
        \Om e_2) = \Om^{-2} |\bB| = b , \qquad
        \beta (e_j, e_3) = 0, \quad j=1,2 \; .
$$
This observation gives rise to the following definition.
\begin{definition}\label{def:constfield} 
Given a field line, the associated coordinate system $\xi$
and the conformal factor $\Omega$ as above such that
magnetic field $\beta$ is constant in the $\rd s_\Om^2$ metric
with strength $b = \beta(e_1, e_2)$.
Then the magnetic field $\beta_c$ given by
$$
        \beta_c : = b\; \rd\xi_1\wedge \rd\xi_2 \; 
$$
is called the approximating constant magnetic field along
the field line. 
\end{definition}

The magnetic field $\beta_c$ is clearly constant 
in the $\rd\xi^2=\sum_{j=1}^3 \rd\xi_j^2$ metric.
A  convenient gauge  is defined as  
 $\a_c: = \sfrac{b}{2}[
\xi_1\rd\xi_2 - \xi_2\rd\xi_1]$, then $\beta_c =\rd\a_c$.

In particular $\beta=\beta_c$ along the central line
and in the regime $|\xi_\perp|\ge 10\ell$.
We compute the norm of the difference field 
$\delta\beta : = \beta
-\beta_c$ and the norm of its derivative
in the $\rd\xi^2$ metric. Using
 (\ref{eq:jacder}),
(\ref{eq:abound}), (\ref{eq:dsom}), (\ref{eq:homtemp})
and (\ref{eq:bi3}), (\ref{eq:bi4}) we obtain
\be
        \delta\beta(\xi)= \e b \cO_2^\ell(|\xi_\perp|) \; 
\label{eq:deltab}
\ee
and $\delta\beta(\xi) \equiv 0$ if $|\xi_\perp|\ge 10\ell$.

\medskip

Next, we define an appropriate gauge $\a$ for the original magnetic
field $\beta$, $\rd \a =\beta$, such that $\a -\a_c$ be small.
The following Lemma was given in \cite{ES-I} (Proposition 2.3).
Although it was stated in a slightly weaker form,
  the explicit formula (2.20) of \cite{ES-I}
gives the following stronger result with a straightforward 
computation:

\begin{lemma}[A-formula]\label{lemma:A}
Given any $C^2$ magnetic 2-form $\beta$ on  $\bR^3$ with
Euclidean cooordinates $(\xi_1,\xi_2,\xi_3)$.
For  $k, m\in \bN$ we define
$$
      b_{k,m}(\xi_\perp): =   \int_0^{\xi_1}
      u^k \sup_{z_2, z_3} \|\nabla^m \beta (u, z_2, z_3)\| \rd u
      +  \int_0^{\xi_2}
      u^k \sup_{z_1, z_3} \|\nabla^m \beta (z_1, u , z_3)\| \rd u \; .
$$
Then
there exists a 1-form $\alpha$ generating $\beta$,  $\rd\a = \beta$,
such that
\bey
          \|\alpha(\xi)\|&\leq& c \Big[ b_{0,0}(\xi_\perp) +
          b_{1,1}(\xi_\perp)\Big]\; ,
\label{eq:Aformest}\\ \nonumber\\
          \|\nabla\alpha(\xi)\|&\leq & 
          c  \Big[\sup \{\|\beta(u)\| \; : \; |u_\perp|\leq
          |\xi_\perp|\}  +
          b_{0,1}(\xi_\perp) + b_{1,2}(\xi_\perp)\Big]\; .
\label{eq:Aformderest}
\eey
\end{lemma}

We apply this lemma to the magnetic 2-form $\delta\beta$ and we
denote by $\delta\a$ the generating 1-form.
We define $\a: =\a_c + \delta \a$, then $\a$ generates 
the original magnetic field $\beta$,
$\rd\a=\beta$ and it is close to the linear gauge
$\a_c$ of the constant field $\beta_c$ using (\ref{eq:deltab})
and Lemma \ref{lemma:A}:
\be 
   (\a-\a_c)(\xi) =  \e b\ell \cO^\ell_1(|\xi_\perp|^2)\; .
\label{eq:adiff} 
\ee
 The norm of the left hand side is computed with respect to
the standard metric.

\medskip

\begin{definition}\label{def:constdirac}
With the notations above, the Dirac operator
\be
      \tcD: = \sum_{k=1}^3 \sigma^k[-i\partial_{\xi_k} +
      \a_c (\partial_{\xi_k} )] 
\label{eq:dbdef}
\ee
with a constant field $\beta_c$ in the $\rd\xi^2$ metric
will be called the approximating constant field Dirac operator
along the field line.
\end{definition}
 
By the properties of the coordinate vectorfields $\partial_{\xi_k}$
and the orthonormal basis $\{ e_1, e_2, e_3\}$
in the $g_\Om$ metric from Lemma \ref{lemma:metric}
and by the definitions (\ref{eq:covdef}), (\ref{eq:nablaalp}),
(\ref{eq:momvector}) we have, for sufficiently small $\e$,
\be
      \D^\a = \bsigma\cdot\bPi^\a = \tcD + \sum_{k=1}^3 \sigma^k
      [\a (e_k) - \a_c(\partial_{\xi_k})] + \sum_{k=1}^3 \cK_k
      (-i\partial_{\xi_k})  + \cM_0\;,
\label{eq:Dblong}
\ee
where $\cK = (\cK_1, \cK_2, \cK_3)$ and $\cK_k, \cM_0$ are 2 by 2 matrix
valued functions.
We use the  bounds (\ref{eq:abound}), (\ref{gttbound}), (\ref{eq:homdertemp})
 and
the estimate (\ref{eq:adiff})
 to obtain,
for $k=1,2,3$,
\bey
       \Big|[\a (e_k) - \a_c(\partial_{\xi_k})](\xi)\Big|&=&
         \e b\ell \cO^\ell_1(|\xi_\perp|^2)\; .
\label{eq:dalpest2}
\eey
We obtain from (\ref{eq:Dblong}) and (\ref{eq:dalpest2}) 
that
\be
      \D^\a  = \tcD +
      \sum_{k=1}^3\cK_k(-i\partial_{\xi_k})
       + \cM
\label{eq:Dbr}
\ee
with matrix valued functions that satisfy
\be
        \cM=  b \ell \cO^\ell_1(|\xi_\perp|) + \ell^{-1}\cO^\ell_1(1)
        \;, \quad \cK_{1,2}  =  \cO^\ell_1(|\xi_\perp|^{\gamma}) \;, \quad
        \cK_3 = \cO^\ell_1(|\xi_\perp|)\; 
\label{eq:mkestell}
\ee
for any $\gamma\ge 0$
if $\e\leq\e(K)$. These estimates follow from Lemma \ref{lemma:metric},
especially from the fact 
 that $e_k=\partial_{\xi_k}$, $k=1,2$,
apart from the region $\ell\leq |\xi_\perp|
\leq 10\ell$, where (\ref{eq:dsomtran}) holds, i.e. $\cK_1, \cK_2$
are supported in this region.

\section{Positive energy  regime: Proof of Proposition
\ref{prop:pos}}
\label{sec:pos}
\setcounter{equation}{0}

We first notice that both sides of (\ref{eq:posprop}) scale 
as $\ell^{-2}$, hence it is sufficient to prove the result for
$\ell:=1$.
We can apply the constructions
of Section \ref{sec:cyl} for the magnetic field $\bB$ to obtain
conformal factors $\Om_j$, orthonormal bases $\{ e_1^{(j)}, 
e_2^{(j)}, e_3^{(j)}\}$,
coordinate functions 
$\xi^{(j)}=(\xi^{(j)}_{1}, \xi^{(j)}_{2},
 \xi^{(j)}_{3})$, spin-up projections $P_j^\uparrow$
 and Gaussian localization functions $v_j$ concentrated
along the field line passing through $y_j$. We recall
that $y_j$ was a lattice with spacing $b^{-1/2}$
on the supporting  plane (see Section \ref{sec:cylpart}).

We first estimate
$$
        \D^2 = \Big(1- \e^{-2}\sfrac{1}{2b}\Big)
         \D^2 + \e^{-2}\sfrac{1}{2b}\big [\bbp_{\bA} ^2 +
        \bsigma\cdot\bB\big]
        \ge \sfrac{1}{2}\D^2 +  \e^{-2}\sfrac{1}{2b}\bbp_{\bA}^2
         -\e^{-2}
$$
since $b= |\bB_\infty|\ge \e^{-2}$
and $\sup |\bB|\le 2b $ if $\e$ is sufficiently small.

Using this estimate and (\ref{eq:moments})--(\ref{eq:alsomoments})
 we have
\bey
        \D^2 + \mu\e^{-5} -M\chi^2V &\ge&
        \sfrac{1}{2}\Big( \D^2 + \e^{-2}\sfrac{1}{b}\bbp_{\bA}^2+
        \mu\e^{-5} -2M\chi^2V\Big)
\label{eq:pap}\\
        &\ge& c \eta\sum_j \Big( \D v_j^4 \D +
         \e^{-2}\sfrac{1}{b}\bbp_{\bA}\cdot v_j^2\bbp_{\bA} +\mu\e^{-5}
         v_j-cM\chi^2Vv_j^6\Big)\nonumber
\eey
if $\e$ is sufficiently small (depending on $\mu$).
Notice from the explicit formula (\ref{eq:gauss}) that
\be
        \Big|[\bbp_\bA , v_j] \Big|^2 = |\nabla v_j|^2\leq c\eta b v_j \; .
\label{eq:pacomm}
\ee
Therefore by Schwarz' inequality 
$\bbp_\bA \cdot v_j^2 \bbp_\bA \ge \sfrac{1}{2} v_j
        \bbp_\bA^2 v_j  - c\eta bv_j$,
and using this estimate 
 in  (\ref{eq:pap}),  including the negative error term
into $\mu\e^{-5}v_j$ and subtracting the pointwise 
inequality (\ref{eq:spinalloc}) we obtain for any $0<\lambda<1$,
 $\eta\leq \eta (\lambda)$ (see Lemma \ref{lemma:moments}) that 
\bey\lefteqn{
        \D^2 + \mu\e^{-5} -M\chi^2V } \label{eq:pap1}\\
        &  \ge& c\eta\sum_j
        \Big( \D v_j^4 \D +
         \e^{-2}\sfrac{1}{2b}  v_j
        \bbp_\bA^2 v_j 
        - v_j^4  b(\lambda-\eta^2 b[\xi^{(j)}_\perp]^2) 
        P_j^\uparrow 
        +\sfrac{\mu}{2}\e^{-5}v_j -cM\chi^2Vv_j^6\Big) \; .
        \nonumber
\eey
The error term $C_0v_j^4$ in (\ref{eq:spinalloc})
has been absorbed into $\mu\e^{-5}v_j$
if $\e$ is small enough depending on $\eta, \mu$.
The inequality (\ref{eq:spinalloc}) has been subtracted
to prepare for a later step.
Hence
\bey\lefteqn{   
        \Big| \Tr \Big(\D^2 + \mu\e^{-5} -M\chi^2V\Big)_-\Big|}
\label{eq:trsum}\\
         &\leq& c\eta\sum_j
        \Big|\Tr \Big( \D v_j^4 \D +
         \e^{-2}\sfrac{1}{2b}  v_j
        \bbp_\bA^2 v_j - v_j^4  b(\lambda-\eta^2 b[\xi_\perp^{(j)}]^2) 
        P_j^\uparrow +
        \sfrac{\mu}{2}\e^{-5}v_j -cM\chi^2Vv_j^6\Big)_-\Big|\nonumber
\eey
by $|\Tr (\sum_j H_j)_-|\leq \sum_j |\Tr (H_j)_-|$ that follows from the
variational
principle.
The following lemma is the  cylindrically
localized version of
Proposition \ref{prop:pos}.

\begin{proposition}\label{prop:cylpos} 
 With the notations above and setting $W:=cM\chi^2V$ 
 we have 
\bey
        \lefteqn{\Big|\Tr \Big( \D v_j^4 \D +
        \e^{-2}\sfrac{1}{2b}   v_j
        \bbp_\bA^2 v_j - v_j^4 b(2^{-7} -\eta^2 b[\xi_\perp^{(j)}]^2)
         P_j^\uparrow+
        \sfrac{\mu}{2}\e^{-5}v_j-Wv_j^6\Big)_-\Big|}
\label{eq:cylpos} \qquad
         \qquad\qquad\qquad \qquad\qquad\qquad\qquad\qquad\qquad \\
        &\leq& c\int \Big( (v_j^2W)^{5/2} 
        +  b(v_j^2W)^{3/2}  \Big)      \nonumber 
\eey
 for each $j$ if $\e$ is small enough depending on $K, \mu, \eta$.
\end{proposition}
Choosing $\lambda:=2^{-7}$ and $\eta:=\eta(2^{-7})$ (see Lemma
\ref{lemma:moments}), 
Proposition \ref{prop:pos}  directly follows from this
proposition, from (\ref{eq:trsum}) and (\ref{eq:moments}). $\;\;\;\Box$

\bigskip

\noindent
{\it Proof of Proposition \ref{prop:cylpos}.}  
The proof contains three transition steps that are performed locally
around each field line from the grid constructed in Section 
\ref{sec:cylpart}.  First we replace $\D$
with a Dirac operator $\D_{\Om}$ in a metric that is conformal
to the Euclidean one. The conformal factor $\Om$ is chosen
such that the strength of the magnetic field becomes constant along a
field line.
Second we replace the volume form $\rd x$
with the volume form $\rd \xi=\rd\xi_1\wedge\rd\xi_2\wedge\rd\xi_3$,
 where $\xi$ 
is the coordinate system associated with the chosen field line.
Then we perform a gauge transformation so that $\D_{\Om}$
becomes close to the Dirac operator $\tcD: = \bsigma\cdot 
(-i\partial_\xi +\a_c(\partial_\xi))$ with a constant
magnetic field $\beta_c=\rd\a_c = b\;\rd\xi_1\wedge\xi_2$
in the linear gauge $\a_c = \sfrac{b}{2}[\xi_1\rd\xi_2-\xi_2\rd\xi_1]$.
Finally, the operator $\tcD$ can be analyzed explicitly.
\medskip

For each fixed $j$ we consider the constructions in Section
\ref{sec:gendirac} with the metric $g_j:=g_{\Om_j} =\Om_j^2\rd x^2$.
We shall apply Section \ref{sec:confch} to the Euclidean metric
with the  standard basis vectors 
$\{\partial_1, \partial_2,\partial_3\}$. 
The vectors $\{\Om^{-1}_j \partial_1, \Om^{-1}_j\partial_2, 
\Om^{-1}_j\partial_3\}$
form an orthonormal basis in $g_j$. Let $\D^a_j$
and $\bPi_j^{a}$ denote the corresponding Dirac operator
and the vector of derivative operators
as defined in (\ref{eq:confdirac}) and (\ref{eq:momvec}). We recall that $a$ 
is the 1-form dual to the vector potential $\bA$ in the standard metric.
Using the estimates in Lemma \ref{lemma:metric} 
to the formula (\ref{eq:piaom}), we obtain
\be
        \bPi_j^{a} = \Om^{-1}_j \bbp_\bA + \e\cO_2 (1) \; ,
        \qquad \mbox{and}\quad \D^a_j =\Om^{-1}_j\D + \e\cO_2(1)\; ,
\label{eq:bpij=bbp}
\ee
where the error terms are functions.
By  Schwarz' inequality 
we obtain the following pointwise bound
\be
     |\bbp_\bA\psi|^2  \ge   
    \sfrac{1}{8}|\bPi_j^a\psi|^2 - c(K)\e^2
    |\psi|^2\; \quad \mbox{and}
    \quad  |\D\psi|^2\ge \sfrac{1}{8} |\D_j^a \psi|^2 - c(K)\e^2
     |\psi|^2 \; 
\label{eq:pointw}
\ee
since  $\sfrac{1}{2}\leq \Om_j \leq 2$
if $\e$ is sufficiently small.
Therefore
\be
          v_j\bbp_\bA^2v_j \ge  \sfrac{1}{8}v_j[\bPi_j^a]^*\cdot
           \bPi_j^a v_j  - c(K)\e^2v_j^2 \qquad   
\label{eq:nabtoom}
\ee 
and
\be
         \D v_j^4 \D 
        \ge\sfrac{1}{8}[\D_j^a]^* v_j^4 \D_j^a-c(K)\e^2 v_j^4 \; ,
\label{eq:dtomega}
\ee
where star denotes  the adjoint  in the standard $L^2$-space.
By applying  the inequalities (\ref{eq:nabtoom}) and (\ref{eq:dtomega}) 
we have
\bey\lefteqn{
        \Big|\Tr \Big( \D v_j^4 \D +
        \e^{-2}\sfrac{1}{2b}  v_j
        \bbp_\bA^2 v_j - v_j^4 b( 2^{-7}-\eta^2 b[\xi_\perp^{(j)}]^2)
         P_j^\uparrow
        +\sfrac{\mu}{2}\e^{-5}v_j-Wv_j^6\Big)_-\Big|}
\label{eq:onj}\\
        &\le&   \sfrac{1}{8}\Big|\Tr \Big( [\D_j^a]^*  v_j^4 \D_j^a +
        \e^{-2}\sfrac{1}{2b}   v_j[\bPi_j^a]^*\cdot
        \bPi_j^a v_j -v_j^4  b(\sfrac{1}{16}-
        8\eta^2 b[\xi_\perp^{(j)}]^2) P_j^\uparrow+
       \sfrac{\mu}{4}\e^{-5}v_j-8Wv_j^6\Big)_-\Big|\; . \nonumber
\eey
The error terms in (\ref{eq:nabtoom}) and (\ref{eq:dtomega})
have been absorbed into the $\sfrac{\mu}{2}\e^{-5}v_j$ 
term using $v_j\leq 1$ if $\e$ is
sufficiently small.

The right hand side of (\ref{eq:onj}) is invariant under an
$SU(2)\times U(1)$ gauge transformation $\cU_{R,\phi}$
as defined in (\ref{eq:gauge}). We shall choose $R$ to be
the rotation from $\{\Om^{-1}_j \partial_1, \Om^{-1}_j\partial_2, 
\Om^{-1}_j\partial_3\}$ to the basis $\{e_1, e_2, e_3\}$
constructed in Lemma \ref{lemma:metric} and $\phi$ to be such
that $\a= a + \rd\phi$, where $\a$ is constructed in Section
\ref{sec:apprfield}. In particular
 $P^\uparrow$ becomes $\sigma^\uparrow:= 
\sfrac{1}{2}[1+\sigma^3]$ according to (\ref{eq:pupgauge}) since $\bn=e_3$
along the central field line.
Therefore the right hand side of (\ref{eq:onj}) continues as 
\be
    (\ref{eq:onj}) =  \sfrac{1}{8}\Big|\Tr \Big( [\D^\a]^*  v^4 \D^\a +
        \e^{-2}\sfrac{1}{2b}   v[\bPi^\a]^*\cdot
        \bPi^\a v -v^4  b(\sfrac{1}{16}-
        8\eta^2 b\xi_\perp^2) \sigma^\uparrow+
       \sfrac{\mu}{4}\e^{-5}v-8Wv^6\Big)_-\Big|\;, 
\label{eq:noj}
\ee
where we also omitted the $j$ index for brevity, i.e.
$\D^\a=\D^\a_j$, $\bPi^\a=\bPi_j^\a$,  
$v=v_j$ and $\xi_\perp = \xi^{(j)}_\perp$ for the rest
of this section.

Now we translate our problem from the standard $L^2(\rd x)$ space
to the $L^2_\xi:=L^2(\rd\xi) $ space.
We change the measure from the volume form $\rd x$ to
$\rd\xi = \rd\xi_1\wedge\rd\xi_2\wedge\rd\xi_3$.
Since these two volume forms are comparable
by a factor of at most 4 
 by (\ref{eq:homtemp})
of Lemma \ref{lemma:metric} if $\e$ is sufficiently small,
 we can use Lemma 
\ref{lemma:snegchange} from the Appendix to obtain 
$$
       (\ref{eq:noj}) \le  \sfrac{1}{8} \Big|\Tr_{L^2_\xi}
       \Big( [\D^\a]^*  v^4 \D^\a +
        \e^{-2}\sfrac{1}{2b}   v[\bPi^\a]^*\cdot
        \bPi^\a v - v^4 b(\sfrac{1}{4}-
        8\eta^2 b\xi_\perp^2) \sigma^\uparrow
        + \sfrac{\mu}{4}\e^{-5}v-32Wv^6\Big)_-\Big|
$$
where the trace and the adjoints are computed in the $L^2(\rd\xi)$
space. 

We remark that already on the right hand side of (\ref{eq:onj}) it
could have been natural to transform the trace on $L^2(\rd x)$
to the trace on $L^2(\Om_j^3\rd x)$, according to Lemma 
\ref{lemma:snegchange}.

We introduce the function
\be 
      G= G(\xi): =  1+\sqrt{b}\min\{ |\xi_\perp|, 1\} \; ,
\label{def:G}
\ee
and we notice that
\be
      \sup_\xi G(\xi)^p v(\xi)^q \leq c(p,q)\eta^{-p/2}\; , \quad
        p\ge 0, \;  q> 0\; .
\label{eq:Gv}
\ee

We recall the definition of $D_b$
and the decomposition (\ref{eq:Dbr}) from  Section
\ref{sec:apprfield}.
Since $\ell=1$, the estimates (\ref{eq:mkestell}) are 
translated into
\be
        \cM= \cO(G^2), \quad \| \nabla \cM\| = \sqrt{b} \cO(G) ,
        \quad \cK_{1,2}  =  \cO_1(|\xi_\perp|^{\gamma}) ,\quad
        \cK_3 = \cO_1(|\xi_\perp|)\; .
\label{eq:mkest}
\ee
for any $\gamma\ge 0$.
Here $\cO(G^k)$ denotes the  class of functions $F(\xi)$ on $\bR^3$
 with $\sup_\xi |F(\xi)|/G^k(\xi)<\infty$.

\medskip

Hence
\be
          \int v^4 |\D^\a\psi|\rd\xi \ge \sfrac{1}{2}
          \int |v^2\tcD\psi|^2\rd\xi
          -c \sum_{k=1}^3 \int |v^2 \cK_k\partial_{\xi_k}
          \psi|^2\rd\xi - \int v^4 O(G^4)|\psi|^2\rd\xi\; .
\label{eq:toco1}
\ee
In the second  term on the right hand side
we first  
commute $v$ through the derivative. Notice that
 $[\partial_{\xi_k}, v] = \eta bv\cO(|\xi_\perp|)$ for $k=1,2$ and
$[\partial_{\xi_3}, v]=0$. Then  we use the estimates (\ref{eq:mkest})
and (\ref{eq:Gv})
to obtain
\bey
     \sum_{k=1}^3\int |v^2\cK_k \partial_{\xi_k}\psi|^2\rd\xi 
     &\leq& 2\sum_{k=1}^3 \int v^2\|\cK_k\|^2|\partial_{\xi_k}v\psi|^2 \rd\xi 
     + 2\sum_{k=1}^2 \int v^2 
     \|\cK_k\|^2 [\eta b O(|\xi_\perp|)]^2|v\psi|^2 \rd\xi 
     \nonumber\\
     &\leq&  cb^{-1} \sum_{k=1}^3
     \int |\partial_{\xi_k}v\psi|^2\rd\xi 
     + c  \int
     v^2 |\psi|^2\rd\xi  \; 
\label{eq:tocont}
\eey
if $\e$ is sufficiently small.
{F}rom the last  term in (\ref{eq:toco1}) we  obtain
 a similar error term as in (\ref{eq:tocont}) using (\ref{eq:Gv}), hence
\be
          \int v^4 |\D^\a\psi|\rd\xi \ge \sfrac{1}{2}
          \int |v^2\tcD\psi|^2\rd\xi -
           cb^{-1} \sum_{k=1}^3
     \int |\partial_{\xi_k}v\psi|^2\rd\xi 
     - c \int
     v^2 |\psi|^2\rd\xi \;.
\label{eq:toco2}
\ee
We also define for $k=1,2,3$
$$
      \Pi_{\eta,k} : = -i\partial_{\xi_k} +
      (1+2\eta)\a_c (\partial_{\xi_k}) 
$$
and
$$
      \D_{\eta}: = \bsigma\cdot \bPi_\eta =\sum_{k=1}^3 \sigma^k \Pi_{\eta,k}
$$
which is a Dirac operator with constant field $(1+2\eta)b \; \rd\xi_1\wedge
\rd\xi_2$
in the $\rd\xi^2$ metric.
Notice that
\be
       \D_\eta v^2 = v^2\tcD
       +2i\eta b v^2  (\sigma^1 \xi_1+\sigma^2\xi_2) \sigma^\uparrow \; .
\label{eq:loc}
\ee
This identity, called the {\it magnetic localization formula},
 was  introduced in \cite{ES-II}.

\medskip
Hence, using (\ref{eq:loc}) to continue (\ref{eq:toco2}), we obtain
\bey
        \int v^4 |\D^\a\psi|\rd\xi 
        &\ge& \sfrac{1}{4}\int |\D_\eta v^2\psi|^2\rd\xi
         - 8\eta^2b^2 \int v^4\xi_\perp^2|\sigma^\uparrow\psi|^2\rd\xi 
\label{eq:Dp} \\
       && - cb^{-1}
       \int \Big(\sum_{k=1}^3 |\partial_{\xi_k}v\psi|^2\Big)\rd\xi
        -c \int
        v^2|\psi|^2\rd\xi\nonumber\\
   &\ge& \sfrac{1}{8}\int |\D_\eta v^2\psi|^2\rd\xi  + 
   \sfrac{b}{4} \int |\sigma^\uparrow v^2 \psi|^2 \rd\xi
        - 8\eta^2b^2 \int v^4\xi_\perp^2 |\sigma^\uparrow\psi|^2\rd\xi
        \nonumber\\
      &&  - cb^{-1}
      \sum_{k=1}^3 \int |\partial_{\xi_k}v\psi|^2\rd\xi
        -c \int
        v^2|\psi|^2\rd\xi \; .\nonumber
\eey
In the last step we used that $\D_\eta^2 \ge 2b(1+2\eta)\sigma^\uparrow \ge 
2b \sigma^\uparrow$, i.e.,
that on the spin-up subspace $\{ \psi \; : \;\sigma^\uparrow \psi=\psi\}$
the constant field Pauli operator is bounded from below by
twice of the constant field.

\medskip
We shall control the second negative error term on the right hand side
of (\ref{eq:Dp}) by the term
$\e^{-2}(2b)^{-1} v[\bPi^\a]^*\cdot \bPi^\a v$. Notice that
the following inequality holds pointwise
$$
        |\bPi^\a \psi|^2\ge \frac{1}{2} \Big(\sum_{k=1}^3
        |\partial_{e_k}\psi|^2\Big)
        - 4 \|\a (\xi)\|^2 |\psi|^2 - 4(\sup\| \bomega\| )^2|\psi|^2
$$
using (\ref{eq:covdef}), (\ref{eq:nablaalp}) and (\ref{eq:momvector}).
We can estimate $\|\alpha(\xi)\|\leq cb^{1/2}G(\xi)$ from (\ref{eq:dalpest2})
and the explicit choice of $\a_c$. We also estimate $\|\bomega\|\leq cK\e$
by Lemma \ref{lemma:metric} and the same lemma is used to
estimate the transition from $\sum_j|\partial_{e_j}\psi|^2$
to $\sum_j |\partial_{\xi_j}\psi|^2$. Therefore
$$
        |\bPi^\a \psi|^2
       \ge \frac{1}{4} \Big(\sum_{k=1}^3
        |\partial_{\xi_k}\psi|^2\Big)
        - cb G^2 |\psi|^2\; ,
$$
if $\e$ is sufficiently small, hence
\be
      \frac{1}{2\e^2 b}
      \int |\bPi^\a v\psi|^2\rd\xi \ge  \frac{1}{8\e^2b} 
      \int \Big(\sum_{k=1}^3
        |\partial_{\xi_k}v\psi|^2\Big)\rd\xi
        - c\eta^{-1} \e^{-2}\int v^2 |\psi|^2\rd\xi
\label{eq:dercon}
\ee
using (\ref{eq:Gv}).

Combining (\ref{eq:Dp}) and (\ref{eq:dercon}) we obtain
$$
 [\D^\a]^* v^4 \D^\a +
        \e^{-2}\sfrac{1}{2b}  v[\bPi^\a]^*\cdot\bPi^\a v
        - v^4  b(\sfrac{1}{4}-8\eta^2 b\xi_\perp^2) \sigma^\uparrow
       +\nu\e^{-5}v^2- 32Wv^6
$$
$$
        \ge \sfrac{1}{8} v^2\D_\eta^2v^2 - 32Wv^6
$$
if $\e$ is sufficiently small depending on
$\nu, \eta$ and $K$.
Since $\Tr( X^*HX)_- \leq \| X^*X\| \Tr H_-$, 
\bey\label{eq:constLT}
        \Big|\Tr\Big(   \sfrac{1}{8} v^2\D_\eta^2v^2 - 32Wv^6 \Big)_-\Big|
        &\leq& \sfrac{1}{8} \Big|\Tr \Big( \D_\eta^2 - 256Wv^2\Big)_-\Big|
        \\
        &\leq& c\int \Big( b(1+2\eta) (256Wv^2)^{3/2} + (256Wv^2)^{5/2}\Big)
        \rd\xi\; , \nonumber
\eey
where in the last step we used the Lieb-Thirring inequality
for a constant  magnetic field \cite{LSY-II}. 
This completes the proof of Proposition  \ref{prop:cylpos}. $\;\;\Box$

\bigskip

{\it Remark.} The reader may have found it confusing
that along the proof of the positive energy regime
we used the Birman-Schwinger principle (\ref{eq:sneg})
 back and forth several times.  
It occured first in (\ref{eq:sneg}), (\ref{eq:BS}), then in (\ref{eq:BSp}),
(\ref{eq:BSfin}), and finally, implicitly, in (\ref{eq:constLT}),
when we referred to the Lieb-Thirring inequality with a constant
magnetic field whose proof also relies on the Birman-Schwinger
principle. 
The frequent changes back to an expression on
the sum of the negative eigenvalues were purely for the purpose
of compact presentation of the intermediate results.
It would have been possible to use only (\ref{eq:sneg})
and stay with the resolvent language all the time since  all estimates
done for the operators are equally valid for the resolvents. 
In this case, of course, we could not have referred directly to the
result of  \cite{LSY-II} on the constant field case, rather
to the details of that proof.

\section{Zero mode regime: Proof of Proposition
 \ref{prop:zero}}\label{sec:zero}
\setcounter{equation}{0}

First notice that the inequalities in Proposition  \ref{prop:zero}
 are scale invariant in powers of $\ell$;
both sides of (\ref{eq:squarediag}) scale like $\ell$
and both sides of  (\ref{eq:sqderdiag}) scale like $\ell^{-1}$.
Therefore we can set $\ell=1$ for the proof.
The arguments for  weak magnetic fields and for extended $(D, 
K)$-regular fields are different.

\subsection{Weak magnetic field}\label{sec:weakmag}

For weak fields  (\ref{eq:squarediag}) and
(\ref{eq:sqderdiag}) will be estimated by a universal constant
$c$ if $\e\leq \e(K)$.
We need the following lemma:

\begin{lemma}\label{lemma:XY}
Let $X, Y$ be self-adjoint operators such that $X\ge 0$, $X+Y\ge 0$
and $\|Y\|\leq M$ for some constant $M>0$. Then
\be
        {1\over (X + Y + 2M)^2}\leq {4\over (X+ M)^2}\leq {4\over X^2 + M^2}
\label{eq:XY}
\ee
\end{lemma}

\noindent
{\it Proof.}  Consider the resolvent expansion
$$
        {1\over X+Y+2M} = {1\over X+M} - {1\over X+M} (Y+M)
        {1\over X+Y+2M}
$$
hence by Schwarz' inequality
\bey
        {1\over (X+Y+ 2M)^2}&\leq& {2 \over (X+ M)^2}
        + 2  {1\over X+M} (Y+M)
        {1\over (X+Y+2M)^2} (Y+M) {1\over X+M}
        \nonumber\\
        &\leq& {2 \over (X+ M)^2}
        + 2 (2M)^{-2}  {1\over X+M} (Y+M)
         (Y+M) {1\over X+M}\nonumber\\
        &\leq& {4\over (X+ M)^2}\nonumber
\eey
since $(Y+M)^2 \leq 2\|Y\|^2 + 2M^2 \leq 4M^2$.  By positivity of $X$
we have $(X+M)^2 \ge X^2 + M^2$ which completes the proof. $\;\;\Box$

\medskip

If $\|\bB\|_\infty \leq c\e^{-2}$ then 
using Lemma \ref{lemma:XY} we obtain
$$
        R[P]^2(u,u) = \Big[ (\bp+\bA)^2 +\bsigma\cdot \bB +
        \e^{-5}\Big]^{-2}(u,u)
        \leq  4\Big[ (\bp+\bA)^2  + \sfrac{1}{2}\e^{-5}\Big]^{-2}(u,u)
$$
if $\e$ is sufficiently small. By the diamagnetic inequality we can
continue
this estimate as
$$
        R[P]^2(u,u)\leq  
        4\Big[ -\Delta  + \sfrac{1}{2}\e^{-5}\Big]^{-2}(u,u) \leq
        c\e^{5} \; .
$$
This proves (\ref{eq:squarediag}).

\medskip

For the proof of (\ref{eq:sqderdiag}) we define a smooth function
$0\leq \chi\leq 1$ such that $\chi(u)=1$, $|\nabla \chi|\leq c$,
$|\nabla^2\chi|\leq c$  and
$\mbox{supp}(\chi)\cap \mbox{supp}(\varphi) =\emptyset$. For
brevity we  set $R:= R[P]$ with $P=\e^{-5}$. Since the inequality
 (\ref{eq:sqderdiag}) is gauge invariant, we can choose the
Poincar\'e gauge $\wh \bA$ centered at $z_0$ to generate the magnetic
field. In particular, $\|\wh\bA\|_\infty \leq c\e^{-2}$ since
by assumption $\bB$ is 
supported on $\tD$ and $\|\bB\|_\infty\leq c\e^{-2}$.

Let $\{ X, Y\}: = XY+YX$ denote the anticommutator.
Notice that  $[R,\chi] = R\{\D, [\D,\chi]\}R$ and 
$$
        \{\D , [\D ,\chi]\}
         = (-i)\{\bsigma\cdot (\bp),\bsigma
        \cdot\nabla \chi\} + 
        (-i)\{\bsigma\cdot\wh \bA, \bsigma\cdot\nabla \chi)\}\; .
$$
We can compactly write
$$      
        \{\D , [\D ,\chi]\} = \sum_{j=1}^3
            (\bp + \wh \bA)_j \wh\cK_j +\wh\cM \; ,
$$
where the $\wh\cK_j$ and $\wh\cM$ are  2 by 2 matrix
valued functions
and from the estimate on $\wh \bA$ and the derivatives of $\chi$ we 
easily obtain that $\sup_x \|\wh\cK_j(x)\|, \|\wh\cM(x)\|\leq c\e^{-2}$.

Therefore commuting $\chi$ through first,
  estimating $\varphi^2\leq 1$,
  then using $R\D^2 R \leq R$ and
applying a Schwarz' inequality we get
\bey
        \chi R\D\varphi^2\D R\chi &= & R\{\D, [\D,\chi]\}^*R
        \D\varphi^2\D R\{\D, [\D,\chi]\}R \nonumber\\
        & \leq&  R\{\D, [\D,\chi]\}^*R
        \{\D, [\D,\chi]\}R \nonumber\\
        & \leq&   4 R\wh\cM^*R\wh\cM R + 4 \sum_{j=1}^3
        R\wh\cK^*_j (\bp + \wh \bA)_j R
        (\bp + \wh \bA)_j  \wh\cK_j R\; .
\label{eq:dd}
\eey
In the first term we use $R\leq P^{-1}=\e^5$ for the middle resolvent
then we use the boundedness of $\wh\cM$  to arrive
at the resolvent square, $R^2$, that was estimated above
in the proof of (\ref{eq:squarediag}).

In the second term we use that $\cD^2 + P = (\bp + \wh \bA)^2
+\bsigma\cdot\bB + P \ge (\bp + \wh \bA)^2 + \sfrac{P}{2}$
since $\|\bB\| \leq c\e^{-2} \leq \sfrac{P}{2} = \sfrac{1}{2}\e^{-5}$
if $\e$ is sufficiently small, therefore
$(\bp + \wh \bA)_j R (\bp + \wh \bA)_j \leq 1$.
The estimate of the second term then can be completed by
using $\|\wh\cK_j^*\wh\cK_j\|\leq c\e^{-4}$ 
and referring to the estimate of the square of
the resolvent in (\ref{eq:squarediag}).
This finishes the proof of Proposition \ref{prop:zero}
for the case of weak magnetic field (case (i)).

\subsection{Strong magnetic field}\label{sec:strongmag}

Here we prove Proposition \ref{prop:zero} for the case (ii).
Throughout the proof we fix $u$ and let $z=\pi(u)$ be
its base point on the supporting plane $\cP$.
Consider the construction of Lemma \ref{lemma:metric},
in particular the coordinate functions $\xi=(\xi_1,\xi_2,\xi_3)$
and the conformal factor $\Om$.
We know that $\sfrac{1}{2}\leq \Om\leq 2$ and $\|\nabla\Om\|_\infty\leq 1$
if $\e$ is sufficiently small. Recall that we set $\ell=1$,
therefore  the bounds
on the right hand side of (\ref{eq:squarediag}) and
(\ref{eq:sqderdiag}) 
saturate to $c(\e,K)b$ since $P=\e^{-5}$
and $|\bB(u)|$ is comparable with $b:=|\bB_\infty|\ge 1$.

\subsubsection{Transformation into good coordinates}

Similarly to the positive energy regime in Section \ref{sec:pos},
we perform three transition steps.
We first change $\D$ into $\D_\Om^a :=\Om^{-2}\D \Om$
and the underlying measure to $\rd x$ to $\Om^3\rd x$, then we change
the measure from $\Om^3\rd x$ to $\rd\xi$
and finally we perform a gauge transformation.
Recall  that $\D_\Om^a$ is self-adjoint on $L^2(\rd s_\Om^2)$
(see \cite{ES-III}). We set $R_\Om^a[P]: = ( [\D_\Om^a]^2 + P)^{-1}$
and we assume that $\e$ is sufficiently small so
that $P\ge 2^9$.

Then Lemma \ref{lemma:conf} from Appendix \ref{sec:compconf}
states that
\be
        \tr \; R[P]^2(u,u)\leq 2^9 \;\tr\; (R_\Om^a[P])^2_{L^2(\Om)}(u,u)
\label{eq:rtor}
\ee
$$ 
    \tr \; \Big( R[P]\D\varphi^2\D R[P]\Big)(u,u)
$$
\be
    \leq 2^{12} \tr\; 
    \Big( R_\Om^a[P]\D_\Om^a\varphi^2\D_\Om^a R_\Om^a[P]\Big)_{L^2(\Om)}(u,u)
    + 2^{12} P \, \tr \; (R_\Om^a[P])^2_{L^2(\Om)}(u,u)
\label{eq:rdtor}
\ee
where the operator kernels on the right hand side are computed
in the $L^2(\Om):=L^2(\Om^3 \rd x)\otimes \bC^2$ space.

Gauge transformation  of the form (\ref{eq:gauge})
leaves the diagonal elements
of operator kernels invariant hence 
 we can use the basis $\{e_1, e_2, e_3\}$
constructed in Lemma \ref{lemma:metric}
and vector potential $\a$ 
constructed in Section \ref{sec:apprfield}
to express the right hand sides of (\ref{eq:rtor}), (\ref{eq:rdtor})
using  $\D_\Om^\a$ instead of $\D_\Om^a$
 as in the proof of Proposition \ref{prop:cylpos}.

We recall that in this gauge the decomposition (\ref{eq:Dbr}) holds, i.e.,
\be
        \D_\Om^\a = \tcD +\sum_{k=1}^3 \cK_k\partial_{\xi_k} + \cM
\label{eq:decompo}
\ee
with $\tcD$ given in (\ref{eq:dbdef}) and $\cK_k$, $\cM$ satisfy the
estimates (\ref{eq:mkest})
 with $\ell=1$.

We apply Lemma \ref{lemma:changenorm} from Appendix \ref{sec:change}
to compare operator kernels on the measure spaces 
with volume forms $\rd\mu:=\Om^3 \rd x$ and
$\rd\nu:=\rd\xi =\rd\xi_1\wedge\rd\xi_2\wedge\rd\xi_3$.
Since these two volume forms are comparable at every point
by Lemma \ref{lemma:metric},
we obtain from (\ref{kernelcomp}) that
\bey
 \tr\; (R_\Om^\a[P])_{L^2(\Om)}^2(u,u)  &\leq& c \;\tr
\; (R_\Om^\a[P]^* R_\Om^\a[P])_{L^2(\rd\xi)}(u,u)
        \nonumber\\
         \tr\; 
          \Big( R_\Om^\a[P]\D_\Om^\a\varphi^2\D_\Om^\a R_\Om^\a[P]
          \Big)_{L^2(\Om)}(u,u)
        &\leq& c\; 
          \tr\;  \Big( \Big[ \varphi \D_\Om^\a R_\Om^\a[P]\Big]^*
          \varphi \D_\Om^\a R_\Om^\a[P]\Big)_{L^2(\rd\xi)}(u,u)\; , \nonumber
\eey
where the adjoints and the operator kernels 
 on the right hand sides are computed in the $L^2(\rd\xi)\otimes\bC^2$
space.

 Therefore  case (ii) of Proposition \ref{prop:zero} has been
reduced to proving that 
for $\e\leq\e(K)$
\bey
        \tr\; \Big[  R_\Om^\a[P]^* R_\Om^\a[P] \Big](u,u)&\leq& cb\; ,
\label{eq:sqre} \\
        \tr \;\Big[ 
        ( \varphi\D_\Om^\a R_\Om^\a[P])^*\varphi\D_\Om^\a
        R_\Om^\a[P]\Big] (u,u)&\leq& cb \; ,
\label{eq:sqdre}
\eey
where the adjoints and the operator kernels are computed on
$L^2(\rd\xi)\otimes\bC^2$.

{\it Remark:} The operator $\D_\Om^\a$ in general is not
self-adjoint in $L^2(\rd\xi)\otimes\bC^2$, but $\tcD$
is.

\subsubsection{Proof of (\ref{eq:sqre}) and (\ref{eq:sqdre})}
\label{sec:propzeroproof}

In this proof
we will omit $\Om$ and $\a$ from the notation
of $\D_\Om^a$ and we will simply use $\D$ for this operator.
This should not be confused with the notation $\D$ (see
(\ref{eq:dirac}) with $h=1$) used
elsewhere in the paper.

We recall that  $\tcD$ denotes the Dirac operator with 
a constant field (see
(\ref{eq:dbdef})).
We also need the notations $\bPi = (\Pi_1, \Pi_2, \Pi_3)$,
 $\tbPi = (\tPi_1, \tPi_2, \tPi_3)$ from
$\D = \bsigma\cdot \bPi$, $\tcD = \bsigma\cdot \tbPi$.
We note that $\tcD$ can be decomposed as 
\be
       \tcD = \tcD_\perp +N
\label{eq:decom}
\ee
with 
\be
        \tcD_\perp: = \sum_{j=1}^2\sigma^j\tPi_j
       \qquad \mbox{and}\qquad
      N:=\sigma^3 \tPi_3  =\sigma^3(-i\partial_{\xi_3})\; .
\label{def:dperp}
\ee
These operators are self-adjoint on  $L^2(\rd\xi^2)$ and
\be
       [\tcD_\perp, \tPi_3]=0, \qquad
       \{ \tcD_\perp, N\} = 0, \qquad \mbox{and}\qquad
       \tcD^2 = \tcD^2_\perp + N^2 =\tcD_\perp^2 + \tPi_3^2 \; .
\label{eq:dperpprop}
\ee

We need two decompositions of the error term $\cE: = \cD - \tcD$
as in (\ref{eq:Dbr})
\be
        \cE= \bPi \cdot \cK + \cM
\label{eq:dec1}
\ee
\be
        \cE =  \tbPi \cdot \tcK + \tcM\; .
\label{eq:dec2}
\ee
Here $\cK= (\cK_1, \cK_2,\cK_3)$ is a vector
of 2 by 2 matrices and
 $\bPi\cdot \cK: =\sum_{j=1}^3 \Pi_j \cK_j$.
The matrices $\cK_j, \cM$  satisfy
(\ref{eq:mkest})
 with $\ell=1$ and the same estimates
hold for $\tcK_j, \tcM$ as well.
The following estimates are straightforward from (\ref{eq:mkest})
and (\ref{eq:adiff}) if $\e\leq \e(K)$
\be
        [\Pi_j, \cK_k] = \cO(1),\qquad
        [\Pi_j, \cM] = b^{1/2}\cO(G), \qquad j,k=1,2,3\; ,
\label{eq:commut}
\ee
and the same holds for $\wt\cK$ and $\wt\cM$.
In particular
the following relations also hold:
\be
        \cE= \cK\cdot\bPi  + \cM_0\; , \quad
        \cE =\tcK\cdot \tbPi + \tcM_0
\label{eq:dec2new}
\ee
with $\cM_0,\tcM_0 = \cO(G^2)$.

The following lemma collects various operator inequalities
related to the diamagnetic inequality.
The proof is  postponed to Section \ref{sec:diamagproof}.

\begin{lemma}\label{lemma:opineq}
With the notations above 
we have the following
operator inequalities in the space $L^2(\rd\xi)\otimes\bC^2$
if $\e\leq \e(K)$:
\bey
        \Pi_j^* \Big( \res\Big)^*
        \Big( \res\Big)
        \Pi_j &\leq& cb \; ,
\label{eq:1}\\
        \Pi_j^* \Big( \frac{\D}{\D^2+P}\Big)^*
        \Big( \frac{\D}{\D^2+P}\Big)
        \Pi_j  &\leq& cb \; ,  \qquad j=1,2,3
\label{eq:1.5}\\
        \Pi_k^*\Pi_j^* \Big( \res\Big)^*\res \Pi_j\Pi_k &\leq& cb^2, 
        \qquad j,k=1,2,3
\label{eq:2}\\
        \cE^*\Big( \res\Big)^* \res \cE&\leq&  
        \cO(G^4)
\label{eq:5} \\
        \cE^*\Big(\frac{\D}{\D^2+P} \Big)^*\frac{\D}{\D^2+P}
        \cE&\leq&  \cO(G^4) \; .
\label{eq:3}
\eey

For the constant field operator we have
\be
        \tPi_j \tres\tPi_j \leq cb \;, \quad j=1,2,3.
\label{eq:11}
\ee
\end{lemma}

The next lemma estimates the diagonal elements
of explicitly computable operators with a constant magnetic field.
The proof is given in Section \ref{sec:constres}.
We set $H(\xi): = \min\{ |\xi_\perp|, 1\}$
and we recall that $G=1+\sqrt{b}H$.

\begin{lemma}\label{lemma:expl} 
With the notations above and for any constants $P\ge 1$, $b\ge 1$
with $P\leq cb$  we have
\be
        \tr\; {1\over (\tcD^2 + P)^2}(u,u)\leq cb \; .
\label{eq:expl1}
\ee
For any $k=1,2\ldots$, $m=1,2,\ldots$
and for any 2 by 2 matrix valued function $\cF$ with $\| \cF(x)\| =\cO(G(x))$
we also have
\bey
        \tr \; \tres G^{2m} \tres (u,u) &\leq& c(m)b \;, 
\label{eq:expl2} \\
        \tr \; \tdres  [\cF^k]^* \tres \cF^k \tdres (u,u)&\leq& c(k)b \; ,
\label{eq:expl3} \\
        \tr\;  \tdres H^{2m} \tdres (u,u)&\leq&
        c(m)b^{1-m} \; ,
\label{eq:expl7}\\
        \tr\;  \tPijres  H^{2m} \tPijres 
        (u,u) &\leq&
        c(m) b^{2-m}\;, \qquad j=1,2,3 \; .
\label{eq:expl8} 
\eey
Let $\tcW$ denote either the identity $I$, or $\tcD$, or $\wt\Pi_3$, then
\be
        \tr \; \tdres [\cF^k]^* \twres G^{2m} 
        \twres \cF^k \tdres (u,u)\leq c(k,m)b\; 
\label{eq:expl4}
\ee
(we recall that $[\wt\Pi_3, \tcD]=0$).
Let $\tcU$ denote either the identity $I$ or $\tcD$ or $\wt\Pi_j$, $j=1,2,3$,
and let $0\leq \varphi\leq 1$ be a
function with $\mbox{dist}(u, \mbox{supp}(\varphi))
\ge 1$, then
\bey
        \tr\; \tures \varphi^2 \tures (u,u)&\leq& ce^{-c\sqrt{b}}\;,
\label{eq:expl11}\\
        \tr\; \tdres  [\cF^k]^*  \tures  G^m\varphi^2 G^m
        \tures \cF^k \tdres (u,u)&\leq& c(k,m)e^{-c\sqrt{b}} \; .
\label{eq:expl12}
\eey
\end{lemma}

\bigskip

Armed with these lemmas, we complete the proof of
 (\ref{eq:sqre}) and (\ref{eq:sqdre}).
 We start with (\ref{eq:sqre}).
We introduce the notation $(\cdots)^*A$ for $A^*A$ if $A$ is a 
long expression. All adjoints are computed in the $L^2_\xi$ space.

We use $\cD^2 =\tcD^2 + \cD\cE + \cE\tcD$ in the following
resolvent expansion:
$$
        \ad \res \leq 3\Big( (A) + (B) + (C)\Big)
$$
with
\bey
        (A) &:  =& \ad\tres \nonumber\\
        (B) &:  =&\ad \res \D \cE \tres\nonumber\\
        (C) &:  =& \ad \res\cE\tcD\tres\; .\nonumber
\eey
Term (A) is explicit from (\ref{eq:expl1}).
In term (B) we first use (\ref{eq:3}) 
in the middle to arrive at (\ref{eq:expl2}) with $m=2$.

In term (C) we use $\cE=\tbPi\cdot \tcK + \tcM$ and we expand the
resolvent in the middle once more. The result is
$$
        (C)\leq 4\Big( (C1) + (C2) + (C3) + (C4)\Big)
$$
with
\bey
        (C1)&: =& \ad\tres \tcM \tdres\nonumber\\
        (C2)&: =& \ad \res (\D\cE  + \cE \tcD) \tres\tcM\tdres\nonumber\\
        (C3)&: =& \ad \tres \tbPi \cdot\tcK \tdres\nonumber\\
        (C4)&: =& \ad\res (\D\cE  + \cE \tcD) \tres \tbPi \cdot\tcK \tdres
         \; .
        \nonumber
\eey
Term (C1) is explicit from (\ref{eq:expl3})
after estimating one of the resolvents in the middle by $P^{-1}\leq 1$.
Term (C2) is split into two terms, 
$$
        (C2)\leq 2\Big( (C21)+ (C22)\Big)\; ,
$$
with
\bey
        (C21)&: =& \ad \res \D\cE  \tres\tcM\tdres\nonumber\\
        (C22)&: = &\ad \res  \cE  \tdres\tcM\tdres\nonumber
\eey
In (C21) we use  (\ref{eq:3}) and finally
(\ref{eq:expl4}) with $\tcW=I$, $m=k=2$.
In (C22) we use (\ref{eq:5}) first then 
(\ref{eq:expl4}) with $\tcW=\tcD$, $m=k=2$.
In the term (C3) we use (\ref{eq:11}) then (\ref{eq:expl7})
with $m=1$
together with the estimates (\ref{eq:mkest}) used for $\tcM$.

Finally, for the term (C4) we estimate
$$
        (C4)\leq 2\Big( (C41)+ (C42)\Big)
$$
$$
        (C41): = \ad\res \D\cE \tres \tbPi \cdot\tcK \tdres
$$
$$
        (C42):= \ad\res \cE \tcD \tres \tbPi \cdot\tcK \tdres
$$
In (C41) we first use  (\ref{eq:3}) to arrive at
\be
        (C41)\leq c\ad G^2  \tres \tbPi \cdot\tcK \tdres \; .
\label{eq:c41}
\ee
For the term (C42) we first observe the following
inequality
\begin{lemma}\label{lemma:bg} With the notations above
\be
        \ad \res \cE\tcD \leq b\cO(G^8)\; .
\label{eq:bg}
\ee
\end{lemma}

\noindent
{\it Proof of Lemma \ref{lemma:bg}.} 
 We can write
 $\tcD = \bPi \cdot \wh\cN + \wh \cM$, with $\wh\cN, \wh\cM = O(G^2)$.
Therefore
$$
        \cE\tcD = (\bPi\cdot\cK + \cM)( \bPi \cdot\wh\cN + \wh \cM)
        = \sum_{j,k=1}^3
        \Pi_j\Pi_k b^{-1/2} \cO(G^4) + \sum_{j=1}^3\Pi_j \cO(G^4)
         + b^{1/2} \cO(G^4)
$$
after commuting $\bPi$ through using (\ref{eq:commut}) and (\ref{eq:mkest}).
Therefore (\ref{eq:bg}) follows from (\ref{eq:2}) and (\ref{eq:1}).
$\;\;\Box$

Armed with Lemma \ref{lemma:bg}, we see that
\be  
     (C42)\leq  (C43):= cb\ad G^4  \tres \tbPi\cdot \tcK \tdres \; .
\label{eq:c42}
\ee
Since $G^2\leq G^4\leq G^8$ and $b\ge 1$, it is sufficient
to estimate (C43) 
that will complete the estimate of (C41) from  (\ref{eq:c41}) as well.

To estimate (C43), we
first  separate the  term $\tbPi\cdot\tcK$ into terms containing
$\tPi_3 \cK_3$ and $\tPi_\perp \cK_\perp:=\tPi_1\cK_1+\tPi_2\cK_2$ 
by Schwarz' inequality. We  arrive at
$$
        (C43)\leq c \ad G^4 \tres \tPi_3 (b^{1/2}\tcK_3) \tdres
        + cb \ad G^4 \tres \tPi_\perp \tcK_\perp \tdres \; .
$$
The first term is explicit from (\ref{eq:expl4}) with $\tcW=\tPi_3$,
$m=2$, $k=1$, using that
$b^{1/2}\tcK_3 =\cO(G)$.
In the second term we estimate $G^8\leq b^4$ in the middle,
then estimate one of the resolvents by $P^{-1}\leq 1$
 and first use (\ref{eq:11}) and finally
(\ref{eq:expl7}) with  $m=5$, together with (\ref{eq:mkest})
for $\tcK$.
This completes the proof of (\ref{eq:sqre}).

\bigskip

Now we prove (\ref{eq:sqdre}). We need the following lemma.
\begin{lemma} With the notations above, we have
\bey
        \tr\; \ad \varphi \cE\tres (u,u)&\leq& c\;  e^{-c\sqrt{b}}\; ,
\label{eq:ed} \\
        \tr\; 
         \ad\varphi\cE\tres \cM \tdres (u,u)&\leq& c\;  e^{-c\sqrt{b}} \; .
\label{eq:pe}
\eey
\end{lemma}

\noindent
{\it Proof.}  For the proof of both inequalities
(\ref{eq:ed}) we write $\cE =  \tcK\cdot\tbPi
 +\tcM$, then we separate the terms
by a Schwarz' inequality and
 we use (\ref{eq:expl11}) and (\ref{eq:expl12}), respectively, with
appropriately chosen $\tcU$. $\;\;\Box$

\medskip

For the operator on the left hand side of (\ref{eq:sqdre}) we use
a resolvent expansion and a Schwarz' inequality to obtain
$$
        \ad \varphi \dres \leq 3 \Big( (D) + (E) + (F)\Big)
$$
with
\bey
        (D) &:=& \ad\varphi \D\tres \nonumber\\
        (E) &:=&\ad \varphi \dres \D\cE \tres \nonumber\\
        (F) &:=&\ad\varphi\dres\cE\tdres \; . \nonumber
\eey
The estimate of (D) is trivial by $\D = \tcD +\cE$
 applying a Schwarz' inequality and using  (\ref{eq:expl11}) and (\ref{eq:ed})
for these two terms, respectively.

In term (E) we use
\be
        \dres\D = I - P\res
\label{eq:id}
\ee
and separate it by a Schwarz' inequality:
$$
        (E)\leq 2 \ad \varphi \cE \tres + 2P^2
        \ad \varphi \res \cE \tres\; .
$$
For the first term we can use (\ref{eq:ed}), for the second one we use
$\varphi\leq 1$, (\ref{eq:5}) then (\ref{eq:expl2})
with $m=2$.

Finally for the term (F) we write
$$
        (F)\leq 2 \Big( (F1) + (F2)\Big)
$$
with
$$
        (F1): = \ad\varphi \dres\bPi\cdot \cK \tdres
$$
$$
        (F2): = \ad\varphi \dres \cM \tdres \; .
$$
For (F1) we first estimate $\varphi\leq 1$, then  use (\ref{eq:1.5}) and
(\ref{eq:expl7})  with  $m=1$ together
with (\ref{eq:mkest}).

For (F2) we need one more resolvent expansion:
$$
        (F2)\leq 2\Big( (F21) + (F22) \Big)
$$
with
\bey
        (F21) &: =& \ad\varphi\D\tres \cM \tdres \nonumber\\
        (F22) &: =& \ad\varphi \dres(\D\cE + \cE\tcD) \tres \cM\tdres \; .
        \nonumber
\eey
We split (F21) further by using $\D = \tcD + \cE$:
$$
        (F21)\leq 2  \ad\varphi\tdres \cM \tdres 
        +2 \ad\varphi\cE\tres \cM \tdres 
$$
The first term is explicit by
(\ref{eq:expl3}) after $\varphi\leq 1$ and estimating $\cD^2$ by
the resolvent. The second term was estimated
in  (\ref{eq:pe}).

Finally, to estimate  (F22), we use again (\ref{eq:id}) and we 
split it as follows
\bey
        (F22)&\leq& 3  \ad\varphi \cE  \tres \cM\tdres 
\nonumber\\
        &&+ 3P^2 \ad\varphi \res\cE  \tres \cM\tdres
        \nonumber\\
        &&+3 \ad\varphi \dres\cE \tdres \cM\tdres \; . \nonumber
\eey
The first term is estimated in  (\ref{eq:pe}).
For the second term we use $\varphi\leq 1$
and  (\ref{eq:5}) first, then (\ref{eq:expl4}) with $\tcW=1$, $m=k=2$.
For the third term we use $\varphi\leq 1$ and
 (\ref{eq:3}) first, then (\ref{eq:expl4})
with $\tcW=\tcD$, $m=k=2$.

This completes the proof of (\ref{eq:sqdre}). $\;\;\Box$

\bigskip\bigskip
\newpage
\appendix
\section{Proof of the technical lemmas}\label{sec:techproof}
\setcounter{equation}{0}

\subsection{Proof of Proposition \ref{prop:L} 
on the tempered lengthscale}\label{sec:varsc}

\noindent
{\it Proof}. 
We recall the definitions of $B_L(x), b_L(x)$ from
(\ref{eq:BL}), (\ref{eq:bL}) and we notice that $B_L(x)$ is increasing
in $L$, while $b_L(x)$ is decreasing. Since $\bB$ and its derivatives
are locally bounded and $\bB$ is not constant everywhere,
 we easily obtain that the sets appearing
on the right hand sides of (\ref{eq:magnscale}) and  (\ref{eq:Lv})
 are non-empty and bounded.
Therefore $L_m$ and $L_c$ are
positive finite valued functions.

We notice that if  $B_{L(x)}(x)> L^{-2}(x)$ then $L_m(x)< L(x)<L_c(x)$,
i.e. $L_c(x)=L_v(x)$. We claim that $L_c(x)=L_v(x)$ implies
that (\ref{eq:tsupnablaB}) and (\ref{eq:tsupnablan}) hold
even if $L(x)$ is replaced with $L_c(x)$ which is a stronger statement.
The validity of this stronger form of (\ref{eq:tsupnablaB}) follows
directly from (\ref{eq:Lv}).
This also implies that
$$
      B_{L_c(x)}(x)-b_{L_c(x)}(x) 
      \leq 2L_c(x) \cdot \sup\{ |\nabla \bB(y)|\; : \;
      |x-y|\leq L_c(x) \} \leq 2b_{L_c(x)}(x)\; ,
$$
 therefore $b_{L_c(x)}(x) \ge \sfrac{1}{3} B_{L_c(x)}(x)$, in particular
$\bB(y)\neq 0$ and  $\bn(y)$ is well defined
 for all $y$ with $|y-x|\leq L_c(x)$. 
Thus (\ref{eq:tsupnablan}) with $L(x)$ replaced with $L_c(x)$
follows from (\ref{eq:Lv}). We also proved that if $L_c(x)=L_v(x)$
then the suprema in (\ref{eq:magnscale})
and  (\ref{eq:Lv}) are actually maxima by the 
continuity of $\bB$.

Finally, we have to show that $L(x)$ is tempered.
Notice that it is sufficient to show that
\be
        |x-y|\leq L(x) \Longrightarrow {1\over 2} \leq{ L (y)\over L(x)}
\label{eq:telltempered1}
\ee
for any $x, y\in \bR^3$ because the inequality $L(y)/L(x)\leq 2$
 easily follows from this. To see it, we assume that $L(y)> 2L(x)$.
Then $|x-y|\leq L(x)$ implies $|x-y|\leq L(y)$,
so using (\ref{eq:telltempered1}) with $x, y$ interchanged
we arrive at a contradiction.

Now we show that (\ref{eq:telltempered1}) holds.
 Let $x,y$ be two points with $|x-y|\leq L(x)=
\sfrac{1}{2}L_c(x)$ and we have to show that $L(x)
\leq 2L(y)=L_c(y)$. This is obvious if  $B_{L(x)}(y)\leq L(x)^{-2}$,
since then $L(x)\leq L_m(y)$ and $L_m(y)\leq L_c(y)$ by definition.

Thus we can assume that $B_{L(x)}(y)> L(x)^{-2}$.
   Since $|x-y|\leq L(x)$, we know that
\be
     \Big\{ z \; : \; |y-z| \leq L(x)\Big\} \subset
     \Big\{ z \; : \; |x-z|\leq 2L(x)=  L_c(x) \Big\} \; ,
\label{eq:incl}
\ee
thus $B_{L(x)}(y)\leq B_{L_c(x)}(x)$, hence
$B_{ L_c(x)}(x) > L(x)^{-2} >  L_c(x)^{-2}$, i.e. $L_c(x)>L_m(x)$, so
$L_c(x)= L_v(x)$.

We will now check that
for $\gamma=1,\ldots 4$
\be
      L(x)^\gamma \sup\Big\{ \Big| \nabla^\gamma |\bB(z)|\, \Big| \; : \;
      |y-z|\leq L(x)\Big\} \leq b_{L(x)}(y)
\label{eq:Lxy}
\ee
and
\be
       L(x)^\gamma \sup\Big\{ \Big| \nabla^\gamma \bn(z)\, \Big| \; : \;
      |y-z|\leq L(x)\Big\}\leq 1 \; 
\label{eq:Lxyn}
\ee
hold, which will imply $L(x)\leq L_v(y)$, hence $L(x)\leq L_c(y)$.
 But as we showed above, $L_c(x)= L_v(x)$ implies
 that (\ref{eq:tsupnablaB}) and (\ref{eq:tsupnablan}) hold with
$L(x)$ replaced by $L_c(x)$. From (\ref{eq:incl})
and $L(x)< L_c(x)$ we therefore immediately conclude (\ref{eq:Lxy}),
(\ref{eq:Lxyn}).
$\;\;\Box$

\subsection{Proof of the covering Lemma 
\ref{lemma:cover}.}\label{section:coverproof}

Introduce the notation  $D_x^* : = B\Big( x, 40\ell(x)\Big)$
and $D_i^*:= D_{x_i}^*$.
Let $S$ be any compact subset of $\bR^3$.
First we show how to find a finite set of points
within $S$ so that the balls $\hD_i$ cover $S$
and they enjoy the finite overlapping property.
Let $\ov D_x: = B(x, \ell(x)/20)$ and we cover $S$
by the collection of balls $\ov D_x$, $x\in S$.
By compactness, we can choose  points $\{ x_\a \}\subset
S$, with a finite index set $\a\in A$, such that the balls 
$\{ \ov D_\a \}_{\a\in A}$  cover $S$. 
Now we discard certain points from the collection $\{ x_\a \}$
and relabel the rest by $\{ x_i \}$.

Let $x_1$ be the point
with the biggest value $\ell(x_1)$ among all values
$\{ \ell(x_\a)\; : \; \a\in A\}$.
 Then let $x_2$ be the point
with the biggest  value $\ell(x_2)$ among all values $\ell(x_\a)$
 such that
$x_\a\in \bR^3\setminus \ov{D}_1$. Then let $x_3$ be
 the point with biggest value $\ell(x_3)$ 
among all values $\ell(x_\a)$ such that
$x_\a\in \bR^3\setminus (\ov{D}_1\cup \ov{D}_2)$, etc.
until all $x_\a$'s are covered by $\ov{D}_i$'s.
This selects a subcollection of the points $\{ x_\a\}$
and they are relabelled to $x_1, x_2,\ldots$.

We claim that the collection of $\hD_i$'s cover $S$. Consider any $y\in S$,
then  $y\in \ov D_\a$ for some $\a$.
But $x_\a$ is covered by some $\ov{D}_i$. We choose the smallest such
index $i$.
 By the maximality of the radii
in the selection procedure, we know that $\ell(x_\a)\leq \ell(x_i)$,
so $|y-x_i|\leq |y-x_\a|+ |x_\a - x_i|\leq (\ell(x_\a) + \ell(x_i))/20
\leq \ell(x_i)/10$,
hence $y\in \hD_i$.

We claim that the union of the $D_i^*$
balls  have the finite
covering property with a sufficiently big universal $N$.
{F}rom construction, the balls $D^{\#}_i: =  B(x_i, \ell(x_i)/40)$
are disjoint. Fix a point $y\in \bR^3$ 
and let $I$ be the set of indices $i$ such
that $y\in D_i^*$, $i\in I$.
 Choosing $\e <1/40$, we see that $y\in D_i^*$, i.e.,
$|x_i-y|\leq 40\ell(x_i)$ implies
$1/2\leq \ell(y)/\ell(x_i)\leq 2$ for all $i\in I$.
Hence the balls $D^{\#}_i$, $i\in I$, all have radius at least
$\ell(y)/80$ and they are within a ball of radius $81\ell(y)$
about $y$. From their disjointness it follows that their
number is universally bounded, i.e. the number of $D_i^*$'s
covering any $y$  is bounded by a universal number $N$.

This completes the construction of the covering balls for
any compact set $S$ satisfying (i) with a universal 
covering property.

 We  denote by $P(S)$
the points $\{ x_1, x_2,\ldots\}$ obtained in this procedure and note
that $P(S)\subset S$.
Let $\wt H(S): = \bigcup_{i\in P(S)} \wt D_i$
and $H^*(S): = \bigcup_{i\in P(S)} D_i^*$

Now we show how to choose points in the whole space. Fix an arbitrary
point $x$, and let $A_k : = \{ y \; : \; 4^{k}\ell(x)\leq |y-x| \leq 4^{k+1}
\ell(x)\}$, $k=1, 2,\ldots$,
 be a sequence of annuli. Clearly $\wt D_x\cup
\bigcup_k A_k = \bR^3$. For each annulus we construct the
points $P(A_k)$ defined above and we let
$$
        P := P(\wt D_x) \cup \bigcup_{k=1}^\infty P(A_k).
$$
This will be our final set of points $\{ x_1, x_2, \ldots \}$ after 
relabelling. It is clear that the balls $\hD_i = B(x_i,\ell(x_i)/10)$,
$x_i\in P$, cover the space.

\bigskip

Next we prove the finite covering property for the balls
$\{ \tD_{x_i} \; : \; x_i\in P\}$, i.e. that
the number of balls that cover any 
given point of $\bR^3$ is universally bounded.
 We need a lemma
whose proof is given later.

\begin{lemma}\label{lemma:inters} Fix any point $x\in\bR^3$.

(i)  Let
$L_k: = \sup\{ \ell(u)\; : \; u\in A_k\}$, $k\ge 1$, then
 $L_k\leq 4^{k+1}\ell(x)$.

(ii) If $y\in A_k$, then $y\not\in H^*(A_m)$
for any $|m-k|\ge 5$.
\end{lemma}

Recall that  for each $m$ every point in $\bR^3$
is covered  by at most $N$ balls $\wt D$ with center  $z\in P(A_m)$
and similarly for balls with center in $P(\tD_x)$.
Hence (ii) of  Lemma \ref{lemma:inters}
shows that any $y$ is covered by at most $12N$ balls
with center from $P$. This completes the proof of the finite
covering property  of the balls $\{ \tD_{x_i} \; : \; x_i\in P\}$.

\bigskip

Finally we show property (ii) of the Definition \ref{def:unif}. 
If $\tD_i\cap \tD_j\neq \emptyset$, then $\ell_i, \ell_j$
are comparable by (\ref{eq:elltempered}). Therefore $D_i^*$ covers
$x_j$, but any point is covered only by finitely many $D_i^*$'s,
hence $\tD_j$ can be intersected by finitely many $\tD_i$'s.

\noindent
{\it Proof of Lemma \ref{lemma:inters}}.
(i) Suppose that there exists $u\in A_k$
with $\ell(u)> 4^{k+1}\ell(x)$. Then $|x-u|\leq 4^{k+1}\ell(x) < \ell(u)$,
hence $\ell(u)\leq 2 \ell(x)$ by (\ref{eq:elltempered}) which
is a contradiction.

(ii) Suppose that there is a point $z\in P(A_m)$
such that $y\in D_z^*$, i.e., $|y-z|\leq 40\ell(z)$.
Using (\ref{eq:elltempered}) this implies
$\ell(z)\leq 2\ell(y)$ assuming $\e < 1/40$. Hence
$|y-z|\leq 80\ell(y)\leq 80\cdot 4^{k+1}\ell(x)$  by (i).
But $z\in A_m$, so $|y-z|\ge (4^{m}- 4^{k+1})\ell(x)$
which is a contradiction if $m\ge k+5$.
Suppose now that $m\leq k-5$. Then $|y-z|\ge (4^{k} - 4^{m+1})\ell(x)$
which contradicts to $|y-z|\leq 40\ell(z)\leq 40\cdot 4^{m+1}
\ell(x)$. $\;\;\Box$

\subsection{Proof of the localization Proposition \ref{prop:pullin}}
\label{sec:pullinpr}

As a preparation for the proof we define a distance function 
on the collection $\{ x_i\}_{i\in I}$ obtained in Lemma \ref{lemma:cover}.
Let $i\in I$ be a fixed index.
 We define the following compact sets successively
$$
        S_0(i) : = D_i\; ,
$$
$$
        S_{k+1}(i) : = \bigcup_{j\; : \; D_j\cap S_k(i)\neq\emptyset}
         \wt D_j \; ,
$$
and we denote
$$
        m_k : = \mbox{card} \{ j \in I\; : \; D_j\cap S_k(i)\neq\emptyset\} 
        \; .
$$

\begin{lemma}\label{lemma:S}
(i) The sets $S_0(i), S_1(i), \ldots $ are increasing.

(ii) Let $u\in  S_{k}(i)$, $v\not\in \mbox{int}(S_{k+1}(i))$, then
$|u-v|> 4\ell(u)$.

(iii) $\bigcup_k S_k(i)= \bR^3$.

(iv) $m_k\leq N^{k+1}$ with the universal constant $N$ from Lemma \ref{lemma:cover}.

(v) For sufficiently small $\e$ and for
any  nonnegative function $G$  such that
$G(x)$ and $G(y)$ are comparable whenever $|x-y|\leq \e^{-1}\ell(x)$,
we have
\be
        \sup_{S_k(i)} G \leq 2^k \sup_{S_0(i)} G \; .
\label{eq:Fgrow}
\ee

\end{lemma} 

\noindent
{\it Proof.} For simplicity, we omit $i$ from  the arguments since
$i$ is fixed.

(i) Since the balls $\{ D_j\}$ cover $\bR^3$ and
$D_j\subset\wt D_j$,
we see that $S_k\subset S_{k+1}$.

(ii) Let $u\in D_j$ for some $j$. From $|u-x_j|\leq \ell(x_j)$,
it follows that  $\ell(x_j)$, $\ell(u)$ are comparable.
Since $\tD_j\subset S_{k+1}$,
we have $|v-x_j|\ge 10 \ell(x_j)$, so $|v-u|\ge 9\ell(x_j)> 4\ell(u)$.

(iii) Suppose that $S:= \bigcup_k S_k$ is not the whole
$\bR^3$ and select a point $z\in
\partial S$.
Then we can find a sequence of points $z_k\in \partial S_k$
converging to $z$ such that $|z_k-z|$ monotonically
decreasing (for example, we can choose the point $z_k\in S_k$
closest to $z$). Since $\ell(z)>0$, we see that $|z-z_n|\leq
\ell(z)$ for some $n$, hence $\ell(z_n)$ and $\ell(z)$ are 
comparable. We have
$$
        |z_{n+1} -z_n|\leq |z_{n+1}-z|+|z-z_n|\leq 2\ell(z) \leq 4\ell(z_n)
$$
which contradicts (ii).

(iv) It is clear that $m_0\leq N$ by Definition \ref{def:unif} (ii).
By induction we show that $m_{k+1}\leq Nm_k$. This again
follows from Definition \ref{def:unif} (ii), since each $\tD_j$ in
the definition of $S_{k+1}$ may intersect at most $N$ balls
from the collection $\{ D_j \}_{j\in I}$.

(v)  Straightforward by induction on $k$ and by the definition
of $S_k(i)$.
$\;\;\;\Box$

\bigskip

This lemma gives an integer valued distance on the collection 
$\{ x_j \}_{j\in I}$:
$$
        d_{ij}: = \min\{ k \; : \; x_j\in S_k(i)\}\; .
$$
Clearly $d_{ii}=0$, and $d_{ij}+d_{jk}\ge d_{ik}$,
but the distance function is not symmetric. However, we
have
\begin{lemma}\label{lemma:distsymm}
For sufficiently small $\e$ the distance function satisfies
\be
        d_{ji} \leq 7d_{ij} +1 \; .
\label{eq:distsymm}\ee
\end{lemma}

\noindent
{\it Proof.} The proof goes by induction on the value of $d_{ij}$.
If $d_{ij}=0$, $x_j\in D_i$, then $x_i \in \wt D_j$, i.e., $d_{ji}\leq 1$.
Suppose that (\ref{eq:distsymm}) is proven for all $(i,j)$ pairs
with $d_{ij}\leq d$ and let now $d_{ij}=d+1$. Then there exists $m$
such that $x_m\in S_d(i)$,
$d_{im}\leq d$ and $x_j \in \wt D_a$ for some index $a$
with $|x_a- x_m|\leq \ell_a
+10\ell_m$. If $\e$ is sufficiently small,
the radii $\ell_a$, $\ell_j$ and $\ell_m$
are comparable, and it easily follows
that $d_{ja}\leq 3$ and $d_{am}\leq 3$.
Therefore $d_{ji}\leq d_{ja}+ d_{am} + d_{mi} \leq 6 + 7d + 1 < 7d_{ij}
+1$. $\;\;\;\Box$

\bigskip

For every $i\in I$, $k\in \bN$ we define
\be
        u_k^{(i)} := \sum_{j\; : \; D_j\cap S_k(i)\neq\emptyset}  
        \theta_j^2\; .
\label{def:u}\ee
Notice that $\mbox{supp} (u_k^{(i)})\subset S_{k+1}(i)$,
and $u_k^{(i)}\equiv 1$ on $S_k(i)$.  Moreover
\be
        |\nabla u_k^{(i)}(x)|\leq cN\ell(x)^{-1}.
\label{eq:uder}
\ee
To see this, we notice that every $x$ is covered by not more than $N$
balls $D_j$ and only these support those $\theta_j$'s which do not
vanish at $x$. Moreover $\ell_j$ is comparable to $\ell(x)$
for all these $j$ indices, hence $\|\nabla \theta_j\|_\infty \leq
c\ell(x)^{-1}$.

\medskip

In the rest of this section
we set 
$$
        R_f=R(f): = (T + f)^{-1}, \qquad
        R_i[f]: = (T_i + f)^{-1}
$$
for simplicity, in accordance with the notations (\ref{eq:resolv}),
(\ref{eq:iresolv}).

\medskip

\noindent 
{\it Proof of Proposition \ref{prop:pullin}}.
We start with an auxiliary lemma.

\begin{lemma}\label{lemma:rescomm} For any number $\mu\ge 0$ and real
function $\chi$ on $\bR^3$,
\bey
        \Big\| R_{P+\mu}^{1/2} (P+\mu) R_{P+\mu}^{1/2}
        \Big\|&\leq& 1\; ,
\label{eq:norm1}\\
        \Big\| R_{P+\mu}^{1/2} [T, \chi]
          R_{P+\mu}^{1/2}
        \Big\|&\leq& c_0\Big\| P^{-1/2} |\nabla\chi| \; \Big\|_\infty \; .
\label{eq:norm2}
\eey
\end{lemma}

\noindent
{\it Proof of Lemma \ref{lemma:rescomm}.} The first inequality is trivial by 
inserting $P+\mu\leq T + P+\mu$.
For the second inequality we use
\be
        [T, \chi]  = A^*[A, \chi] + [A^*, \chi]A\; ,
\label{eq:Tcomm}
\ee
and it is sufficient to estimate one of these terms;
$$
        \Big\| R_{P+\mu}^{1/2}  A^*[A, \chi] 
         R_{P+\mu}^{1/2}
        \Big\|
        \leq \Big\| R_{P+\mu}^{1/2} A^* \Big\|
        \;  \Big\| [A, \chi]  R_{P+\mu}^{1/2}  \Big\|\; .
$$
Using $\|M \| = \| MM^*\|^{1/2}$, we obtain that the first
factor is bounded by 1 (again, using $A^*A= T\leq T +P+\mu$).
For the second factor we need pointwise commutator
bounds
\be
        [A, \chi][A,\chi]^* \leq c_0 |\nabla \chi|^2, \;\;\;
        [A^*, \chi][A^*,\chi]^* \leq c_0 |\nabla \chi|^2 \; ,
\label{eq:commutator}
\ee
that follows from (\ref{eq:Abound}).

Hence, we estimate the second factor as
$$
       \Big\| [A, \chi]  R_{P+\mu}^{1/2}  \Big\|
        \leq c_0 \Big\| R_{P+\mu}^{1/2} |\nabla\chi|^2 
         R_{P+\mu}^{1/2} \Big\|^{1/2}
         \leq c_0  \Big\| R_{P+\mu}^{1/2}P R_{P+\mu}^{1/2}
          \Big\|^{1/2}
         \Big\| P^{-1}|\nabla\chi|^2 \Big\|_\infty^{1/2}\; ,
$$
and we use (\ref{eq:norm1}). $\;\;\;\Box$

\bigskip

The key lemma is the following (recall the definition
of $\wt\chi_i$ from Section \ref{sec:cutoff}):

\begin{lemma}\label{lemma:pullin} For sufficiently small  $\e$ we have
the following estimates for any $i,j\in I$
\bey
        \wt\chi_i R_{P+\mu}\t_j F^2\theta_j R_{P+\mu}\wt\chi_i &\leq &
        c(4c_0\e)^{2(d_{ij}-1)_+}  F_i^2
         (P_i+\mu)^{-1} \wt\chi_i R_{P+\mu}\wt\chi_i\; ,
\label{eq:ij}\\
        \wt\chi_i R_{P+\mu} F^2 R_{P+\mu}\wt\chi_i
        &\leq &  c F_i^2
          (P_i+\mu)^{-1}\wt\chi_i R_{P+\mu}\wt\chi_i \; ,
\label{eq:pullin}\\
        \wt\chi_i R_{P+\mu}^2\wt\chi_i &\leq & (P_i+\mu)^{-2}\; ,
\label{eq:Rpest}\\
        \chi_i A R_{P+\mu}^2 A^* \chi_i  &\leq & c(P_i+\mu)^{-1} \; .
\label{eq:Rpder}\eey
\end{lemma}

\noindent
{\it Proof of Lemma \ref{lemma:pullin}}. For brevity, we
denote $R:= R_{P+\mu}$.
First we show (\ref{eq:ij}).
We assume that  $\e_0$ is small enough
so that $F(x_i)$ and $F(x_j)$ are comparable
as long as $\tD_j\cap \tD_i\neq\emptyset$ (see (\ref{eq:Freg})).

We first consider the case $d_{ij}\leq 1$. Then
$F^2\leq c F_j^2\leq cF_i^2$ on the support of $\theta_j$.
We also use
\be
        \t_j^2\leq c(P_j+\mu)^{-1}(P+\mu)\leq c(P_i+\mu)^{-1}(P+\mu)\; .
\label{eq:thetaest}\ee
Hence (\ref{eq:ij}) follows from
$$
        \wt\chi_i R\t_j F^2 \t_j R \wt\chi_i \leq cF_i^2
        \wt\chi_i R\t_j^2 R \wt\chi_i \leq cF_i^2(P_i+\mu)^{-1}
        \wt\chi_i R (P+\mu) R \wt\chi_i
$$
and finally we use (\ref{eq:norm1}) to estimate $R(P+\mu)R\leq R$.

\medskip

To prove (\ref{eq:ij}) for $d=d_{ij}\ge 2$,
we recall the definition
of the functions $u_k^{(i)}$ (\ref{def:u}). For brevity, we omit the
superscript $i$.
We successively insert the functions $u_1, u_2,
\ldots u_{d-1}$ where $d=d_{ij}$:
\bey
        \wt\chi_i R\theta_j &= &\wt\chi_i u_1 R\theta_j 
        = \wt\chi_i R [T, u_1]R\theta_j
        = \wt\chi_i R [T, u_1] u_2 R\theta_j
        = \wt\chi_i R [T, u_1] R [ T, u_2] R\theta_j
        \nonumber\\
        &=& \ldots 
        = \wt\chi_i R [T, u_1] R [T, u_2] R
         \ldots R [T, u_{d-1}] R \theta_j\; . \nonumber
\eey
We used that $u_1\equiv 1$ on the support of $\wt\chi_i$,
 $u_{k+1}\equiv 1$ on the support of $\nabla u_k$ and $\mbox{supp} (u_{d-1})
\cap \mbox{supp}(\theta_j)=\emptyset$.
Therefore we can first estimate $\t_j F^2\theta_j \leq cF_j \t_j^2$, then
use the succesive insertions to obtain
\bey
        \lefteqn{\wt\chi_i R \t_j F^2\theta_j  R\wt\chi_i}
\label{eq:manycomm} \\
       & \leq& cF_j^2 \wt\chi_i R^{1/2}
         \Bigg[ \prod_{k=1}^{d-1}\Big( R^{1/2} 
        [T, u_k] R^{1/2}\Big) \Bigg] R^{1/2} \theta_j^2 R^{1/2} 
        \Bigg[ \prod_{k=1}^{d-1}\Big( R^{1/2} 
        [T, u_k] R^{1/2}\Big) \Bigg]^* R^{1/2} \wt \chi_i\; .
        \nonumber
\eey

First we use that
$$
        R^{1/2} \theta_j^2 R^{1/2} 
        \leq c(P_j+\mu)^{-1} R^{1/2} (P+\mu) R^{1/2}
        \leq c(P_j+\mu)^{-1}
$$
by (\ref{eq:thetaest})
and (\ref{eq:norm1}). Then we use (\ref{eq:norm2}) to estimate the
commutator norms and  we use (\ref{eq:uder}) to get
$$
        \Big\| P^{-1/2}|\nabla u_k|\; \Big\|_\infty \leq c\e^{5/2}\leq
        \e
$$
for sufficiently small $\e$.
We obtain
\be
        \wt\chi_i R \t_j F^2\theta_j  R \wt\chi_i
        \leq c(c_0\e)^{2(d_{ij}-1)}
        F_j^2(P_j+\mu)^{-1} 
         \wt\chi_i R 
         \wt\chi_i
\label{eq:manycomm1}
\ee
By (\ref{eq:Fgrow}), we see that 
$F_j^2 (P_j+\mu)^{-1}\leq 16^d F_i^2 (P_i+\mu)^{-1}$
because part (v) of Lemma \ref{lemma:S} 
applies both to the function $G=F$ and $G=(P+\mu)^{1/2}$.
 This completes the proof of (\ref{eq:ij}).

\bigskip

To prove (\ref{eq:pullin}) 
we insert a partition of unity 
$$
        \wt\chi_i R F^2 R\wt\chi_i 
        =\sum_{j\in I} \wt\chi_i R \t_j F^2\theta_j R\wt\chi_i\; .
$$
We use (\ref{eq:ij}), (iv) of Lemma \ref{lemma:S} and that
\be
        \sum_{j\in I} (4c_0\e)^{2(d_{ij}-1)_+} 
        \leq 1+N +\sum_{p=1}^\infty (4c_0\e)^{p} N^p 
        \leq N+2
\label{eq:dsum}\ee
if $\e\leq \e_0$,
where the universal constant $N$ is from Lemma \ref{lemma:cover}.
This proves 
(\ref{eq:pullin}).

\medskip

The proof of (\ref{eq:Rpest}) is straight-forward by applying 
(\ref{eq:pullin}) with $F\equiv 1$,
\be
        \wt\chi_i R_{P+\mu}^2\wt\chi_i\leq 
        c(P_i+\mu)^{-1} \wt\chi_i R_{P+\mu}\wt \chi_i, 
\label{eq:2to1}
\ee
and then using $R_{P+\mu}\leq  (P+\mu)^{-1}$ which is bounded by
$c(P_i+\mu)^{-1}$ on the support of $\wt\chi_i$
since $P$ and $P_i$ are comparable on this set. 

\medskip

For the proof of (\ref{eq:Rpder}) we insert $\wt\chi_i$ that is
identically 1 on the support of $\chi_i$ and use (\ref{eq:2to1})
$$
        \chi_i A R_{P+\mu}^2A^*\chi_i = 
        \chi_i A \wt\chi_i R_{P+\mu}^2\wt\chi_iA^*
        \chi_i \leq c(P_i+\mu)^{-1} \chi_i A \wt\chi_i R_{P+\mu}\wt \chi_i
        A^*\chi_i \; .
$$
We can remove $\wt\chi_i$ and use $A R_{P+\mu}A^*\leq 1$
to finish the proof. $\;\;\Box$

\bigskip

The next lemma is a strengthening of (\ref{eq:ij})
in Lemma \ref{lemma:pullin}.
Notice that in (\ref{eq:ij}) we lost a resolvent, and the right hand
side is not locally trace class in the high momentum regime.
The following lemma remedies this:

\begin{lemma}\label{lemma:ijj}
For sufficiently small $\e$
\be
        \theta_i^2 R_{P+\mu}  \t_j F^2\theta_j R_{P+\mu}
        \theta_i^2 \leq  c(4c_0\e)^{2(d_{ij}-1)_+}
         F_i^2\theta_i^2
        \Big( R_i^2[P_i] + P_i^{-1}R_i[P_i]A_i^*
        \varphi_i^2 A_iR_i[P_i]\Big)
        \theta_i^2 \; .
\label{eq:ijj}\ee
\end{lemma}

\noindent
{\it Proof of Lemma \ref{lemma:ijj}}. For simplicity, we let $R:= R_{P+\mu}$
and $R_i: = R_i[P_i]$ in this proof.
We start with the identity
\be
        \wh\chi_i R  = R_i \wh\chi_i
        +  R_i \Big(\wh \chi_i (P_i-P-\mu) + A^*_i
        [A,\wh\chi_i] + [A^*,\wh\chi_i]
        A\Big)R\; ,
\label{eq:chiR}
\ee
since $A$ and $A_i$ coincide on the support of $\wh\chi_i$ by 
(\ref{eq:coinc}). After a Schwarz' inequality
\bey
        \lefteqn{\wh\chi_iR \t_j F^2\theta_j R\wh\chi_i}\label{eq:negy}\\
        &\leq& c \Bigg( R_i \wh\chi_i  \t_j F^2\theta_j
        \wh\chi_iR_i +
         R_i \wh\chi_i (P_i-P-\mu)R \t_j F^2\theta_j
        R(P_i-P-\mu)\wh\chi_iR_i\nonumber\\
       && + R_i A_i^* [A,\wh\chi_i] R \t_j F^2\theta_j
         R [A,\wh\chi_i]^*A_iR_i
         + R_i[A^*,\wh\chi_i]A R \t_j F^2\theta_j
         RA^* [A^*,\wh\chi_i]^*R_i\Bigg)\;.\nonumber
\eey
The first term is estimated as
\be
         R_i \wh\chi_i  \t_j F^2\theta_j
        \wh\chi_iR_i \leq cF_i^2 R_i^2 {\bf 1}(d_{ij}\leq 1)\; ,
\label{eq:oner}
\ee
since $F\leq cF_i$ on the support of $\t_j\wh\chi_i$.

Since $\chi_i\equiv 1$ on the support of $\wh\chi_i$, we
can freely insert $\chi_i$ in the last three terms of (\ref{eq:negy})
replacing $ R\t_j F^2\theta_j R$ with $\chi_i R\t_j F^2\theta_j R\chi_i$
everywhere and apply (\ref{eq:ij}) after multiplying it by $\chi_i$
from both sides:
$$
        \chi_i R\t_j F^2\theta_j R\chi_i\leq c(4c_0\e)^{2(d_{ij}-1)_+}
        F_i^2 (P_i+\mu)^{-1} \chi_i R \chi_i\; .
$$

In the second term of (\ref{eq:negy}) we use $R\leq (P+\mu)^{-1}$
and  that $P+\mu$ and $P_i+\mu$ are comparable on the support
of $\wh\chi_i$. Hence
\be
        R_i \wh\chi_i (P_i-P-\mu)R \t_j F^2\theta_j
        R(P_i-P-\mu)\wh\chi_iR_i \leq c(4c_0\e)^{2(d_{ij}-1)_+}
        F_i^2  R_i^2\; ,
\label{eq:twor}
\ee
using $(P_i-P-\mu)^2\leq (P_i+\mu)^2$ on the support of $\wh\chi_i$.

In the third term of (\ref{eq:negy})
we again estimate  $R\leq (P+\mu)^{-1}\leq c P_i^{-1}$
on the support of $\wh\chi_i$, we
use (\ref{eq:commutator}) and
that $|\; [A,\wh\chi_i][A,\wh\chi_i]^*|=
|\nabla\wh\chi_i|^2\leq c \e^2 P_i \varphi_i^2$ to obtain
\be
        R_i A_i^* [A,\wh\chi_i]  R\t_j F^2\theta_j
         R  [A,\wh\chi_i]^*A_iR_i \leq
        c(4c_0\e)^{2d_{ij}}
        F_i^2 P_i^{-1} R_i A_i^* 
         \varphi_i^2A_iR_i\; .
\label{eq:threer}
\ee

Finally, the fourth term of (\ref{eq:negy}) satisfies
$$
         R_i [A^*, \wh\chi_i] A R \t_j F^2\theta_j
         RA^* [A^*, \wh\chi_i]^* R_i \leq c(4c_0\e)^{2(d_{ij}-1)_+}
        F_i^2 P_i^{-1}
          R_i  [A^*, \wh\chi_i] A \chi_i R \chi_i
         A^* [A^*, \wh\chi_i]^*R_i
$$
\be
        \leq c(4c_0\e)^{2d_{ij}} F_i^2 R_i^2\;,
\label{eq:fourr}
\ee
since $\chi_i\equiv 1$ on the support of $\wh\chi_i$, we
can omit it, and we used $AR A^*\leq 1$ and (\ref{eq:commutator}).
Lemma \ref{lemma:ijj} follows from
(\ref{eq:negy})--(\ref{eq:fourr}) using that $\theta_i\wh\chi_i =\theta_i$.
$\;\;\;\Box$

\bigskip
\medskip

Finally, we complete  the proof of Proposition \ref{prop:pullin}.
We use $R=R_{P+\mu}$ for brevity.
We insert three partitions of unity and perform
a weighted Schwarz' inequality
\bey
        RF^2R &=& \sum_{i,j,k\in I}
         \theta_i^2 R \theta_j F^2 \theta_j R \theta_k^2
\label{eq:threesum}\\
    &\leq& \sum_{i,j,k\in I} \Bigg( \e^{(d_{kj}-1)_+- (d_{ij}-1)_+} 
         \theta_i^2 R \theta_j F^2 \theta_j R \theta_i^2
        +  \e^{(d_{ij}-1)_+- (d_{kj}-1)_+} 
         \theta_k^2 R \theta_j F^2 \theta_j R \theta_k^2\Bigg)\; .
        \nonumber
\eey
Using (\ref{eq:distsymm})  we see as in (\ref{eq:dsum}) that
$$
        \sum_{k\in I} \e^{(d_{kj}-1)_+} \leq \sum_{k\in I}
        \e^{(d_{jk}-8)_+/7} \leq N^9
$$
if $\e$ is small enough. Here $N$ is from 
Lemma \ref{lemma:cover}. Therefore (\ref{eq:threesum}) implies
$$
        RF^2R \leq   cN^9 \sum_{i,j} \e^{- (d_{ij}-1)_+} 
         \theta_i^2 R \theta_j F^2 \theta_j R \theta_i^2 \; .
$$

We use (\ref{eq:ijj}) and sum up the index $j$ similarly to (\ref{eq:dsum})
with a possible smaller $\e$
\bey
        RF^2R &\leq &
        cN^8  \sum_{i,j} (16c^2_0\e)^{ (d_{ij}-1)_+}F_i^2
        \theta_i^2
        \Big( R_i^2 + P_i^{-1}R_iA_i^* \varphi_i^2 A_iR_i\Big)
        \theta_i^2 \nonumber\\
        & \leq& c N^{10}\sum_i F_i^2 \theta_i^2
        \Big( R_i^2 + P_i^{-1}R_iA_i^* \varphi_i^2 A_iR_i\Big)
        \theta_i^2\; . \qquad \Box \nonumber
\eey

\subsection{Proof of Lemma \ref{lemma:metric} on
the magnetic coordinates}\label{sec:metricproof}

We give the construction of the new coordinates and conformal metric
 but we do not follow the explicit bounds along the proof as they
 easily
follow by scaling.

We consider $z\in\cP$ fixed in the proof and omit the notation $z$
in the sub- and superscripts.
We define the function
\be
        \kappa(\tau):=- \frac{1}{2} \frac{\rd}{\rd \tau} \log        
        |\bB(\varphi(\tau))| \; 
\label{eq:Kdef}
\ee
where $\kappa(\tau)\in C^3(\bR)$.
At each point $\varphi(\tau)$ we
consider a small spherical cap of the sphere $\cS(\tau)$,
going through $\varphi(\tau)$, orthogonal to the field line $\varphi$ and
 having  curvature $|\kappa(\tau)|$. The different spheres should
curve in a direction determined by the sign of $\kappa(\tau)$:
positive curvature means a sphere with outward normal
pointing in the direction of $\dot\varphi$.
We consider now a small cylindrical 
tubular neighborhood, $\wt\cN: = \{ x \; : \; \inf_\tau
\mbox{dist}(x, \varphi(\tau))\leq 10\ell\}$, along the $C^5$-curves
$\varphi$, in which the  spherical caps define a $C^3$-foliation. 
Since $\bB$ is extended $(D, K)$-regular,
 such neighborhood exists 
if $\e\leq \e(K)$ is sufficiently small.
The foliation naturally extends the function $\tau$ onto 
 $\wt\cN$ such that it is constant on the leaves
and $\tau\in C^3(\bR^3)$.

This foliation can be extended to the whole of $\bR^3$ in the
following way. Since $\bB=\bB_\infty$ outside of $D= B(z_0,\ell)$,
for some $z_0\in\bR^3$,
in case of $|z-z_0|\ge 2\ell$  we have
$\kappa(\tau)\equiv0$  and $\varphi$ is a straight line,
so the foliating spherical caps are parallel flat discs
and they can be trivially extended to a foliation of $\bR^3$
with parallel planes.

Now we consider the case $|z-z_0|< 2\ell$, where we only know
$\kappa(\tau)\equiv 0$ for $|\tau|\ge 3\ell$. In this region
 the spherical caps are again parallel flat discs and
they can be extended to parallel planes. 
That leaves a parallel slab unfoliated
 between the planes passing through $\varphi(-3\ell)$
and  $\varphi(3\ell)$. The width of the slab is $(6 \pm cK\e)\ell$.

 We consider the smooth function $F:\bR^3\mapsto \bR$ 
of the form $F(x)= \lambda (\bn_\infty\cdot x)$,
where $\lambda:\bR\to\bR$ is smooth and is chosen
such that  $F(\varphi(\tau))=\tau$ for $|\tau|\ge 3\ell$,
 $\|\lambda' - 1\|_\infty\leq cK\e\ell$
 and $\|\lambda^{(\gamma)}\|_\infty\leq cK\e \ell^{-\gamma}$, $\gamma=2,3,4$.
The level sets of $F$ define a parallel foliation of $\bR^3$ which
coincides with the previous foliation outside of the slab.

Let $\chi$ be a smooth cutoff function supported 
on the ball
$\wt D :=B(z_0, 6\ell)$, $\| \nabla^\gamma \chi\|_\infty\leq c\ell^{-\gamma}$,
$\gamma=1,\ldots, 4$,  and $\chi\equiv 1$ on $B(z_0, 5\ell)$. 
Since $|z-z_0|\leq 2\ell$, we 
note that $\wt D\subset\wt\cN$, so $\tau$ is already defined 
on $\mbox{supp}(\chi)$.
We define the function
$$
        t : = \chi \tau + (1-\chi)F\; .
$$
An easy calculation shows that $t\in C^3(\bR^3)$,
 $\|\nabla t - \bn_\infty\|\leq cK\e$ and
the level sets of $t$ define a regular foliation of $\bR^3$.
This is clearly an extension of the foliation given by $\tau$ on $D$
and the leaves are planes on $\wt D^c$.
Moreover, if we define a smaller tubular neighborhood $\cN$ 
of the central field line as $\cN: = \{ x \; : \; \inf_\tau
\mbox{dist}(x, \varphi(\tau))\leq 2\ell\}$, then
we note that $\cN\subset B(z_0, 5\ell)\cup \{ |\tau|\ge 3\ell\}$,
therefore $t\equiv \tau$ on $\cN$. Let $N:=\|\nabla t\|^{-1}\nabla t$
be the unit  $C^2$-vectorfield orthogonal to the foliation. 
We remark that 
the integral curves of $N$ typically 
do not coincide with the field lines except
on the field line $\varphi$ and in the region far away from $z_0$.

Armed with this foliation,
we introduce new coordinates on $\bR^3$.
On the plane $\cP$ we choose Euclidean orthonormal
 coordinates $\xi_1, \xi_2$ with
origin at $z$ and dual to the basis $\{ p_1, p_2\}$, i.e.
$x-z = \xi_1 p_1 + \xi_2 p_2$. Clearly 
$(\partial_{\xi_j}, \partial_{\xi_k} )=\delta_{jk}$ for $j,k=1,2$.
For simplicity we set $\partial_j := \partial_{\xi_j}$
and $\nabla_j: =\nabla_{\partial_j}$ in this proof.

We extend the coordinate system $\xi_1, \xi_2$ defined on the plane
$\cP$ by setting $\xi_1, \xi_2$ constant on the integral curves
of $N$.  It is easy to check that $\xi_1, \xi_2\in C^3(\bR^3)$.
Together with $t$ they
define a regular set of coordinates on $\bR^3$. 
The central line is given by $(0,0,t)$ in these coordinates.
Let $b(t): = |\bB(0,0,t)|\in C^3(\bR)$ be the strength of the
magnetic field along the central line, note
that $b(t)$ is comparable with $b$ for all $t$ by (\ref{eq:Btem}).
We define $\xi_3=\xi_3(t)$ to be the solution of $\sfrac{\rd}{\rd t}\xi_3 =
[b(t)/b]^{1/2}$ with $\xi_3(0)=0$, clearly $\xi_3\in C^4(\bR)$.
We reparametrize the coordinate $t$ with $\xi_3:= \xi_3(t)$.
In this way we defined a new coordinate system, $\{ \xi_1, \xi_2, \xi_3\}$,
with origin at $z$.
We shall view the coordinates $\xi_1, \xi_2, \xi_3$ as $C^3$ functions
of $x \in \bR^3$ and whenever the dependence on $z$ is
relevant, we use the notation $\xi^z= (\xi_1^z, \xi_2^z, \xi_3^z)$.
It is easy to check that the function $(\hat z, \xi) \mapsto \xi^z$
belongs to $C^3(\bR^5)$ and the vectorfields $\partial_i
=\partial_{\xi_i}$
are $C^2$.

We note that in the trivial case, $|z-z_0|\ge 2\ell$,
we simply have $\xi^z(x)= R^t(x-z)$,
where  $R:=[p_1 |  p_2 | \bn_\infty]$ is the
3 by 3 matrix with columns $p_1, p_2, \bn_\infty$,
and all statements 
of Lemma \ref{lemma:metric} are trivial with $\Om \equiv h \equiv 1$.

{F}rom now on we shall assume that $|z-z_0|< 2\ell$.
The relations (\ref{eq:x=xi})--(\ref{eq:xizero}) and (\ref{eq:dsomtran}), 
i.e., the fact that $\rd s^2$ has no $\rd\xi_j\, \rd \xi_3$ ($j=1,2$)
components follow directly from the construction.
{F}rom the regularity of the magnetic field (Definition
\ref{def:Dreg}) it easily follows that the Jacobian of the change of
coordinates $x\mapsto\xi^z(x)$ 
is close to the matrix $R^t$ and it varies regularly in $z$.
This proves
that the function $(\hat z, \xi)\mapsto
x^z(\xi)$ is well defined and $C^3(\bR^5)$, it
 also proves 
(\ref{eq:jacder})
and (\ref{eq:abound}) by the inverse function theorem.

The metric is diagonal in the $\xi$ coordinate system 
on the plane $\cP$, i.e. for $t=0$.
The key point is to show that it remains diagonal
within the tubular neighborhood $\cN$. The diagonal metric
elements will define the functions $\Om$ and $h$.

We derive a differential equation for the metric components
$g_{jk}: = (\partial_j, \partial_k)_g
 \in C^2(\bR^3)$ where $j,k=1, 2$ within $\cN$.
We have $\partial_t g_{jk} =\partial_t(\partial_j,\partial_k)_g
=(\nabla_t\partial_j,\partial_k)_g+(\partial_j,\nabla_t\partial_k)_g$.
Using that $\partial_t,\partial_j,\partial_k$, are coordinate fields,
i.e, have vanishing Lie derivatives we have $\nabla_t\partial_j=
\nabla_j\partial_t$. Recall that
 $N=g_{tt}^{-1/2}\partial_t$
is the unit normal to the spherical foliation, where  $g_{tt}:= (\partial_t,
 \partial_t)_g$.
Then $\nabla_j\partial_t=g_{tt}^{1/2}\nabla_jN+\partial_j(g_{tt}^{1/2})N$
and therefore we have 
$$
        \partial_t g_{jk} = g_{tt}^{1/2} [(\nabla_j N,\partial_k)_g
        +(\partial_j,\nabla_kN)_g]=2g_{tt}^{1/2} K_{jk}\; ,
$$
where $K_{jk}$ is the second fundamental form of the leaves of the foliation.
For a sphere immersed in $\bR^3$ we have $K_{jk}=\kappa g_{jk}$,
where $\kappa$ is the curvature.  We recall the choice
of $\kappa$ from (\ref{eq:Kdef}) and that $t\equiv \tau$ on $\cN$.
Thus 
\be
        \partial_t g_{jk} = 2g_{tt}^{1/2}\kappa(t)g_{jk}.
\label{eq:gdiff}
\ee

This proves that since $g_{12}$ is zero on the 
supporting plane $t=0$, it is zero everywhere in $\cN$. It also proves that
$g_{11}=g_{22}$ everywhere in $\cN$ since they satisfy the same
equation and initial condition. The same relations trivially hold
for the region $|\xi_\perp|\ge 10\ell$, where $g_{11}=g_{22}=g_{12}=1$.
Moreover, we define $\Om : = g_{11}^{-1/2}\in C^2(\bR^3)$
 and we obtain that within $\cN$ as well as in the regime
$|\xi_\perp|\ge 10\ell$ the conformal metric
can be written as 
$$
\rd s_\Om^2
= \rd\xi_1^2 + \rd\xi_2^2 + \Om^2 g_{tt} \rd t^2
=  \rd\xi_1^2 + \rd\xi_2^2 + \Om^2 g_{tt}f(\xi_3)^{-2} \rd \xi_3^2
$$
using the definition of $f$ and $\xi_3$. This proves (\ref{eq:dsom})
with $h: = \Om g_{tt}^{1/2} f(\xi_3)$.
Since the new coordinates form an orthonormal system
for $|\xi_3|\ge 3\ell$ and also for $|\xi_\perp|\ge 10\ell$
modulo a change of variables in the third direction,
the identities (\ref{eq:Omhup}) and (\ref{eq:omout}), respectively,
follow from the definitions.

Along the central  line
 we have $g_{tt}(\varphi(t))=1$. Thus (\ref{eq:gdiff}), (\ref{eq:Kdef})
and $g_{11} \equiv 1$ for $|\tau|\ge 3\ell$ implies
that $g_{11} = b/|\bB|$, i.e., $\Om = f(\xi_3)$ and $h\equiv 1$
along the central line. Then 
(\ref{eq:jacder})
 implies (\ref{gttbound}).
The global bounds (\ref{eq:homtemp})--(\ref{eq:homdertemp})
also follow from the smoothness of the contstruction, i.e.
from   (\ref{eq:jacder}). The details are left
to the reader.

Finally, the orthonormal basis $\{ e_1, e_2, e_3\}$ in $\rd s_\Om^2$
is defined by first constructing $e_1':=\partial_1$, $e_2':=\partial_2$,
$e_3':= h^{-1}\partial_3$ which are
automatically orthonormal apart from the region $\{ \xi \; : \; \sfrac{3}{2}
\ell\leq |\xi_\perp|\leq 9\ell, \; |\xi_3|\leq 3\ell\}$.
On this region we apply a Gram-Schmidt orthonormalization procedure
to obtain  $\{ e_1, e_2, e_3\}$ from
 $\{ e_1', e_2', e_3'\}$.  $\;\;\;\Box$

\subsection{Comparison of operators on equivalent
$L^2$-spaces}\label{sec:change}

Let $\rd \mu$ and $\rd \nu$ be two positive measures on $\bR^d$,
that are mutually and uniformly absolutely continuous, i.e.
$\rd \nu (x) = F(x) \rd \mu(x)$ with a 
 a positive bounded function $F$
with bounded inverse $F^{-1}$. We let 
$$
        C_F: =\| F \|_\infty \| F^{-1} \|_\infty
        = {\max F\over \min F}\; .
$$
Consider the spaces $L^2(\rd \mu)$ and
 $L^2(\rd \nu) =L^2(F\rd \mu)$ and let $A$ be any operator
defined on $L^2(\rd \mu)$. Since these two spaces are the same
as sets, we can consider $A$ acting on $L^2(F\rd \mu)$ as well.
We denote this operator by $A_F$.
 Let 
$(\cdot ,\cdot ):= (\cdot,\cdot)_{L^2(\rd \mu)}$ and 
and $(\cdot ,\cdot )_F:= (\cdot, \cdot)_{L^2(F\rd\mu)}$.
Similar convention is used for $\| \cdot \|$ and $\| \cdot\|_F$
and for the traces over these $L^2$-spaces: $\Tr : =
\Tr_{L^2(\rd\mu)}$
and $\Tr_F:= \Tr_{L^2(F\rd\mu)}$

\begin{lemma}\label{lemma:changenorm}
Let $A$ be a Hilbert-Schmidt
 operator on $L^2(\rd \mu)$ with a kernel $A(x,y)$.
 Then  $A_F$ is also Hilbert-Schmidt on $L^2(\rd\nu)$, and
the kernels of these operators satisfy
\be
        A(x,y) = A_F(x,y)F(y)\; .
\label{AFkernel}
\ee
Moreover, the diagonal kernels of $A^*A$ and $A_F^*A_F$ are comparable:
\be         
        C_F^{-1} \| F\|^{-1}_\infty 
        (A^*A) (x,x)\leq \Big( A_F^* A_F \Big)(x, x)
        \leq C_F  \| F^{-1}\|_\infty  (A^*A)(x,x) \; .
\label{kernelcomp}\ee
Furthermore, if $0< \a \leq A^*A\leq \beta$ for some constants $\a,
\beta$,
then 
\be
        C_F^{-1}\a\leq A_F^* A_F \leq C_F\beta\; .
\label{kernelcomp2}\ee
If, in addition, $A$ is of trace class, then so is $A_F$ and
their diagonal kernels satisfy
\be
        A(x,x)=A_F(x,x)F(x) \; .
\label{AFdiagkernel}
\ee
\end{lemma}

\noindent
{\it Proof.} Recalling the conventions at the end of Section
\ref{sec:structure},
the identities (\ref{AFkernel}) and (\ref{AFdiagkernel})
are obvious.
For (\ref{kernelcomp}) we estimate
\bey
        (A_F^*A_F)(x,x) &=&\int A_F^*(x,y)A_F(y,x)F(y)\rd\mu( y)\nonumber\\
        &=& F^{-2}(x)\int |A(x,y)|^2 F(y)\rd\mu (y) \nonumber \\
        &\leq&
        C_F \| F^{-1}\|_\infty (A^*A)(x,x)\; . \nonumber
\eey
The lower bound is proven similarly.

For (\ref{kernelcomp2}) we notice that
$$
        (\psi, A_F^*A_F\psi)_F
         = \| F^{1/2}A\psi\|^2
        \leq (\max F) \| A \psi\|^2 \leq \beta
         (\max F) \|\psi\|^2
$$
and 
$$
        \| \psi \|^2 \leq \| F^{-1/2} \psi \|^2_F \leq (\max F^{-1})
         \| \psi\|_F^2\; ,
$$
which proves the upper bound. The proof of the
lower bound in  (\ref{kernelcomp2})
is  similar. $\;\;\;\Box$

\medskip

\begin{lemma}\label{lemma:snegchange}
Let $A_k$ be a finite collection
of closed operators on $L^2(\rd\mu)$, let $W_1$, $W_2$
 be nonnegative functions on
$\bR^d$.
 Then 
\be
        \Big|\;
        \Tr_F \Big( \sum_k (A_k)_F^*(A_k)_F + W_1- W_2\Big)_-\Big|
        \leq \Big|\; \Tr\Big(\sum_k A_k^*A_k + W_1- C_FW_2\Big)_- \; \Big|\; .
\label{eq:snegchange}
\ee
\end{lemma}

\noindent
{\it Proof of Lemma \ref{lemma:snegchange}.} By the variational
principle
$$
        \Tr_F \Big( \sum_k (A_k)_F^*(A_k)_F + W_1-W_2\Big)_-
        =\inf\Big\{ \Tr_F\Big( \sum_k (A_k)_F^*(A_k)_F +W_1
        -W_2\Big)\gamma \; : \; 0\leq \gamma \leq 1 \Big\}
$$
where the infimum is over all finite rank
density matrices $\gamma$ on $L^2(F\rd\mu)$.
We can write $\gamma = \sum_n \lambda_n (f_n, \cdot)_Ff_n$
with $0\leq \lambda_n\leq 1$ and $\{ f_n\}$ being orthonormal in $L^2(F)$.

Define the  operator 
$\wt\gamma := (\min F)\sum_n \lambda_n (f_n, \cdot)f_n$ on $L^2$.
Since
$$
        (\phi, \wt\gamma\phi) = (\min F)\sum_n \lambda_n |(f_n,
        F^{-1}\phi)_F|^2 = (\min F)(F^{-1}\phi, \gamma F^{-1}\phi)_F
        \leq  \|\phi\|^2,
$$
$\wt\gamma$ is 
a density matrix on $L^2$.
Furthermore, for any $A=A_k$
\bey
        \Tr_F A_F^*A_F\gamma &=& \sum_n\lambda_n \| A_Ff_n\|_F^2
        = \sum_n\lambda_n \| Af_n\|_F^2 \ge (\min F)\sum_n\lambda_n \|
        Af_n\|^2\nonumber\\
        &=& (\min F) \sum_n\lambda_n \Tr |A^*Af_n\rangle\langle f_n|
        =  \Tr A^*A\wt\gamma \; . \nonumber
\eey
The potential term is estimated as
\bey
        \Tr_F (W_1-W_2)\gamma &=& \sum_n\lambda_n (f_n, (W_1-W_2)f_n)_F
        \nonumber\\
        &\ge&  (\min F)\sum_n\lambda_n  (f_n, (W_1-C_FW_2)f_n)
        \nonumber\\
        &= & \Tr (W_1-C_FW_2)\wt\gamma \; .\nonumber
\eey
Therefore
$$
        \Tr_F\Big( \sum_k (A_k)_F^*(A_k)_F +W_1
        -W_2\Big)\gamma \ge
         \Tr\Big( \sum_k A_k^*A_k +W_1
        -C_FW_2\Big)\wt\gamma \; ,
$$
and (\ref{eq:snegchange}) follows from the variational
principle.
$\;\;\;\Box$

\subsection{Comparison of Dirac operators under a conformal
transformation}\label{sec:compconf}

Let $\Om : \bR^3\to \bR_+$ be a $C^1$-function satisfying
\be
        \frac{1}{2} \leq \Omega (x) \leq 2 \; 
\label{omegabound}\ee
and 
\be
        \| \nabla \Om \|_\infty \leq \ell^{-1}
\label{omegaderbound}\ee
with some constant $\ell>0$.
We define the metric $\rd s^2_\Omega := \Omega^2\rd s^2$ that is conformally
equivalent to the   Euclidean metric $\rd s^2$.
Let $\D$ be a Dirac operator in the $\rd s^2$ metric, then
a Dirac operator  in the $\rd s_\Om^2$ metric is
given by $\D_\Omega := \Omega^{-2}\D \Omega$.
Notice that $\D_\Om$ is self-adjoint on $L^2(\rd s_\Om^2)\otimes\bC^2$
(see \cite{ES-III}). 
The following lemma compares certain resolvent kernels of
$\D$ and $\D_\Om$ on $L^2(\rd s^2)\otimes\bC^2$ and on $L^2(\rd s_\Om^2)
\otimes\bC^2$,
respectively.

\begin{lemma}\label{lemma:conf}  Let $P\ge 2^9 \ell^{-2}$ be a number.
Under the conditions (\ref{omegabound}), (\ref{omegaderbound}) 
we have
\be
        \tr \; \Bigg[
        {1\over (\D^2 + P)^2}\Bigg]_{L^2}(x,x)
        \leq 
        2^{9} \; \tr \; \Bigg[ {1\over (\D_\Omega^2+
        P)^2}\Bigg]_{L^2_\Om}
        (x,  x)   \qquad x\in \bR^3 \; .
\label{eq:sqconf}\ee
The left hand side is the diagonal of an
 operator kernel on $L^2(\rd s^2)\otimes\bC^2$,
the right hand side is the diagonal of 
an operator kernel on $L^2(\rd s^2_\Om)\otimes\bC^2$.
Moreover, if $0\leq \varphi\le 1$ is a bounded function then
\bey\lefteqn{
                \tr\; \Bigg[ {1\over \cD^2 + P} \cD \varphi^2 \cD 
        {1\over \cD^2 +P}\Bigg]_{L^2}(x,x)} \qquad\qquad
\label{eq:offconf}\\
         &\leq& 2^{12}\;
         \tr\Bigg[ {1\over \D^2_\Om + P}\D_\Om
        \varphi^2
        \D_\Om {1\over \D^2_\Om +P} 
         +   P 
         {1\over (\D_\Omega^2+ P)^2} \Bigg]_{L^2_\Om} (x,x)
\nonumber
\eey
for any $x\in \bR^3$.
\end{lemma}

\noindent
{\it Proof of Lemma \ref{lemma:conf}}. 
Let $\cV : L^2(\rd s^2)\otimes\bC^2\to 
L^2(\rd s_\Om^2)\otimes\bC^2$ be a unitary map
given by $\cV\psi : = \Om^{-3/2}\psi$.
Notice that
$$
        \D
        =\cV^*(\Omega^{1/2}\D_\Omega\Omega^{1/2})\cV\; ,
$$
therefore the unitary operator $\Om^{1/2} \D_\Om \Om^{1/2}$
on $L^2(\rd s_\Om^2)\otimes\bC^2$ is unitarily equivalent to $\D$ on
$L^2(\rd s^2)\otimes\bC^2$. In particular, for any real function $f$
\be
          f(\D)
        =\Om^{3/2}f(\Omega^{1/2}\D_\Omega \Omega^{1/2})\Om^{-3/2}\; .
\label{uniteq}
\ee

  {F}rom  (\ref{uniteq}) we obtain
$$
        {1\over (\D^2 +P)^2}\leq
         {4 \over (\D^2 +P)^2+3P^2}
        = \Omega^{3/2}
        {4\over ( [\Omega^{1/2}\D_\Omega \Omega^{1/2}]^2
        +P)^2+3P^2}\Omega^{-3/2} \; .
$$
In particular,
$$
         {1\over (\D^2 +P)^2}(x,x) \leq
         \Bigg({4\over (\Omega^{1/2}\D_\Omega \Omega \D_\Omega \Omega^{1/2}
        +P)^2+3P^2}\Bigg)_{L^2} (x,x) \; .
$$      
Here the right hand side is the $L^2(\rd s^2)\otimes\bC^2$ kernel of
the corresponding bounded non self-adjoint operator. However,
the same operator can be viewed on $L^2(\rd s_\Om^2)\otimes\bC^2$ as
well, where it is self-adjoint.
Using (\ref{AFdiagkernel})
from Lemma \ref{lemma:changenorm} we know that the two diagonal kernels
differ by a factor $\Om^3(x)$.

To conclude (\ref{eq:sqconf}), it is therefore  sufficient to show that
\be
        {1\over (\Omega^{1/2}\D_\Omega \Omega \D_\Omega \Omega^{1/2}
        +P)^2+3P^2} \leq  \Om^{-1/2}  {32\over (\D_\Omega^2+ P)^2}\Om^{-1/2}
\label{eq:Dome}\ee
as self-adjoint operators on $L^2(\rd s_\Om^2)\otimes\bC^2$.
Using
$$
        \Omega^{1/2}\D_\Omega \Omega \D_\Omega \Omega^{1/2}     
        = \Omega^{3/2} \D_\Omega^2 \Omega^{1/2}
        + \Omega^{1/2}[\D_\Omega, \Omega]\D_\Omega \Omega^{1/2}
$$
and  a Schwarz' inequality, we obtain
\bey\lefteqn{
         (\Omega^{1/2}\D_\Omega \Omega \D_\Omega \Omega^{1/2}
        +P)^2+3P^2 }
\label{omreso}\\
        &\ge& {1\over 2}(\Omega^{3/2} \D_\Omega^2 \Omega^{1/2})^*
        (\Omega^{3/2} \D_\Omega^2 \Omega^{1/2}) -
        2 \Big( \Omega^{1/2}[\D_\Omega, \Omega]
        \D_\Omega \Omega^{1/2}\Big)^*
        \Big(\Omega^{1/2}[\D_\Omega, \Omega]
        \D_\Omega \Omega^{1/2}\Big)
        +P^2
\nonumber\\
        &=& \frac{1}{2}\Bigg[ \Omega^{1/2} \D_\Omega^2 \Omega^3
         \D_\Omega^2 \Omega^{1/2} -
        4 \Omega^{1/2}\D_\Omega [\D_\Omega, \Omega]^*
        \Omega [\D_\Omega, \Omega]
        \D_\Omega \Omega^{1/2}
        +2P^2\Bigg]
\nonumber\\
        &\ge& \frac{1}{2}\Bigg[ \frac{1}{8} \Omega^{1/2} \D_\Omega^4
         \Omega^{1/2} -
         8 \Omega^{1/2}\D_\Omega [\D_\Omega, \Omega]^*
         [\D_\Omega, \Omega]
        \D_\Omega \Omega^{1/2}
        +2P^2\Bigg]\; . \nonumber
\eey
Using that $[\D_\Omega, \Omega]=\Omega^{-2}[\D,\Omega]\Omega$
and  (\ref{omegaderbound}) we can estimate
\be
        \Big\|[\D_\Omega, \Omega]^* [\D_\Omega, \Omega] \Big\|_\infty
       \leq 4\|\nabla\Omega\|_\infty^2 \leq 2^{-7} P\; ,
\label{eq:commb}
\ee
so we can continue
$$
        (\ref{omreso})
        \ge \frac{1}{2}\Omega^{1/2} \Bigg[ \frac{1}{8}  \D_\Omega^4
          -  \frac{1}{16} P \D_\Omega^2 
        +P^2 \Bigg]\Omega^{1/2} 
         \ge \frac{1}{32}\Omega^{1/2}( \D_\Omega^2 + P)^2
         \Omega^{1/2} \; .
$$
 This completes the proof of (\ref{eq:Dome}) and hence (\ref{eq:sqconf}).

\bigskip

For the proof of (\ref{eq:offconf}) we can use the argument above
to reduce the problem to estimating the diagonal element of the
self-adjoint operator
$$
        T:= R
         \Om^{1/2}\D_\Om \Om^{1/2} \varphi^2\Om^{1/2}\D_\Om \Om^{1/2}R \quad
         \mbox{with} \quad R: = {1\over (\Om^{1/2}\D_\Om \Om^{1/2})^2 + P}
$$
viewed on $L^2(\rd s_\Om^2)\otimes\bC^2$, where $\D_\Om$  is self-adjoint.
The resolvent can be written as
$R = \Om^{-1/2} R_1 \Om^{-1/2}$ with
$$
        R_1:= {1\over \D_\Om \Om D_\Om + P\Om^{-1}}
        = \Om^{-1}{1\over \D_\Om^2 - \D_\Om [\D_\Om, \Om] \Om^{-1}
        + P\Om^{-2}}\; ,
$$
and we can expand
$$
        R_1=\Om^{-1}{1\over \D_\Om^2 + P}
        + R_1 \Big[  \D_\Om [\D_\Om, \Om] \Om^{-1}
        - P(\Om^{-2}-1)\Big]{1\over \D_\Om^2+P}\; .
$$
Therefore, using a Schwarz' inequality,  (\ref{omegabound}) and
$0\leq\varphi\leq 1$, we can estimate
\bey
        T &\leq& 2 \Om^{-1/2} R_1 \D_\Om \varphi^2\D_\Om  R_1 \Om^{-1/2}
\label{eq:doms}\\
         &\leq& 4\Om^{-1/2} {1\over \D_\Om^2 + P}
         \Om^{-1} \D_\Om \varphi^2
         \D_\Om \Om^{-1}{1\over \D_\Om^2 + P}\Om^{-1/2}
\nonumber\\
  &&      + 8 \Bigg( \cdots\Bigg)^*
        \Bigg(  \D_\Om R_1\D_\Om [\D_\Om, \Om] \Om^{-1}
         {1\over \D_\Om^2 + P}\Om^{-1/2}\Bigg)
\nonumber\\
&&
        +  8 \Bigg( \cdots\Bigg)^*
        \Bigg(  \D_\Om R_1  P(\Om^{-2}-1)
         {1\over \D_\Om^2 + P}\Om^{-1/2}\Bigg)\; .
\nonumber
\eey
Here we used the shorthand notation $(\cdots )^* A$ for the operator
$A^*A$ where $A$ is a long expression.

In the first term on the right hand side
of (\ref{eq:doms}) we use $\D_\Om \Om^{-1} =
\Om^{-1} \D_\Om + [\D_\Om,\Om^{-1}]$ and (\ref{eq:commb}) to obtain
$$
        {1\over \D_\Om^2 + P}
         \Om^{-1} \D_\Om \varphi^2
         \D_\Om \Om^{-1}{1\over \D_\Om^2 + P}
         \leq   {8\over \D_\Om^2 + P}
         \D_\Om \varphi^2
         \D_\Om {1\over \D_\Om^2 + P}
         + 2^{-8}P  {1\over (\D_\Om^2 + P)^2}\; ,
$$
 and both terms 
 explicitly appear on the right hand side of
(\ref{eq:offconf}). For the other two terms
it is sufficient to show that
\bey
         R_1 \D_\Om^2 R_1  &\leq& 4P^{-1}\; ,
\label{eq:1suff} \\
        \D_\Om R_1 \D_\Om^2 R_1  \D_\Om &\leq&  4\; ,
\label{eq:2suff}
\eey
and then the last two terms in (\ref{eq:doms}) can be estimated
by the second term on the right hand side of (\ref{eq:offconf})
using (\ref{omegabound}), (\ref{omegaderbound}).

For the proof of (\ref{eq:1suff}) and (\ref{eq:2suff})
 we first use $\D_\Om^2 \leq 2 \D_\Om
\Om \D_\Om + 2P\Om^{-1} = 2R_1^{-1}$ to cancel one of the resolvents.
The proof of (\ref{eq:1suff}) is then completed by
estimating the other $R_1$
by $2P^{-1}$. For the proof of (\ref{eq:2suff}) we notice that
$$
        \D_\Om R_1  \D_\Om = \D_\Om {1\over \D_\Om \Om
        \D_\Om + P\Om^{-1}}  \D_\Om \leq  \D_\Om {2\over \D_\Om^2
        + 2P\Om^{-1}}  \D_\Om \leq 2\; .
$$
This completes the proof of Lemma
 \ref{lemma:conf}. $\;\;\;\Box$.

\subsection{Proof of Lemma \ref{lemma:opineq}:
apriori bound on the full resolvent}\label{sec:diamagproof}

Using (\ref{kernelcomp2}) from Lemma \ref{lemma:changenorm} and
since the volume forms $\rd\nu$ and $\rd\mu=\Om^3\rd x$
are comparable at every point,
it is sufficient to prove (\ref{eq:1})--(\ref{eq:3})
in the space $L^2(\rd \mu)=L^2(\Om^3 \rd x)$. In this space
$\cD= \cD_\Om^\a$ and the components of $\bD=\bD^\a_\Om = \bPi_\Om^\a
- \sfrac{i}{2}( \mbox{div}_\Om f_1, \mbox{div}_\Om f_2, \mbox{div}_\Om
f_3)$ 
 are self-adjoint. We recall that $\bPi_\Om^\a$ was given
in (\ref{eq:momvec}) and $\bD$ was defined in general in (\ref{def:D}).
 Throughout the proof we will work in the space $L^2(\Om^3 \rd x)$,
and we adapt the notation $\cD=\cD^\a_\Om$, $\bD=\bD^\a_\Om$  in this section.
 We also recall that $\Pi_j = D_j + id_j$ with
$d_j:=\sfrac{1}{2}\mbox{div}_\Om\; e_j$.

Using Lichnerowicz' formula (\ref{eq:lichD}),
$\sup_x \| \star \beta(x) \| = \sup_x \|\bB(x) \|\leq cb$
and that all geometric terms are bounded 
by (\ref{eq:homtemp}) and
(\ref{eq:homdertemp}), we can
estimate
\be
        \D^2\ge \bD^2 - cb \; .
\label{eq:alul}
\ee
We recall that $\ell=1$,  $b\ge \e^{-2} \ge 1$ and $P = \e^{-5}\ge 1$.

For the proof of (\ref{eq:1}) 
 we start with a Schwarz' inequality
$$
        \Pi_j^* \Big( \res\Big)^2 \Pi_j 
        \leq 2 D_j \Big( \res\Big)^2 D_j + 2\sup |d_j|
$$
and use that $|d_j|\leq c$. We estimate
one of the resolvents trivially and use (\ref{eq:alul})
$$
        D_j \Big( \res\Big)^2 D_j 
        \leq D_j \frac{b}{\D^2 + Pb} D_j
        \leq D_j \frac{b}{\bD^2 - cb + Pb} D_j
        \leq D_j \frac{b}{D_j^2  + Pb/2} D_j \leq b
$$
for sufficiently small $\e$. This completes the proof of
(\ref{eq:1}).

The proof of (\ref{eq:1.5}) is identical just we estimate 
$(\cD/(\cD^2+P))^2$ by $(\cD^2+P)^{-1}$.

\medskip

For the proof of (\ref{eq:2}) we first compute
 $\Pi_j\Pi_k = D_jD_k + i( D_j d_k + D_k d_j) -
(\partial_{e_k}d_k) +d_jd_k$. We use a Schwarz' inequality,
the estimate (\ref{eq:1})
and the boundedness
of $d_j$'s together with their derivatives, we obtain
\be
        \Pi_k^*\Pi_j^* \Big( \res\Big)^2 \Pi_j\Pi_k
        \leq 2D_kD_j \Big( \res\Big)^2D_jD_k
        + cb\; .
\label{eq:tocomo}
\ee
We apply Lemma \ref{lemma:XY} to estimate the resolvent square,
using that $\D^2$ and $\bD^2$ differ only by an operator
bounded by $cb\le Pb/2$ if $\e$ is sufficiently small:
\be
        \Big( \res\Big)^2 \leq \frac{b^2}{ (\D^2 + Pb)^2}
        \leq  \frac{4b^2}{ (\bD^2)^2 + (Pb)^2/4}\; .
\label{eq:ressq}
\ee
We expand $ (\bD^2)^2 = \sum_j D_j^4+ \sum_{j<k} (D_j^2D_k^2+ D_k^2D_j^2)$ 
and use the following commutator identity for $A, B$ self-adjoint operators
$$
        A^2B^2 + B^2A^2 = \Bigg[ AB^2A + AB[A,B] + [B,A]BA 
        +\sfrac{1}{2}\Big( [A,[A,B]]B + B[[B,A],A]\Big)\Bigg]
        + \Bigg[ A\leftrightarrow B\Bigg]
$$
(the second square bracket contains the same expression as the first
one with $A$ and $B$ interchanged). After several Schwarz's inequalities,
we obtain
\bey
        A^2B^2 + B^2A^2 &\ge& \sfrac{1}{2}\Big( AB^2A + BA^2B - A^2 - B^2\Big)
        - c [A,B][B,A]
\label{eq:commu}\\
        && - c[A,[A,B]][[B,A],A] - c [B,[B,A]][[A,B],B] \; .
        \nonumber
\eey
Using the formula given in Theorem 2.12 \cite{ES-III}
for the curvature of the covariant derivative $\nabla=\nabla^{\a,\Om}$
 we obtain
\bey
        [D_j, D_k] &=& -\nabla_{[e_j, e_k]} - \partial_{e_j}d_k + 
        \partial_{e_k}d_j
        \nonumber\\
        &&- \frac{1}{4}\sum_{a,b=1}^3 (e_a, \cR(e_j, e_k)e_b) \sigma^a\sigma^b
        + i \beta(e_j, e_k) \; ,\nonumber
\eey
where $\cR$ is the Riemannian curvature, $j,k =1,2,3$.
In short, we can write
$$
        [D_j, D_k] = \sum_{a=1}^3  U_{jk}^a D_a + W_{jk}\; ,
$$
where $U_{jk}^k, W_{jk}$ are 2 by 2 matrix valued functions
 with $\| U_{jk}^a\|_\infty\leq c$, $ \|\nabla U_{jk}^a\|_\infty \leq c$
and $\|W_{jk}\|_\infty\leq cb$, $\|\nabla W_{jk}\|_\infty\leq cb$
using (\ref{eq:homdertemp}). These estimates guarantee
bounds on the double commutators as well.

{F}rom these estimates and (\ref{eq:commu}) it follows that
$$
        D_j^2D_k^2+ D_k^2D_j^2 \ge
        \sfrac{1}{2}\Big( D_jD_k^2D_j + D_kD_j^2D_k\Big)
        - cb \bD^2 - cb^2 \; .
$$
Therefore
\bey
        (\bD^2)^2+(Pb)^2/4 &\ge& \sfrac{1}{2}\Bigg[ \sum_j D_j^4
        +\sum_{j<k} \Big( D_jD_k^2D_j + D_kD_j^2D_k\Big)\Bigg]
\nonumber\\
        &&+ \sum_j \Big(\sfrac{1}{2} D_j^4 -cb D_j^2\Big) + (Pb)^2/4-cb^2 \; .
\label{eq:sqbel}
\eey
The second line is bigger than $(Pb)^2/8$
if $\e$ is sufficiently small ($P=\e^{-5}$).
Every term in the first line is nonnegative, so we can complete
the estimate (\ref{eq:tocomo}) using (\ref{eq:ressq}) and (\ref{eq:sqbel})
$$
        D_jD_k\frac{4b^2}{ (\bD^2)^2 + (Pb)^2/4}D_kD_j
         \leq  D_jD_k \frac{8b^2}{ D_kD_j^2D_k + (Pb)^2/8}D_k D_j
        \leq 8b^2 \; .
$$
This completes the proof of (\ref{eq:2}).

\medskip

The proof of (\ref{eq:5}) and (\ref{eq:3}) are 
 straightforward from (\ref{eq:dec1}),
(\ref{eq:1}), (\ref{eq:1.5}) and (\ref{eq:mkest}). Finally (\ref{eq:11})
is proven in the same way as (\ref{eq:1}) but now directly
on the space $L^2(\rd\xi)\otimes\bC^2$. $\;\;\Box$

\subsection{Proof of Lemma \ref{lemma:expl}:
 Estimates on the resolvent with a constant field}
\label{sec:constres}

The proof of (\ref{eq:expl1})--(\ref{eq:expl12}) may be
done by straightforward explicit calculations since
the magnetic Schr\"odinger operator with a constant
field is exactly solvable. We show below
how to obtain these estimates in a reasonably short way.

We work in the $\xi$ coordinate system, and use that $\xi_\perp(u)=0$.
Because of translation invariance in the third direction, we can assume 
$\xi(u)=0$, so in the $\xi$ coordinates
we need to estimate the operator kernels at $(0,0)$.

We recall the decompositions (\ref{eq:decom})--(\ref{eq:dperpprop})
and let 
$$
        -\tdel: = \tPi_1^2 + \tPi_2^2
        = (-i\partial_1 - \sfrac{b}{2}\xi_2)^2+
        (-i\partial_2 + \sfrac{b}{2}\xi_1)^2
$$
be the two dimensional magnetic Laplacian that commutes with $\tPi_3$.
For simplicity, we denoted $\partial_j:=\partial_{\xi_j}$.
By Lichnerowicz' formula (\ref{eq:lich}), $\tcD^2 = -\tdel + \tPi_3^2
+\sigma^3b$ and recall that $\tPi_3 = -i\partial_3$.

The key idea is that the heat kernel of $\tdel$
has a closed form (see, e.g., Chapter 15 in \cite{S79})
\be
        e^{t\tdel} (\xi_\perp, \zeta_\perp)
        = {b\over 4\pi \sinh (bt)}\exp{\Bigg[
        - {b\coth (bt)\over 4}(\xi_\perp-\zeta_\perp)^2 - {ib\over 2}
        (\xi_2\zeta_1 - \xi_1\zeta_2)\Bigg]}\; .
\label{eq:mehler}\ee
Then the resolvent can be expressed as
\be
        \tres = \int_0^\infty e^{-t(P+\sigma^3b)} e^{t\tdel}
        e^{t\partial_3^2} \; \rd t\;.
\label{eq:heatint}
\ee
We define the following norm on $\bR^3$
$$
        \tri \xi \tri : =( b\xi_\perp^2 + P\xi_3^2)^{1/2}\; .
$$
\begin{lemma}\label{lemma:resker}
Let  $P\leq c b$, then the following bounds hold
\bey
        \Big\| \tres (\xi, \zeta)\Big\| &\leq &
        cbP^{-1/2}\; \frac{e^{-c\tri \xi-\zeta\tri} }{\tri \xi-\zeta\tri}\; ,
\label{eq:constresest}\\
        \Big\| \ttres (\xi, \zeta)\Big\| &\leq &
        cb\; \frac{e^{-c\tri \xi-\zeta\tri} }{\tri \xi-\zeta\tri^2}\; ,
\label{eq:constresest2}\\
        \Big\| \tpres (\xi, \zeta)\Big\|  &\leq &
        cb^{3/2}P^{-1/2}\; 
        \frac{e^{-c\tri \xi-\zeta\tri} }{\tri \xi-\zeta\tri^2}\; , \qquad
        j=1,2
\label{eq:constresest3} \\
        \Big\| \tdres (\xi, \zeta)\Big\|  &\leq &
        c \; e^{-c\tri \xi-\zeta\tri} \Big[ \frac{1}{|\xi-\zeta|^2} + b\Big]
        \; ,
\label{eq:constresest4}
\eey
where $\| \cdot \|$ refer to the 2 by 2 matrix norm
of the operator kernel as a function from $\bR^3\times\bR^3$
into the set of 2 by 2 matrices.
\end{lemma}

{\it Remark.} If $b\leq cP$, then the same estimates 
(\ref{eq:constresest})--(\ref {eq:constresest4}) hold with $b$
replaced
by $P$ everywhere, including in the definition of $\tri\cdot\tri$.

\bigskip

\noindent
{\it Proof.} 
{F}rom (\ref{eq:mehler}) and (\ref{eq:heatint}) we estimate
$$
        \Big\| \tres (\xi, \zeta)\Big\| 
        \leq cb \int_0^\infty \frac{e^{-tP+tb}}{\sqrt{t}\sinh (bt)}
        \exp\Big( - \frac{1}{4t}\Big[ bt\coth (bt)(\xi_\perp-\zeta_\perp)^2
        + (\xi_3-\zeta_3)^2\Big]\Big) \; \rd t \; .
$$
We split the integration into two regimes: $bt\leq 1$ and $bt\ge 1$.
In the first regime we use $bt\coth(bt)\ge 1$, $\sin(bt)\ge bt$.
In the second regime we estimate $\coth(bt)\ge 1$ and 
$\sinh(bt)\ge \sfrac{1}{4}e^{bt}$. We obtain
\bey
        \Big\| \tres (\xi, \zeta)\Big\| 
        &\leq& c \Bigg[ \int_0^{1/b}\frac{e^{-tP}}{t^{3/2}} \;
        \exp{\Big(-\frac{(\xi-\zeta)^2}{4t}\Big)} \rd t \nonumber\cr
        &&+ b\; \exp{\Big(-\frac{b}{4}(\xi_\perp-\zeta_\perp)^2\Big)}
        \int_{1/b}^\infty \frac{e^{-tP}}{t^{1/2}}
        \exp{\Big(-\frac{(\xi_3-\zeta_3)^2}{4t}\Big)} \rd t\Bigg]\nonumber\cr
        &\leq& c(P^{1/2}+bP^{-1/2})\;
         \frac{e^{-c\tri \xi-\zeta\tri} }{\tri \xi-\zeta\tri} \; , 
        \nonumber
\eey
after extending both integrations over $(0,\infty)$
and using the resolvent kernels of the one and three dimensional
free Laplacians. 
The proofs of (\ref{eq:constresest2})--(\ref{eq:constresest3})
are similar and left to the reader.

For the proof of (\ref{eq:constresest4}), explicit calculation
and trivial estimates yield
\bey
        \Big\| \tdres (\xi, \zeta)\Big\|
        &\leq& c \int_0^{\infty}
        \frac{e^{-t(P-b)}b^2t(\coth(bt)-1)
        |\xi_\perp-\zeta_\perp|}{t^{3/2} \sinh (bt)}\nonumber\cr
         &&\times \exp{ \Big( - \frac{b\coth(bt)}{4}|\xi_\perp-\zeta_\perp|^2
        - \frac{(\xi_3-\zeta_3)^2}{4t} \Big)} \; \rd t 
        \nonumber\cr
        &\leq& c \int_0^{1/b}
        \frac{e^{-t(P+b)}|\xi_\perp-\zeta_\perp|}{t^{5/2}}
         \; \exp{ \Big( - \frac{(\xi-\zeta)^2}{4t} \Big)} \; \rd t 
        \nonumber \cr
        && +  cb \; e^{-cb|\xi_\perp-\zeta_\perp|^2} \int_{1/b}^\infty 
        \frac{e^{-tP}|\xi_\perp-\zeta_\perp|}{t^{3/2}}
         \; \exp{ \Big( - \frac{(\xi_3-\zeta_3)^2}{4t} \Big)} 
        \; \rd t \nonumber\cr
        &\leq&  c  \int_0^{1/b} 
        \frac{e^{-t(P+b)}}{t^2}
         \; \exp{ \Big( - c\frac{(\xi-\zeta)^2}{t} \Big)} \; \rd t 
        \nonumber \cr
        && +  cb^{1/2} \; e^{-cb| \xi_\perp-\zeta_\perp|^2} \int_{1/b}^\infty 
        \frac{e^{-tP}}{t^{3/2}}
         \; \exp{ \Big( - \frac{(\xi_3-\zeta_3)^2}{4t} \Big)} 
        \; \rd t \nonumber\cr
        & \leq&  c \; e^{-\sqrt{b+P}| \xi-\zeta|} \int_0^\infty
        \frac{1}{t^{2}} 
         \; \exp{ \Big( - c\frac{(\xi-\zeta)^2}{t} \Big)} \; \rd t 
        \nonumber \cr
        && +  cb \; 
        e^{-cb| \xi_\perp-\zeta_\perp|^2-c\sqrt{P} |\xi_3-\zeta_3|}
        \nonumber \cr 
        &\leq&  c \; e^{-c\tri \xi-\zeta\tri} \Big[\frac{1}{
        |\xi-\zeta|^2} + b\Big] \; . \qquad\qquad \;\;\;\Box\nonumber
\eey

\medskip

With the estimates of Lemma \ref{lemma:resker}
at hand, the proof of Lemma \ref{lemma:expl}
is straightforward. For example, the proof of (\ref{eq:expl1}) is
as follows
$$
        \tr \; \frac{1}{(\tcD^2 + P)^2}(0,0)
        =\int_{\bR^3} \Big\|\tres (0,\xi)\Big\|^2 \rd \xi
        \leq cb^2P^{-1}\int_{\bR^3}
         \frac{e^{-c\tri \xi\tri} }{\tri \xi\tri^2} \rd \xi
        \leq cbP^{-3/2}
$$
after  a change of variables. The other inequalities are proved
similarly. $\;\;\Box$

\newpage
\noindent Current addresses of the authors:

\medskip

\noindent  L\'aszl\'o Erd\H os \\
Mathematisches Insititut, LMU\\
Theresienstrasse 39, D-80333 Munich, Germany\\
 lerdos@mathematik.uni-muenchen.de

\bigskip

\noindent Jan Philip Solovej\\
Department of Mathematics, University of Copenhagen
\\ Universitetsparken 5, DK-2100, Copenhagen, Denmark
\\
 solovej@math.ku.dk


\begin{thebibliography}{hhhhhh}

\bibitem[AMN]{AMN}  C. Adam, B. Muratori and C. Nash:
{\sl Multiple zero modes of the Dirac operator in three dimensions},
Phys. Rev. D (3) {\bf 62}, no. 8, 085026 (2000)

\bibitem[BE]{BE} A. Balinsky, W. D. Evans: {\sl
On the zero modes of Pauli operators}, J. Funct. Anal. 
{\bf 179}(1), 120-135 (2001)

\bibitem[BFFGS]{BFFGS} L. Bugliaro, C. Fefferman, J. Fr\"ohlich,
G. M. Graf and J. Stubbe: {\sl A Lieb-Thirring bound for a magnetic
Pauli Hamiltonian}, Commum. Math. Phys. {\bf 187}, 567--582 (1997)


\bibitem[BFrG]{BFrG} L. Bugliaro, J. Fr\"ohlich, and G. M. Graf:
{\sl Stability of quantum electrodynamics with nonrelativistic matter},
Phys. Rev. Lett. {\bf 77}, 3494--3497 (1996)

\bibitem[BFG]{BFG} L. Bugliaro, C. Fefferman and G. M. Graf:
{\sl A Lieb-Thirring bound for a magnetic
Pauli Hamiltonian, II}, Rev. Mat. Iberoamericana, {\bf 15}, 593-619
(1999)

\bibitem[El-1]{El-1} D. Elton: {\sl New examples of zero modes},
J. Phys. A {\bf 33} (41), 7297-7303, (2000)

\bibitem[El-2]{El-2}  D. Elton: {\sl The local structure of
zero mode producing magnetic potentials}. Commun. Math. Phys. 
{\bf 229}, 121-139 (2002).

\bibitem[E-93]{E-93} L. Erd\H os: 
{\sl Ground state density of the Pauli operator 
in the large field limit.}  Lett. Math. Phys. {\bf 29}, 219-240 (1993)


\bibitem[E-1995]{E-1995} L. Erd\H os: {\sl Magnetic Lieb-Thirring
inequalities. \/}  Commun. Math. Phys. {\bf 170}, 629--668 (1995)

\bibitem[ES-I]{ES-I}  L. Erd{\H o}s and J. P. Solovej: {\it Semiclassical
eigenvalue estimates for the Pauli operator with strong
non-homogeneous magnetic fields. I. Non-asymptotic Lieb-Thirring
type estimate.}  Duke J. Math. {\bf 96}, 127-173 (1999)

\bibitem[ES-II]{ES-II} L. Erd{\H o}s and J. P. Solovej: {\it Semiclassical
eigenvalue estimates for the Pauli operator with strong
non-homogeneous magnetic fields. II. Leading order asymptotic estimates.}
Commun. Math. Phys. {\bf 188}, 599--656 (1997)

\bibitem[ES-III]{ES-III} L. Erd{\H o}s and J. P. Solovej:
{\it The kernel of Dirac operators on $S^3$ and $\bR^3$}.
Rev. Math. Phys. {\bf 13} No. 10, 1247-1280 (2001) 

\bibitem[ES-IV]{ES-IV} L. Erd{\H o}s and J. P. Solovej:
{\it Magnetic Lieb-Thirring inequalities with optimal dependence on the field
strength}. Accepted to J. Statis. Phys. (2003).
Available at {\tt http://xxx.lanl.gov/pdf/math-ph/0306066}.




\bibitem[HNW]{HNW} B. Helffer, J. Nourrigat and X. P. Wang,
{\em Sur le spectre de l'equation de Dirac (dans ${\bf R}^2$ ou
${\bf R}^3$) avec champs magn\'etique. \/} Ann. scient. \'Ec. Norm.
Sup. $4^{e}$ serie t. 22 (1989), 515-533.


\bibitem[LLS]{LLS} E. H. Lieb, M. Loss and J. P. Solovej: {\sl 
Stability of Matter in Magnetic Fields}, Phys. Rev. Lett. {\bf 75},
 985--989 (1995)


\bibitem[LSY-I]{LSY-I} E. H. Lieb, J. P. Solovej and J. Yngvason:
{\sl Asymptotics of heavy atoms in high magnetic fields: I. Lowest
Landau band region}, Commun. Pure  Appl. Math. {\bf 47},
513--591 (1994)

\bibitem[LSY-II]{LSY-II} E. H. Lieb, J. P. Solovej and J. Yngvason:
{\sl Asymptotics of heavy atoms in high magnetic fields: II. Semiclassical
regions. \/}  Commun. Math. Phys. {\bf 161}, 77--124 (1994)

\bibitem[LT1]{LT1}  E. H. Lieb, W. Thirring: {\sl Inequalities for moments
of the eigenvalues of the Schr\"odinger Hamiltonian and their relation
to Sobolev inequalities.} In: Studies in Mathematical Physics
(E. Lieb, B. Simon, A. Wightman eds.) Princeton University Press,
269--330 (1975)



\bibitem[LY]{LY} M. Loss and H.-T. Yau: {\sl Stability of Coulomb
systems with magnetic fields: III. Zero energy bound states of
the Pauli operator. \/} Commun. Math. Phys. {\bf 104}, 283--290 (1986)

\bibitem[S79]{S79} B. Simon, {\em Functional Integration and Quantum
Physics. \/} Academic Press, New York, 1979.


\bibitem[Sh]{Sh} Z. Shen: {\sl On the moments of negative eigenvalues
for the Pauli operator}, J. Diff. Eq. {\bf 149}, 292-327 (1998)
and {\bf 151}, 420-455 (1999).

\bibitem[Sob-86]{Sob-86} A. Sobolev, {\em Asymptotic
 behavior of the energy levels of a quantum
particle in a homogeneous magnetic field, perturbed by a decreasing
electric field. \/} J. Sov. Math. {\bf 35} (1986), 2201--2212.



\bibitem[Sob-96]{Sob-96}
 A. Sobolev: {\sl On the Lieb-Thirring estimates
for the Pauli operator, \/} Duke Math. J. {\bf 82}, 607--635 (1996)


\bibitem[Sob-97]{Sob-97}
 A. Sobolev: {\sl  Lieb-Thirring inequalities for the Pauli operator
in three dimensions}, IMA Vol. Math. Appl. {\bf 95}, 155--188  (1997)

\bibitem[Sob-98]{Sob-98} A. Sobolev: {\sl Quasiclassical asymptotics for the
Pauli operator}, Commun. Math. Phys. {\bf 194},  109--134 (1998)

\bibitem[Sol]{Sol} S. N. Solnyshkin, 
{\em The asymptotic behavior of the energy
of bound states of the Schr\"odinger operator in the presence of electric
and magnetic fields. \/} Probl. Mat. Fiz. {\bf 10} (1982), 266--278.



\end{thebibliography}
\end{document}